\documentclass[letterpaper,twocolumn,10pt]{article}

\usepackage{hegroup}
\usepackage{amsmath}
\usepackage{amssymb}
\usepackage{hyperref}
\usepackage{cleveref}
\usepackage{xspace}
\usepackage{bbm}
\usepackage{subcaption}
\usepackage{graphicx}
\usepackage{booktabs}      % \toprule \midrule \bottomrule
\usepackage{multirow}      % \multirow
\usepackage{makecell}      % \makecell for multi-line headers
\usepackage[table]{xcolor} % \textcolor
\usepackage{arydshln}      % \hdashline / \cdashline
\usepackage{pifont}        % \ding symbols (check/cross)
\usepackage{array}
\usepackage{bm}
\usepackage{tikz}
\usepackage[most]{tcolorbox}
\newlength{\adrad}
\DeclareRobustCommand{\attackonly}{%
\tikz[baseline=-0.6ex]{
  \fill (0,0) -- (0,\adrad) arc (90:270:\adrad) -- cycle;
  \draw (0,0) circle (\adrad);
}}

\DeclareRobustCommand{\defenseonly}{%
\tikz[baseline=-0.6ex]{
  \fill (0,0) -- (0,\adrad) arc (90:-90:\adrad) -- cycle;
  \draw (0,0) circle (\adrad);
}}

\DeclareRobustCommand{\attackdefense}{%
\tikz[baseline=-0.6ex]{
  \fill (0,0) circle (\adrad);
  \draw (0,0) circle (\adrad);
}}

\DeclareRobustCommand{\adnone}{%
\tikz[baseline=-0.6ex]{\draw (0,0) circle (\adrad);}
}
\newcommand{\cmark}{\textcolor{mygreen}{\ding{51}}}
\newcommand{\xmark}{\textcolor{myred}{\ding{55}}}

\newcolumntype{L}[1]{>{\raggedright\arraybackslash}m{#1}}
\newcolumntype{C}[1]{>{\centering\arraybackslash}m{#1}}
\newcommand{\mypara}[1]{\noindent{\bf {#1}.}\xspace}
\newcommand{\bench}{\textit{SADBench}\xspace}

\begin{document}

\definecolor{w1}{HTML}{E76F73}
\definecolor{w2}{HTML}{EC868A}
\definecolor{w3}{HTML}{F09EA1}
\definecolor{w4}{HTML}{F4B5B8}
\definecolor{w5}{HTML}{F7CACC}
\definecolor{w6}{HTML}{F9DADD}
\definecolor{w7}{HTML}{FBE8EA}
\definecolor{w8}{HTML}{FCF1F2}
\definecolor{w9}{HTML}{FEF7F7}
\definecolor{w10}{HTML}{FFFBFB}

\definecolor{c1}{HTML}{5B8FD1}
\definecolor{c2}{HTML}{73A0D9}
\definecolor{c3}{HTML}{8CB2E2}
\definecolor{c4}{HTML}{A6C4EA}
\definecolor{c5}{HTML}{BED5F1}
\definecolor{c6}{HTML}{D3E3F7}
\definecolor{c7}{HTML}{E3ECFA}
\definecolor{c8}{HTML}{EEF4FC}
\definecolor{c9}{HTML}{F6FAFE}
\definecolor{c10}{HTML}{FBFDFF}

\definecolor{mygreen}{RGB}{0,153,0}
\definecolor{myred}{RGB}{200,0,0}
\definecolor{myyellow}{RGB}{245,180,0}

\title{Stego Battlefield: Evaluating Image Steganography Attacks and Steganalysis Defenses}
\date{}

\author{
Zhen Sun\textsuperscript{1,2}\thanks{Equal contribution.},
Zongmin Zhang\textsuperscript{2}\footnotemark[1],
Leyi Sheng\textsuperscript{1},
Yule Liu\textsuperscript{2},
Yifan Liao\textsuperscript{2}, \\
Ke Li\textsuperscript{2},
Xinhu Zheng\textsuperscript{2},
Jiaheng Wei\textsuperscript{2},
Wenyuan Yang\textsuperscript{3},
Xinlei He\textsuperscript{1}\thanks{Corresponding author (\href{mailto:xinlei.he@whu.edu.cn}{xinlei.he@whu.edu.cn}).}
\\
\textsuperscript{1}\textit{Wuhan University}
\\
\textsuperscript{2}\textit{The Hong Kong University of Science and Technology (Guangzhou)}
\\
\textsuperscript{3}\textit{Sun Yat-sen University}
}

\maketitle

%-----------------------------
\begin{abstract}
%-----------------------------

Image steganography is widely used to protect user privacy and enable covert communication.
However, it can also be abused by the adversary as a covert channel to bypass content moderation, disseminate harmful semantics, and even hide malicious instructions in images to elicit dangerous outputs from large models, posing a practical security risk that continues to evolve.
To address the lack of a unified and systematic evaluation framework, we propose \bench, a systematic benchmark that assesses the adversary's ability to inject harmful secrets via steganography and the defender's ability to detect such threats through steganalysis.
Crucially, \bench comprises $4$ core tasks, namely steganography attack capability evaluation, steganalysis defense capability evaluation, efficiency evaluation, and transferability evaluation.
It evaluates both image-payload and text-payload steganography across diverse cover distributions, utilizing harmful visual semantics and toxic instructions to simulate malicious attacks.
Across a broad set of attacks and detectors, \bench reveals that (i) INN and autoencoder-based methods demonstrate superior stability compared to other architectures, (ii) in-domain detection is near-perfect and cheaper than generation, (iii) a critical asymmetry exists in transferability where attacks robustly generalize to new distributions while detectors fail to adapt, and (iv) real-world threats persist on social media, where payloads either survive minimal compression or effectively adapt to aggressive compression via simulated training.
Overall, \bench establishes a systematic, reproducible, and extensible framework to quantify risks, paving the way for measurable and security-driven advancements in steganography defense.
\end{abstract}

%-----------------------------
\section{Introduction}
%-----------------------------

Steganography is an information hiding technique that embeds secret data into a cover medium with minimal perceptible changes to avoid detection~\cite{DBLP:journals/mta/DalalJ21}.
As an important branch of this field, image steganography uses images as the cover medium for information hiding, which has been used in a range of scenarios for information protection and secure communication, including covert messaging in military and intelligence settings~\cite{jamil1999steganography,mercuri2004many}, hiding patient information in medical applications~\cite{balu2019secure}, and copyright protection through watermarking for multimedia content~\cite{cox2002digital}.
Conventional image steganography typically embeds information in spatial or transform domains, exemplified by LSB~\cite{DBLP:journals/pr/WangLL01,DBLP:journals/pr/ChangHC03,DBLP:conf/awcc/WuLC04} and DCT-based methods~\cite{DBLP:conf/ih/Westfeld01,patel2012steganography}. 
These methods are often limited in capacity, failing to meet modern demands for high payload and stealthiness.
Deep learning-based image steganography overcomes these limitations and has emerged as a major research focus~\cite{DBLP:journals/csur/KombrinkGW25}.

\begin{figure}
    \centering
    \includegraphics[width=0.75\linewidth]{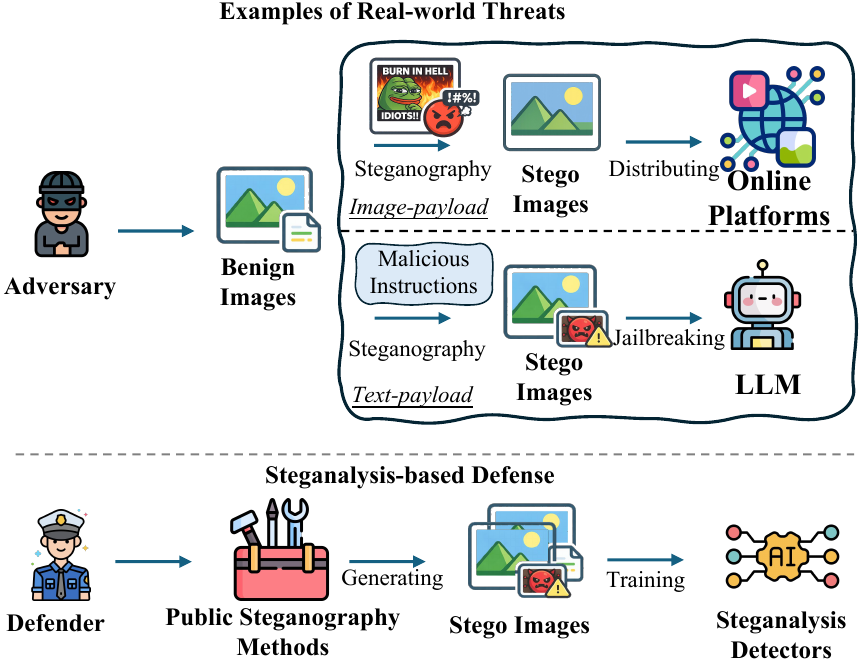}
    \caption{Overview of real-world image-cover steganography misuse scenarios and the steganalysis-based defense.}
    \label{fig:overview}
\end{figure}

However, these advancements also introduce dual-use risks. 
Recent cases show that image steganography can be misused to transmit malicious instructions for remote control~\cite{trendmicro_malicious_memes_2018}, bypass content moderation for data theft~\cite{chaganti2021stegomalware}, or hide harmful text in benign-looking images to enable multimodal jailbreaks of vision-capable LLMs~\cite{li2025odysseus}. 
These threats suggest that evaluating steganography solely by distortion, capacity, or random-bit recovery is insufficient: we should also assess whether safety-relevant payloads can be recovered, detected, and transferred across realistic deployment shifts.

To this end, we introduce \bench, the first attack-defense benchmark for image steganography with safety-relevant image and text payloads.
\bench jointly evaluates steganography attack capability, steganalysis defense capability, efficiency, and cross-distribution transferability under a unified protocol.
\textit{Steganography Attack Capability:} From the adversary's perspective, we evaluate $7$ image-payload and $5$ text-payload methods, assessing both visual stealthiness and payload recoverability.
\textit{Steganalysis Defense Capability:} From the defender's perspective, we evaluate the binary detection performance of $8$ universal steganalysis detectors.
\textit{Efficiency:} We compare the computational overhead and deployment cost of both attack and defense methods during the training and inference stages.
\textit{Transferability:} We analyze the robustness of both sides against various distribution shifts, including changes in cover domains, secret payloads, and steganographic methods.
Additionally, we conduct a real-world transmission test to assess the payload recoverability of stego images when disseminated across different social media platforms.

\mypara{Main Findings}
\bench reveals several critical insights into the real-world steganography arms race:
\textbf{(i) Unmasking the True Threat:} To build targeted defenses, we must identify attacks that pose practical risks.
We find that achieving visual stealth is relatively trivial, but ensuring the semantic survival of the payload is a major bottleneck for adversaries.
While the generative diffusion-based method~\cite{CroSS2023} struggles to maintain payload integrity, specific structural designs such as INN-based~\cite{DeepMIH2023} and autoencoder-based methods~\cite{stegformer2024} effectively enable high-fidelity recovery.
Defenders should strategically prioritize these high-recoverability architectures.
\textbf{(ii) A Critical Asymmetry in Transferability:} A stark contrast exists between attack and defense robustness under distribution shifts.
On the attack side, steganography (particularly image-payload methods) generalizes robustly to new cover distributions, though text payloads remain somewhat vulnerable to secret-domain shifts due to bit-error accumulation.
Conversely, the defender's transferability is alarmingly poor. Detectors heavily overfit to domain-specific traces, leading to catastrophic degradation when facing unseen datasets, cross-modality payloads, or zero-day steganographic methods.  
\textbf{(iii) The Cost Battlefield:} We uncover a massive computational mismatch. Generating high-fidelity stego images requires exorbitant training and inference costs (e.g., up to 153 hours for training), whereas effective detection takes mere milliseconds.
Consequently, the defender's true bottleneck is not computational efficiency, but acquiring representative training distributions to counter unknown threats.  
\textbf{(iv) The Social Media Barricade:} Real-world platform processing (e.g., aggressive image compression) acts as a natural barrier that can effectively destroy hidden payloads.
Nevertheless, this threat is evolving: adversaries can actively bypass these natural defenses by simulating the platform's processing pipeline during training, allowing toxic payloads to adapt and survive.

In summary, we make the following contributions:
\begin{itemize}
    \item We propose \bench, a benchmark and code framework for evaluating image-cover steganography misuse and steganalysis defenses. 
    \bench focuses on two safety-relevant payload settings: harmful image payloads and harmful text instructions hidden in image covers.
    \item We conduct a common-protocol evaluation of $7$ image-payload steganography methods, $5$ text-payload steganography methods, and $8$ steganalysis detectors on two cover-image datasets.
    The evaluation covers attack capability, defense capability, efficiency, and transferability, and further includes selected platform-transmission tests to assess payload recoverability under platform-side image processing.
    \item We identify key attack-defense gaps in this scoped setting, including the mismatch between visual fidelity and payload recovery, the cost-performance asymmetry between stego generation and steganalysis detection, and the transferability gap of current defenses.
    Our results show that detectors can degrade under cover-dataset and steganographic-method shifts, while platform-side compression can substantially affect payload recovery; simulated channel training partially mitigates this degradation in the evaluated setting.
\end{itemize}

%-----------------------------
\section{Preliminary and Related Work}
\label{sec:related}
%-----------------------------

%-----------------------------
\subsection{Real World Threats of Steganography}
\label{sec:threat}
%-----------------------------
Steganography was originally developed to protect information security~\cite{DBLP:journals/mta/DalalJ21}, with legitimate applications ranging from covert military communications~\cite{jamil1999steganography,mercuri2004many} and medical data privacy~\cite{balu2019secure} to digital watermarking for copyright protection~\cite{cox2002digital}.
However, adversaries increasingly exploit steganography to establish ``covert channels'' for malicious activities~\cite{chaganti2021stegomalware}.
In Advanced Persistent Threat (APT) campaigns, the adversaries embed malicious payloads within seemingly benign images hosted on cloud services to disguise communication as normal image access, effectively evading network traffic auditing~\cite{scarcruft_bluetooth_harvester_2019,operation_endtrade_tick_2019}.
Similarly, malware can scrape meme images posted on social media (e.g., Twitter) to extract hidden command-and-control (C2) instructions~\cite{trendmicro_malicious_memes_2018}.
Insiders also leverage these techniques to embed confidential business documents into ordinary photos, bypassing Data Loss Prevention (DLP) systems during data exfiltration~\cite{usdoj_zheng_et_al_indictment_2019}.
Beyond traditional malware, emerging threats demonstrate that image steganography can even be weaponized against modern AI systems.
Recently, Li et al.~\cite{li2025odysseus} successfully jailbroke Large Language Models (LLMs) by encoding malicious queries (e.g., ``how to make a virus'') into innocuous images. By incorporating robust error correction, they enabled the LLM to reliably recover and execute the hidden harmful instructions.
These escalating realities severely complicate security detection and digital forensics~\cite{DBLP:journals/csur/KombrinkGW25}.
Consequently, we adopt an explicit attack-defense perspective to evaluate existing image steganography (adversary) and steganalysis (defender) methods, providing a unified benchmark to measure their effectiveness and robustness under realistic threat scenarios.

%-----------------------------
\subsection{Steganography}
%-----------------------------
Image steganography embeds a secret message $m$ into a cover image $x$ to produce a visually indistinguishable stego image $y$, from which a receiver can reliably extract $\hat{m}$. 
Depending on the payload modality, existing methods fall into two categories: image-payload (image-in-image) and text-payload (text-in-image) steganography.

\mypara{Image-payload Steganography}
Image-payload steganography handles high-capacity visual data and can be broadly categorized into three deep-learning architectures: autoencoders, invertible neural networks (INNs), and diffusion models~\cite{DBLP:journals/csur/KombrinkGW25}.
Early methods, pioneered by Baluja~\cite{ds2017}, utilize autoencoder architectures with separate hiding and revealing networks~\cite{DBLP:journals/corr/abs-2303-13713,UDH2020,stegformer2024}.
Conversely, INN-based approaches~\cite{HiNet2021,DeepMIH2023,PRIS2024} formulate concealment and extraction as inverse problems using shared parameters, minimizing information loss to support higher capacity.
More recently, diffusion-based methods like CRoSS~\cite{CroSS2023} introduce a coverless paradigm. Instead of modifying an existing image, they generate realistic stego images from scratch conditioned on semantic and secret inputs, prioritizing stealthiness and controllability. 
Our benchmark systematically evaluates the trade-offs among these three structural paradigms.

\mypara{Text-payload Steganography}
Text-payload steganography typically encodes text as binary bitstreams, focusing heavily on robust message recovery under channel perturbations~\cite{DBLP:journals/csur/KombrinkGW25}.
End-to-end learning pipelines are prevalent in this domain. For instance, HiDDeN~\cite{HiDDeN2018} utilizes an Encoder-Noise-Decoder architecture to simulate and resist transmission distortions, while SteganoGAN~\cite{SteganoGAN2019} incorporates adversarial training to align the stego image distribution with natural images for enhanced stealthiness.
Beyond these standard frameworks, researchers have also explored normalizing flows (e.g., RMSteg~\cite{RMSteg2025}) and training-free optimization methods (e.g., FNNS~\cite{FNNS2022}). 
Together, these diverse lines of work offer different trade-offs across robustness, capacity, and computational overhead, which we evaluate under our unified framework.

%-----------------------------
\subsection{Steganalysis}
\label{sec:steganalysis}
%-----------------------------
Following prior surveys, we primarily focus on detectors that can generalize across different steganographic methods, commonly referred to as universal steganalysis detectors~\cite{DBLP:journals/csur/KombrinkGW25}.
Since deep learning was introduced into image steganalysis, researchers have quickly realized that generic image-classification backbones are not well-suited for detecting extremely weak steganographic signals.
Instead, architectures must be structured to capture the statistical characteristics of steganographic noise.
Xu et al.~\cite{xunet2016} first propose XuNet, which introduces residual and high-pass style preprocessing at the frontend, together with tailored nonlinear operations to amplify steganographic traces, making it one of the early CNN baselines widely used in spatial-domain steganalysis.
Building on this line, Yedroudj-Net~\cite{yedroudjnet2018} integrates key components such as filter-bank preprocessing, truncation-based activations, and normalization, and achieves stronger detection performance with improved training strategies.
Subsequently, as training deeper models became feasible, SRNet~\cite{SRNet2021} adopts deep residual architectures and emphasizes more end-to-end feature learning, aiming to reduce reliance on hand-crafted heuristics and further advance CNN-based steganalysis.
Beyond these, several effective universal steganalysis detectors have also been proposed~\cite{ZhuNet2020,YeNet2017,stegnet2019,SiaStegNet2021}.
However, a unified benchmarking suite for systematic evaluation is still lacking, which constitutes one of the main contributions of our work.

%-----------------------------
\subsection{Prior Benchmarks}
\label{sec:benchmark}
%-----------------------------

Existing image steganography and steganalysis benchmarks~\cite{DBLP:conf/sswmc/KharraziSM05,DBLP:conf/mmsec/Ker07,DBLP:conf/ih/PevnyF08,DBLP:conf/wifs/CogranneGB20} measure capacity, distortion, and detectability, but mostly use random-bit payloads, limiting their relevance to safety-related misuse, as summarized in~\Cref{tab:benchmark_compare}.

First, random-bit payloads provide a content-agnostic measure of channel reliability, but they do not capture whether the recovered payload preserves harmful semantics. 
In misuse scenarios, the attack objective is not only bit recovery, but also application-level viability: a recovered jailbreak instruction must remain executable by a downstream model, and a recovered harmful image must preserve recognizable visual semantics. 
Thus, realistic harmful payloads complement random-bit tests by evaluating semantic recoverability, not only low-level transmission quality.
Second, prior benchmarks are typically not organized around an explicit attack-defense setting. 
Image steganography becomes security-critical when it is used to bypass moderation or deliver harmful content through image covers. 
This motivates a benchmark that jointly evaluates steganographic attack capability, steganalysis defense capability, efficiency, and transferability under a common protocol.
Third, practical deployment factors remain underexplored. 
In real settings, defenders may face unknown cover distributions, payload sources, and steganographic methods, while platform-side processing such as social-media compression can directly affect payload recovery. 
However, these transfer and transmission factors are only partially covered in prior benchmarks.

\bench complements prior random-bit benchmarks by focusing on image-cover steganography misuse with safety-relevant image and text payloads. 
It evaluates whether harmful visual or textual payloads can be embedded, recovered, and detected under selected distribution shifts and platform-processing conditions, without claiming to cover all steganography scenarios.

\begin{table}[t]
\centering
\setlength{\tabcolsep}{1.6pt}
\renewcommand{\arraystretch}{1.03}
\setlength{\aboverulesep}{0pt}
\setlength{\belowrulesep}{0pt}
\fontsize{7.0}{8.0}\selectfont

\caption{Comparison of our \bench with prior benchmarks.
Std. denotes in-distribution evaluation.
In A/D Coverage, we mark which evaluation side a benchmark provides:
(\protect\attackonly) steganography-side evaluation,
(\protect\defenseonly) steganalysis-side evaluation, and
(\protect\attackdefense) both.}
\label{tab:benchmark_compare}

\begin{tabular*}{\columnwidth}{@{\extracolsep{\fill}} l l c c c c c c}
\toprule
\multirow{2}{*}[-0.4ex]{Benchmark} &
\multirow{2}{*}[-0.4ex]{Payload} &
\multicolumn{5}{c}{Key Evaluation Scenarios} &
\multirow{2}{*}[-0.4ex]{Tasks} \\
\cmidrule(lr){3-7}
\addlinespace[0.55ex]
& & Std. & \makecell[c]{A/D\\Metrics} & \makecell[c]{Transfer\\ability} &
\makecell[c]{Real\\Platform} & \makecell[c]{Effi-\\ciency} & \\
\midrule

Kharrazi et al.~\cite{DBLP:conf/sswmc/KharraziSM05}
& Random bits
& \cmark
& \defenseonly
& \xmark
& \xmark
& \xmark
& 1 \\

Ker et al.~\cite{DBLP:conf/mmsec/Ker07}
& Random bits
& \cmark
& \adnone
& \xmark
& \xmark
& \xmark
& 1 \\

Pevn\'y et al.~\cite{DBLP:conf/ih/PevnyF08}
& Random bits
& \cmark
& \attackonly
& \xmark
& \xmark
& \xmark
& 1 \\

ALASKA\#2~\cite{DBLP:conf/wifs/CogranneGB20}
& Random bits
& \cmark
& \defenseonly
& \cmark
& \xmark
& \xmark
& 1 \\

\addlinespace[0.65ex]
\midrule

\textbf{\bench (Ours)}
& \textbf{Image\&Text}
& \cmark
& \attackdefense
& \cmark
& \cmark
& \cmark
& 4 \\

\bottomrule
\end{tabular*}
\end{table}

\begin{figure*}[t]
    \centering
    \includegraphics[width=\textwidth]{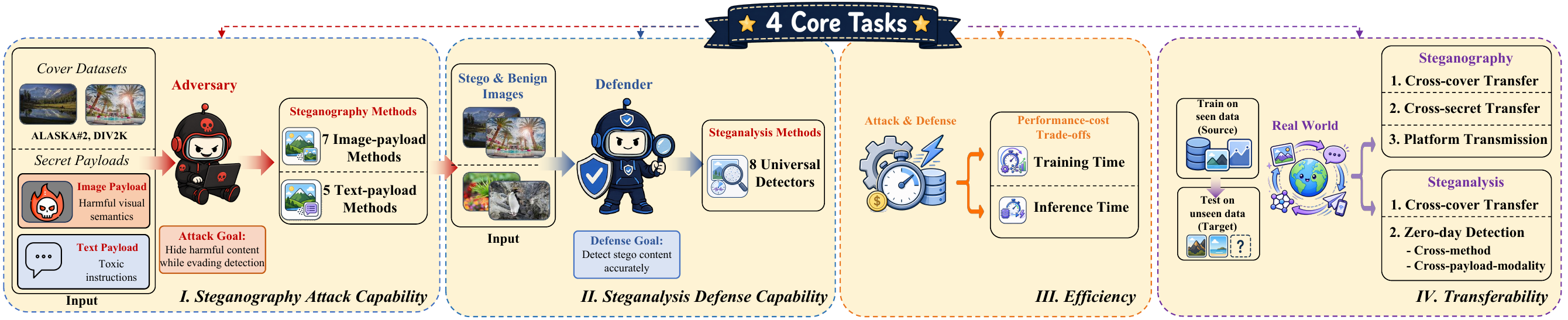}
    \caption{
    Overview of \bench framework. The framework consists of four core tasks.}
    \label{fig:framework}
\end{figure*}

%-----------------------------
\section{Threat Model}
%-----------------------------
As shown in \Cref{fig:overview}, we model image steganography as an adversarial interaction between a steganography sender and a steganalysis detector: the adversary hides payloads in images, while the defender detects stego images to prevent misuse.

%-----------------------------
\subsection{Adversary}\label{sec:adversary}
%-----------------------------
\mypara{Goal}
We consider an adversary who misuses image-cover steganography to hide safety-relevant payloads in benign-looking cover images. 
The payload can be either a harmful image payload or a harmful textual instruction. 
The resulting stego image may be disseminated through social media or messaging platforms~\cite{DBLP:journals/csur/KombrinkGW25}, or provided as input to vision-capable LLMs or LLM agents. 
The attack succeeds when the hidden payload remains recoverable and application-level meaningful, e.g., a harmful visual payload remains recognizable or a hidden instruction can be decoded and used in a downstream jailbreak setting~\cite{li2025odysseus}.
The adversary aims to satisfy four requirements: 
(i) \textbf{Usability:} the hidden payload can be reliably recovered under the target channel or downstream model setting; 
(ii) \textbf{Stealthiness:} the stego image remains visually close to the cover image and avoids obvious detection by human inspection, moderation systems, or steganalysis detectors; 
(iii) \textbf{Resource efficiency:} the adversary prefers methods with lower training or inference cost when they provide comparable recovery and stealthiness; and 
(iv) \textbf{Transferability:} the adversary benefits when a steganography method trained on one cover or payload distribution remains usable on another distribution or after platform processing.

\mypara{Capability}
We assume that the adversary controls the steganographic generation process, including the choice of cover images, payloads, embedding method, extraction model or key, and message length. 
The adversary may use public or self-trained image steganography methods and can tune the trade-off among recoverability, stealthiness, and cost. 
Before uploading images to a platform, the adversary may have partial knowledge of common image-processing operations, such as resizing or compression, and may simulate such transformations to improve robustness.
However, we do not assume a fully adaptive white-box adversary against the defender. 
In particular, the adversary does not necessarily know the exact steganalysis detector, training data, decision threshold, or ensemble strategy deployed by the defender. 
This reflects a practical setting in which the adversary controls the steganographic pipeline but only has partial knowledge of the defense.

%-----------------------------
\subsection{Defender}
%-----------------------------
\mypara{Goal}
We consider a defender who inspects images at the platform side, such as during image uploading, transmission, or storage. 
The defender aims to detect stego images that may contain hidden payloads while maintaining a low false-positive rate on benign images. 
In practice, the defender also needs detectors that are efficient enough for large-scale screening and robust to distribution shifts, since the deployed system may encounter unseen cover sources, payload types, and steganographic methods.

\mypara{Capability}
We assume that the defender operates in a black-box setting with respect to the adversary. 
The defender can observe the final uploaded images but does not know the exact steganographic algorithm, payload, key, extraction model, or training distribution used by the adversary. 
The defender can use public steganography methods to synthesize labeled Cover/Stego pairs and train steganalysis detectors. 
The defender may also deploy multiple detectors or ensemble different detection strategies when resources allow.

Nevertheless, we primarily evaluate single-detector transferability for two reasons. 
First, single-detector evaluation isolates whether the learned steganalysis features generalize across cover datasets, payload sources, and steganographic methods, without conflating this property with ensemble design choices. 
Second, ensembles can improve coverage but do not eliminate the transferability problem: they increase deployment cost, still require representative training data, and may remain vulnerable to unseen steganographic methods or shifted cover distributions. 
Thus, poor single-detector transferability reveals a fundamental weakness that an ensemble may mitigate but not automatically solve.

Given this capability asymmetry, transferability emerges as a critical security concern.
The adversary can reuse a trained steganography model across multiple cover sources or adapt it with simulated platform processing, whereas the defender must detect stego images without knowing the attack method or payload distribution. 
Therefore, a gap between attack transferability and defense transferability reveals a practical blind spot for image-cover steganalysis in open-world deployment.

%-----------------------------
\section{Design of \bench}
%-----------------------------
\Cref{fig:framework} presents the overall framework of our \bench, which consists of $4$ core tasks covering both the attack and defense sides.
%-----------------------------
\subsection{Task Definition}
%-----------------------------

\mypara{Task 1: Steganography Attack Capability}
This task evaluates the adversary's ability to embed safety-relevant payloads into image covers.
We consider two payload settings: image payloads that contain harmful visual semantics and text payloads that contain harmful instructions.
Attack quality is measured from two perspectives: \textbf{stealthiness}, i.e., whether the stego image remains visually close to the original cover image, and \textbf{recoverability}, i.e., whether the embedded payload can be accurately recovered and remain application-level meaningful.

\mypara{Task 2: Steganalysis Defense Capability}
This task evaluates the defender's detection capability against image steganography threats, with an emphasis on universal steganalysis detectors.
Under a unified evaluation protocol, we formalize detection as a binary classification problem of whether an input image is a stego.

\mypara{Task 3: Efficiency}
This task quantifies the resource overhead and practical deployability of methods on both the attack and defense sides.
For steganography methods, we evaluate the computational cost of generating stego images and recovering secrets, including training and inference time.
For steganalysis methods, we likewise measure efficiency during both training and inference.
By adopting unified efficiency metrics and measurement procedures, this task provides a comparable reference for analyzing performance-cost trade-offs across methods.

\mypara{Task 4: Transferability}
This task analyzes the out-of-distribution generalization capabilities of both steganographic attacks and steganalysis defenses. 
We systematically evaluate robustness across three critical transfer dimensions: cross-cover dataset shifts, cross-method shifts (i.e., generalizing to unseen steganographic methods), and cross-secret payload shifts. 
By quantifying the performance degradation under these distribution mismatches, this task exposes the stability boundaries and structural failure modes of current methods when confronting zero-day threats and diverse real-world data pipelines.

%-----------------------------
\subsection{Framework}
%-----------------------------

\mypara{Datasets}
We select two cover-image datasets representing distinct visual distributions. 
First, ALASKA\#2~\cite{DBLP:conf/wifs/CogranneGB20}, a highly diverse steganalysis benchmark. To manage computational costs, we randomly sample 7,200 images, applying a 7:1:2 train/validation/test split. 
Second, DIV2K~\cite{Agustsson_2017_CVPR_Workshops}, a high-quality dataset standard in steganography research, utilizes its official split.

To simulate realistic security threats, we construct safety-relevant payloads across two modalities. 
For image payloads, we use 10,000 samples from Hateful Memes~\cite{DBLP:conf/nips/KielaFMGSRT20} to model the covert transmission of harmful visual semantics and moderation evasion. 
For text payloads, we employ 520 malicious instructions from AdvBench~\cite{li2025odysseus, DBLP:journals/corr/abs-2307-15043} to simulate multimodal LLM jailbreak scenarios. 
Crucially, replacing traditional random-bit payloads with these semantic datasets allows us to assess whether the extracted content retains sufficient integrity to trigger downstream misuse. 
Both payload datasets follow a 7:1:2 split and are dynamically resampled during training to match the larger volume of cover images.

\mypara{Steganography Methods}
For image-payload steganography, \bench evaluates $7$ representative methods spanning $3$ architectural families commonly studied in learning-based image steganography.
Autoencoder-style methods include DS~\cite{ds2017}, UDH~\cite{UDH2020}, and StegFormer~\cite{stegformer2024}; INN-based methods include HiNet~\cite{HiNet2021}, DeepMIH~\cite{DeepMIH2023}, and PRIS~\cite{PRIS2024}.
A representative diffusion-based method is CRoSS~\cite{CroSS2023}.
For text-payload steganography, \bench evaluates $5$ public methods including HiDDeN~\cite{HiDDeN2018}, SteganoGAN~\cite{SteganoGAN2019}, FNNS~\cite{FNNS2022}, CLPSTNet~\cite{clpstnet}, and RoSteALS~\cite{RoSteALS}.
These methods cover both classical encoder-decoder pipelines and more recent text/bitstream-oriented image steganography approaches, allowing us to compare recoverability and visual stealthiness under the same safety-relevant text-payload setting.

For implementation and training settings, we follow the experimental configurations of the original papers as closely as possible and adopt their reported best-performing hyperparameters.
When a paper or public implementation does not fully specify a training setting, we use a fixed configuration and report it for reproducibility; for example, the DS paper does not explicitly specify the number of training epochs, and we therefore set it to $1000$.
For all trainable models, we perform model selection on the validation set and use the checkpoint with the best validation performance for testing and all subsequent experiments.

\mypara{Steganalysis Detectors}
As described in \Cref{sec:steganalysis}, we primarily evaluate universal steganalysis detectors.
Specifically, we consider eight detectors that can be used to jointly detect both image-payload and text-payload steganography.
These include one hand-crafted feature-based baseline, SRM+EC~\cite{couturier2016steganalysis}, and $7$ deep learning-based detectors, namely XuNet~\cite{xunet2016}, Yedroudj-Net~\cite{yedroudjnet2018}, ZhuNet~\cite{ZhuNet2020}, SRNet~\cite{SRNet2021}, YeNet~\cite{YeNet2017}, StegNet~\cite{stegnet2019}, and SiaStegNet~\cite{SiaStegNet2021}.
All methods are configured using the best hyperparameters reported in the corresponding papers or public implementations, and are evaluated under the same protocol.

%-----------------------------
\subsection{Experimental Setup}
%-----------------------------

\mypara{Steganography Metrics}
We organize steganography metrics around two evaluation goals: \textbf{stealthiness} and \textbf{recoverability}. 
For image-payload steganography, both goals are evaluated with image similarity and perceptual metrics on two types of image pairs: Cover/Stego pairs for stealthiness and Secret/Recovery pairs for recoverability.
We report MAE, PSNR, SSIM, and LPIPS, which capture complementary aspects of image quality.
MAE measures average pixel-level distortion, while PSNR emphasizes pixel-wise reconstruction fidelity and is sensitive to large reconstruction errors.
SSIM measures structural similarity by comparing luminance, contrast, and local structure, making it useful for assessing whether the recovered or stego image preserves visual organization.
LPIPS measures perceptual discrepancy in a deep feature space and better reflects perceptual differences that may not be captured by pixel-wise metrics.
Together, these metrics allow us to distinguish low-level distortion, structural preservation, and perceptual fidelity.

For text-payload steganography, stealthiness is still evaluated on Cover/Stego image pairs using MAE, PSNR, SSIM, and LPIPS.
For recoverability, because the payload is a natural-language instruction, we report text-specific metrics including Exact Match Rate (EMR), Character Error Rate (CER), and Bit Error Rate (BER).
EMR is the strictest metric and measures whether the recovered instruction exactly matches the original payload, which is important when a downstream system requires precise prompt recovery.
CER measures fine-grained character-level corruption and reflects how severely the recovered text deviates from the original instruction.
BER measures bit-level transmission reliability after text encoding, making it comparable to traditional bitstream-oriented steganography evaluation.
This combination allows us to evaluate both application-level exact recovery and lower-level decoding errors.
Formal metric definitions are provided in~\Cref{sec:metrics}.

\begin{table*}[t]
\centering
\footnotesize
\caption{Comparison of steganography methods with capability metrics. \small{Deeper cell shading indicates better performance.}}
\label{tab:attack_capability}

\begin{subtable}[t]{\textwidth}
\centering
\caption{Image-payload steganography.}
\label{tab:attack_capability_image_payload}

\begin{minipage}[t]{0.49\textwidth}
\centering
\textbf{ALASKA\#2}
\resizebox{\linewidth}{!}{%
\begin{tabular}{lccccccc}
\toprule
Metric & DS & UDH & StegFormer & HiNet & DeepMIH & PRIS & CRoSS \\
\midrule
\multicolumn{8}{c}{\emph{Cover/Stego image pair}} \\
PSNR$\bm{\uparrow}$ & \cellcolor{w3}32.610 & \cellcolor{w2}36.100 & \cellcolor{w1}42.020 & \cellcolor{w3}34.850 & \cellcolor{w1}\textbf{44.130} & \cellcolor{w4}30.750 & \cellcolor{w6}19.888 \\
SSIM$\bm{\uparrow}$ & \cellcolor{w2}0.868 & \cellcolor{w1}0.927 & \cellcolor{w1}0.983 & \cellcolor{w1}0.927 & \cellcolor{w1}\textbf{0.988} & \cellcolor{w2}0.828 & \cellcolor{w3}0.761 \\
MAE$\bm{\downarrow}$ & \cellcolor{c8}4.504 & \cellcolor{c7}3.307 & \cellcolor{c2}1.442 & \cellcolor{c7}3.423 & \cellcolor{c1}\textbf{1.261} & \cellcolor{c8}5.443 & \cellcolor{c10}17.080 \\
LPIPS$\bm{\downarrow}$ & \cellcolor{c10}0.013 & \cellcolor{c5}0.002 & \cellcolor{c5}0.002 & \cellcolor{c9}0.008 & \cellcolor{c1}\textbf{0.001} & \cellcolor{c8}0.005 & \cellcolor{c10}0.284 \\
\midrule
\multicolumn{8}{c}{\emph{Secret/Recovery image pair}} \\
PSNR$\bm{\uparrow}$ & \cellcolor{w3}33.930 & \cellcolor{w4}31.190 & \cellcolor{w1}43.200 & \cellcolor{w2}41.980 & \cellcolor{w1}\textbf{47.780} & \cellcolor{w4}30.280 & \cellcolor{w6}19.393 \\
SSIM$\bm{\uparrow}$ & \cellcolor{w1}0.948 & \cellcolor{w1}0.925 & \cellcolor{w1}0.992 & \cellcolor{w1}0.987 & \cellcolor{w1}\textbf{0.997} & \cellcolor{w1}0.932 & \cellcolor{w3}0.739 \\
MAE$\bm{\downarrow}$ & \cellcolor{c9}3.699 & \cellcolor{c9}5.506 & \cellcolor{c5}1.272 & \cellcolor{c6}1.423 & \cellcolor{c1}\textbf{0.668} & \cellcolor{c9}5.427 & \cellcolor{c10}18.120 \\
LPIPS$\bm{\downarrow}$ & \cellcolor{c2}0.041 & \cellcolor{c2}0.049 & \cellcolor{c1}0.004 & \cellcolor{c1}0.002 & \cellcolor{c1}\textbf{0.000} & \cellcolor{c2}0.044 & \cellcolor{c10}0.302 \\
\bottomrule
\end{tabular}%
}
\end{minipage}
\hfill
\begin{minipage}[t]{0.49\textwidth}
\centering
\textbf{DIV2K}
\resizebox{\linewidth}{!}{%
\begin{tabular}{lccccccc}
\toprule
Metric & DS & UDH & StegFormer & HiNet & DeepMIH & PRIS & CRoSS \\
\midrule
\multicolumn{8}{c}{\emph{Cover/Stego image pair}} \\
PSNR$\bm{\uparrow}$ & \cellcolor{w4}29.880 & \cellcolor{w2}35.582 & \cellcolor{w1}41.730 & \cellcolor{w4}30.240 & \cellcolor{w1}\textbf{43.960} & \cellcolor{w4}30.750 & \cellcolor{w6}19.400 \\
SSIM$\bm{\uparrow}$ & \cellcolor{w2}0.891 & \cellcolor{w1}0.936 & \cellcolor{w1}0.990 & \cellcolor{w2}0.884 & \cellcolor{w1}\textbf{0.994} & \cellcolor{w2}0.867 & \cellcolor{w3}0.756 \\
MAE$\bm{\downarrow}$ & \cellcolor{c9}6.381 & \cellcolor{c7}3.569 & \cellcolor{c3}1.495 & \cellcolor{c8}5.473 & \cellcolor{c1}\textbf{1.132} & \cellcolor{c8}5.378 & \cellcolor{c10}17.547 \\
LPIPS$\bm{\downarrow}$ & \cellcolor{c3}0.070 & \cellcolor{c1}0.001 & \cellcolor{c1}0.001 & \cellcolor{c1}0.021 & \cellcolor{c1}\textbf{0.000} & \cellcolor{c1}0.004 & \cellcolor{c10}0.294 \\
\midrule
\multicolumn{8}{c}{\emph{Secret/Recovery image pair}} \\
PSNR$\bm{\uparrow}$ & \cellcolor{w4}31.090 & \cellcolor{w5}25.670 & \cellcolor{w3}33.240 & \cellcolor{w3}36.830 & \cellcolor{w1}\textbf{47.190} & \cellcolor{w4}29.860 & \cellcolor{w6}19.070 \\
SSIM$\bm{\uparrow}$ & \cellcolor{w1}0.914 & \cellcolor{w2}0.846 & \cellcolor{w1}0.957 & \cellcolor{w1}0.969 & \cellcolor{w1}\textbf{0.997} & \cellcolor{w1}0.925 & \cellcolor{w3}0.741 \\
MAE$\bm{\downarrow}$ & \cellcolor{c9}5.245 & \cellcolor{c10}9.673 & \cellcolor{c9}4.468 & \cellcolor{c8}2.480 & \cellcolor{c1}\textbf{0.688} & \cellcolor{c9}5.622 & \cellcolor{c10}18.179 \\
LPIPS$\bm{\downarrow}$ & \cellcolor{c2}0.054 & \cellcolor{c4}0.122 & \cellcolor{c1}0.025 & \cellcolor{c1}0.009 & \cellcolor{c1}\textbf{0.000} & \cellcolor{c2}0.045 & \cellcolor{c10}0.309 \\
\bottomrule
\end{tabular}%
}
\end{minipage}

\end{subtable}

\begin{subtable}[t]{\textwidth}
\centering
\caption{Text-payload steganography.}
\label{tab:attack_capability_text_payload}

\begin{minipage}[t]{0.49\textwidth}
\centering
\textbf{ALASKA\#2}
\resizebox{\linewidth}{!}{%
\begin{tabular}{lccccc}
\toprule
Metric & HiDDeN & SteganoGAN & FNNS & CLPSTNet & RoSteALS \\
\midrule
\multicolumn{6}{c}{\emph{Cover/Stego image pair}} \\
PSNR$\bm{\uparrow}$ & \cellcolor{w3}33.035 & \cellcolor{w1}40.151 & \cellcolor{w2}35.606 & \cellcolor{w1}\textbf{41.610} & \cellcolor{w4}27.663 \\
SSIM$\bm{\uparrow}$ & \cellcolor{w1}0.965 & \cellcolor{w1}0.974 & \cellcolor{w1}0.935 & \cellcolor{w1}\textbf{0.995} & \cellcolor{w1}0.924 \\
MAE$\bm{\downarrow}$ & \cellcolor{c8}4.561 & \cellcolor{c5}1.986 & \cellcolor{c7}3.327 & \cellcolor{c1}\textbf{1.050} & \cellcolor{c9}7.300 \\
LPIPS$\bm{\downarrow}$ & \cellcolor{c5}0.031 & \cellcolor{c1}0.004 & \cellcolor{c3}0.018 & \cellcolor{c1}\textbf{0.000} & \cellcolor{c10}0.072 \\
\midrule
\multicolumn{6}{c}{\emph{Secret/Recovery text pair}} \\
EMR$\bm{\uparrow}$ & \cellcolor{w10}0.000 & \cellcolor{w2}0.791 & \cellcolor{w1}0.927 & \cellcolor{w1}\textbf{0.983} & \cellcolor{w10}0.000 \\
CER$\bm{\downarrow}$ & \cellcolor{c10}0.698 & \cellcolor{c10}0.209 & \cellcolor{c10}0.073 & \cellcolor{c1}\textbf{0.001} & \cellcolor{c10}0.695 \\
BER$\bm{\downarrow}$ & \cellcolor{c10}0.219 & \cellcolor{c10}0.209 & \cellcolor{c4}0.073 & \cellcolor{c1}\textbf{0.000} & \cellcolor{c10}0.224 \\
\bottomrule
\end{tabular}%
}
\end{minipage}
\hfill
\begin{minipage}[t]{0.49\textwidth}
\centering
\textbf{DIV2K}
\resizebox{\linewidth}{!}{%
\begin{tabular}{lccccc}
\toprule
Metric & HiDDeN & SteganoGAN & FNNS & CLPSTNet & RoSteALS \\
\midrule
\multicolumn{6}{c}{\emph{Cover/Stego image pair}} \\
PSNR$\bm{\uparrow}$ & \cellcolor{w3}29.403 & \cellcolor{w1}\textbf{40.059} & \cellcolor{w2}35.400 & \cellcolor{w2}33.944 & \cellcolor{w3}30.677 \\
SSIM$\bm{\uparrow}$ & \cellcolor{w1}0.956 & \cellcolor{w1}\textbf{0.985} & \cellcolor{w1}0.951 & \cellcolor{w1}0.980 & \cellcolor{w1}0.948 \\
MAE$\bm{\downarrow}$ & \cellcolor{c8}7.022 & \cellcolor{c1}\textbf{1.909} & \cellcolor{c5}3.362 & \cellcolor{c4}2.760 & \cellcolor{c7}5.325 \\
LPIPS$\bm{\downarrow}$ & \cellcolor{c10}0.042 & \cellcolor{c1}\textbf{0.002} & \cellcolor{c8}0.007 & \cellcolor{c4}0.003 & \cellcolor{c10}0.028 \\
\midrule
\multicolumn{6}{c}{\emph{Secret/Recovery text pair}} \\
EMR$\bm{\uparrow}$ & \cellcolor{w10}0.000 & \cellcolor{w2}0.740 & \cellcolor{w1}0.890 & \cellcolor{w1}\textbf{0.920} & \cellcolor{w10}0.000 \\
CER$\bm{\downarrow}$ & \cellcolor{c10}0.691 & \cellcolor{c10}0.260 & \cellcolor{c10}0.110 & \cellcolor{c1}\textbf{0.001} & \cellcolor{c10}0.960 \\
BER$\bm{\downarrow}$ & \cellcolor{c8}0.212 & \cellcolor{c9}0.260 & \cellcolor{c4}0.110 & \cellcolor{c1}\textbf{0.000} & \cellcolor{c10}0.295 \\
\bottomrule
\end{tabular}
}
\end{minipage}

\end{subtable}

\end{table*}

\mypara{Steganalysis Metrics}
We formulate steganalysis as a binary classification problem between cover and stego images.
In the main text, we report F1-score as the primary metric for compactness and comparability across methods and detectors.
F1-score summarizes the trade-off between false alarms and missed detections, and is appropriate for our balanced Cover/Stego test sets.
We additionally report accuracy and AUC in~\Cref{sec:appendix_detection_metrics}, where accuracy reflects threshold-dependent correctness and AUC measures threshold-independent separability.

\section{Evaluation and Analysis}

\subsection{Steganography Attack Capability} 

We evaluate attack capability on ALASKA\#2 and DIV2K under both image- and text-payload settings; quantitative results are shown in~\Cref{tab:attack_capability}, with qualitative examples in~\Cref{fig:ablation_study_examples}.

\begin{table*}[t]
\centering
\footnotesize
\caption{Results of steganalysis defense capability evaluation measured by F1-score.}
\label{tab:steganalysis_f1_methods_rows}

\resizebox{\textwidth}{!}{
\begin{tabular}{lllcccccccc}
\toprule
\multirow{2}{*}{Payload} & \multirow{2}{*}{Dataset} & \multirow{2}{*}{Steganography} & \multicolumn{8}{c}{Detector} \\
\cmidrule(lr){4-11}
& & & SRM+EC & XuNet & Yedroudj-Net & SRNet & YeNet & ZhuNet & StegNet & SiaStegNet \\
\midrule
\multirow{14}{*}{Image} & \multirow{7}{*}{ALASKA\#2} & DS
& \cellcolor{w3}0.764 & \cellcolor{w1}0.970 & \cellcolor{w1}0.976 & \cellcolor{w1}0.918 & \cellcolor{w1}0.968 & \cellcolor{w1}0.957 & \cellcolor{w1}\textbf{0.986} & \cellcolor{w1}0.981 \\
& & UDH
& \cellcolor{w2}0.892 & \cellcolor{w1}0.998 & \cellcolor{w1}0.998 & \cellcolor{w2}0.809 & \cellcolor{w1}0.998 & \cellcolor{w1}0.995 & \cellcolor{w1}\textbf{1.000} & \cellcolor{w1}\textbf{1.000} \\
& & StegFormer
& \cellcolor{w5}0.589 & \cellcolor{w1}0.929 & \cellcolor{w1}0.954 & \cellcolor{w6}0.459 & \cellcolor{w1}0.935 & \cellcolor{w1}0.897 & \cellcolor{w1}0.986 & \cellcolor{w1}\textbf{0.996} \\
& & HiNet
& \cellcolor{w2}0.848 & \cellcolor{w1}0.997 & \cellcolor{w1}0.999 & \cellcolor{w1}0.979 & \cellcolor{w1}0.999 & \cellcolor{w1}0.993 & \cellcolor{w1}\textbf{1.000} & \cellcolor{w1}\textbf{1.000} \\
& & DeepMIH
& \cellcolor{w5}0.588 & \cellcolor{w1}0.969 & \cellcolor{w1}0.977 & \cellcolor{w6}0.447 & \cellcolor{w1}0.955 & \cellcolor{w7}0.323 & \cellcolor{w1}0.997 & \cellcolor{w1}\textbf{0.999} \\
& & PRIS
& \cellcolor{w1}0.971 & \cellcolor{w1}\textbf{1.000} & \cellcolor{w1}\textbf{1.000} & \cellcolor{w1}0.999 & \cellcolor{w1}0.999 & \cellcolor{w1}0.999 & \cellcolor{w1}\textbf{1.000} & \cellcolor{w1}\textbf{1.000} \\
& & CRoSS
& \cellcolor{w1}0.923 & \cellcolor{w1}\textbf{1.000} & \cellcolor{w1}\textbf{1.000} & \cellcolor{w1}\textbf{1.000} & \cellcolor{w1}0.999 & \cellcolor{w1}0.998 & \cellcolor{w1}\textbf{1.000} & \cellcolor{w1}\textbf{1.000} \\
\cmidrule(lr){2-11}
& \multirow{6}{*}{DIV2K} & DS
& \cellcolor{w5}0.536 & \cellcolor{w2}0.829 & \cellcolor{w1}0.877 & \cellcolor{w2}0.766 & \cellcolor{w2}0.816 & \cellcolor{w2}0.777 & \cellcolor{w1}\textbf{0.942} & \cellcolor{w1}0.920 \\
& & UDH
& \cellcolor{w5}0.589 & \cellcolor{w1}0.992 & \cellcolor{w1}0.990 & \cellcolor{w5}0.528 & \cellcolor{w1}0.987 & \cellcolor{w3}0.786 & \cellcolor{w1}\textbf{0.997} & \cellcolor{w1}\textbf{0.997} \\
& & StegFormer
& \cellcolor{w5}0.544 & \cellcolor{w1}0.976 & \cellcolor{w1}\textbf{0.993} & \cellcolor{w5}0.563 & \cellcolor{w1}0.913 & \cellcolor{w5}0.556 & \cellcolor{w1}0.982 & \cellcolor{w1}\textbf{0.993} \\
& & HiNet
& \cellcolor{w1}0.937 & \cellcolor{w1}\textbf{1.000} & \cellcolor{w1}0.997 & \cellcolor{w1}0.958 & \cellcolor{w1}\textbf{1.000} & \cellcolor{w1}0.974 & \cellcolor{w1}\textbf{1.000} & \cellcolor{w1}\textbf{1.000} \\
& & DeepMIH
& \cellcolor{w5}0.525 & \cellcolor{w1}0.987 & \cellcolor{w1}0.988 & \cellcolor{w6}0.474 & \cellcolor{w1}0.962 & \cellcolor{w5}0.542 & \cellcolor{w1}\textbf{0.995} & \cellcolor{w1}0.993 \\
& & PRIS
& \cellcolor{w2}0.854 & \cellcolor{w1}0.997 & \cellcolor{w1}0.998 & \cellcolor{w1}0.922 & \cellcolor{w1}0.998 & \cellcolor{w1}0.973 & \cellcolor{w1}\textbf{1.000} & \cellcolor{w1}\textbf{1.000} \\
& & CRoSS
& \cellcolor{w3}0.768 & \cellcolor{w1}\textbf{1.000} & \cellcolor{w1}\textbf{1.000} & \cellcolor{w1}0.980 & \cellcolor{w1}0.995 & \cellcolor{w1}0.978 & \cellcolor{w1}\textbf{1.000} & \cellcolor{w1}\textbf{1.000} \\
\midrule
\multirow{10}{*}{Text} & \multirow{5}{*}{ALASKA\#2}
& CLPSTNet
& \cellcolor{w5}0.509 & \cellcolor{w2}0.851 & \cellcolor{w1}0.921 & \cellcolor{w6}0.456 & \cellcolor{w3}0.749 & \cellcolor{w3}0.785 & \cellcolor{w1}\textbf{0.982} & \cellcolor{w1}0.981 \\
& & HiDDeN
& \cellcolor{w5}0.516 & \cellcolor{w1}0.961 & \cellcolor{w1}0.992 & \cellcolor{w1}0.990 & \cellcolor{w1}0.991 & \cellcolor{w1}0.989 & \cellcolor{w1}\textbf{0.998} & \cellcolor{w1}\textbf{0.998} \\
& & SteganoGAN
& \cellcolor{w4}0.567 & \cellcolor{w1}0.815 & \cellcolor{w1}0.855 & \cellcolor{w5}0.525 & \cellcolor{w1}0.831 & \cellcolor{w1}0.803 & \cellcolor{w1}\textbf{0.886} & \cellcolor{w1}0.883 \\
& & FNNS
& \cellcolor{w4}0.658 & \cellcolor{w1}0.919 & \cellcolor{w1}0.946 & \cellcolor{w1}0.894 & \cellcolor{w1}0.919 & \cellcolor{w1}0.954 & \cellcolor{w1}0.976 & \cellcolor{w1}\textbf{0.983} \\
& & RoSteALS
& \cellcolor{w4}0.657 & \cellcolor{w1}0.963 & \cellcolor{w1}0.990 & \cellcolor{w1}0.951 & \cellcolor{w1}0.992 & \cellcolor{w1}0.943 & \cellcolor{w1}0.998 & \cellcolor{w1}\textbf{0.999} \\
\cmidrule(lr){2-11}
& \multirow{5}{*}{DIV2K}
& CLPSTNet
& \cellcolor{w5}0.573 & \cellcolor{w2}0.860 & \cellcolor{w1}0.893 & \cellcolor{w6}0.473 & \cellcolor{w3}0.744 & \cellcolor{w5}0.555 & \cellcolor{w1}\textbf{0.987} & \cellcolor{w1}0.977 \\
& & HiDDeN
& \cellcolor{w4}0.594 & \cellcolor{w3}0.735 & \cellcolor{w1}0.949 & \cellcolor{w1}0.923 & \cellcolor{w2}0.843 & \cellcolor{w2}0.786 & \cellcolor{w1}0.970 & \cellcolor{w1}\textbf{0.981} \\
& & SteganoGAN
& \cellcolor{w4}0.570 & \cellcolor{w1}0.865 & \cellcolor{w1}0.849 & \cellcolor{w6}0.402 & \cellcolor{w2}0.746 & \cellcolor{w5}0.538 & \cellcolor{w1}0.892 & \cellcolor{w1}\textbf{0.904} \\
& & FNNS
& \cellcolor{w5}0.554 & \cellcolor{w1}0.945 & \cellcolor{w1}0.955 & \cellcolor{w5}0.532 & \cellcolor{w1}0.922 & \cellcolor{w4}0.682 & \cellcolor{w1}\textbf{0.982} & \cellcolor{w1}0.978 \\
& & RoSteALS
& \cellcolor{w4}0.692 & \cellcolor{w1}0.953 & \cellcolor{w1}0.971 & \cellcolor{w2}0.883 & \cellcolor{w1}0.924 & \cellcolor{w2}0.889 & \cellcolor{w1}0.990 & \cellcolor{w1}\textbf{0.995} \\
\bottomrule
\end{tabular}
}
\end{table*}

\mypara{Image-payload Steganography}
As shown in~\Cref{tab:attack_capability_image_payload}, DeepMIH exhibits the strongest overall performance on both datasets.
It achieves the highest Cover/Stego fidelity, with PSNR values of $44.130\,\mathrm{dB}$ on ALASKA\#2 and $43.960\,\mathrm{dB}$ on DIV2K, and maintains near-zero perceptual perturbation with LPIPS no larger than 0.001.
This suggests that its invertible design can constrain the embedding distortion to a very low level while preserving a reliable information path for accurate secret reconstruction.
StegFormer and HiNet also show highly competitive Secret/Recovery quality.
For instance, StegFormer reaches a Secret/Recovery PSNR of $43.200\,\mathrm{dB}$ on ALASKA\#2.
We hypothesize that these methods benefit from their structural designs: StegFormer may leverage transformer-based representation modeling to facilitate payload reconstruction, while HiNet's invertible architecture explicitly encourages information-preserving transformations.
Other methods, such as UDH, DS, and PRIS, generally lag behind the top-performing baselines.
Their lower Cover/Stego fidelity and/or weaker Secret/Recovery quality indicate a less favorable fidelity–recovery trade-off, suggesting that they may be less effective in allocating capacity between cover preservation and secret extraction.

In contrast, the diffusion-based method CRoSS performs substantially worse under this fixed-cover evaluation protocol, showing consistent and severe degradation across both datasets.
Its Cover/Stego PSNR drops to $19.888\,\mathrm{dB}$ on ALASKA\#2 and $19.400\,\mathrm{dB}$ on DIV2K, with Secret/Recovery PSNR similarly hovering around $19\,\mathrm{dB}$.
This is because its generative conditional diffusion mechanism is not naturally aligned with strict cover-preserving steganography.
Unlike residual or invertible methods that perturb a given cover locally, CRoSS relies on an iterative generative denoising process that may alter the overall image appearance and semantics to satisfy conditioning constraints.
Such generative deviations can substantially reduce Cover/Stego consistency, while information transfer and inversion errors accumulated during generation may further degrade the quality of secret reconstruction.

\mypara{Text-payload Steganography}
As shown in~\Cref{tab:attack_capability_text_payload}, text-payload steganography imposes a stricter recoverability requirement than image-payload hiding, since textual messages are represented as long discrete bit sequences and small bit errors can accumulate into message-level failures.
CLPSTNet achieves the strongest recoverability on both datasets, with EMR of 0.983 on ALASKA\#2 and 0.920 on DIV2K, and nearly zero decoding errors, i.e., CER/BER of 0.001/0.000 on both datasets.
It also maintains strong Cover/Stego quality, with PSNR of $41.610\,\mathrm{dB}$ on ALASKA\#2 and $33.944\,\mathrm{dB}$ on DIV2K, suggesting that it better supports the long binary sequences induced by text payloads.
Its consistently low CER/BER further indicates that the recovered bit sequence remains globally well aligned with the original message, rather than only preserving partial local fragments.
This is particularly important for text payloads, since successful recovery requires stable decoding over the entire sequence instead of merely reducing average bit-level errors.
SteganoGAN preserves high visual quality, achieving Cover/Stego PSNR of $40.151\,\mathrm{dB}$ on ALASKA\#2 and $40.059\,\mathrm{dB}$ on DIV2K, but its EMR drops to 0.791 and 0.740, respectively.
FNNS provides a more balanced alternative, with EMR of 0.927 and 0.890 while maintaining moderate PSNR of $35.606\,\mathrm{dB}$ and $35.400\,\mathrm{dB}$ on the two datasets.
In contrast, HiDDeN and RoSteALS fail to recover text reliably in our evaluation, with EMR remaining at 0.000 on both datasets, even though their Cover/Stego quality is not always the worst.
This suggests that methods originally designed for relatively short bitstreams may not provide sufficient reliable capacity for realistic text payloads, where even a single sentence can correspond to thousands of bits.
As the payload length increases, small per-bit errors are accumulated and amplified into complete message-level recovery failures.
Additional ablations on training-data size and payload/embedding rate are provided in~\Cref{sec:payload_ablation,sec:data_size}.

\begin{tcolorbox}[
  colback=black!5,     
  colframe=black,
  boxrule=0pt,
  toprule=0.4pt, 
  bottomrule=0.4pt,
  leftrule=0.4pt,
  rightrule=0.4pt,
  left=4pt,right=4pt, 
  top=2pt,bottom=2pt,
  boxsep=0pt,
  before skip=4pt,
  after skip=4pt
]
\textbf{Takeaway:} 
For image-payload steganography, cover-conditioned methods with explicit reconstruction paths, such as invertible or autoencoder-based designs, are more effective because they can jointly preserve Cover/Stego consistency and Secret/Recovery fidelity.
Diffusion-based generation is less aligned with fixed-cover preservation and therefore suffers from larger consistency loss under this protocol.
For text-payload steganography, the decisive factor is reliable capacity for long bitstreams: methods such as CLPSTNet and FNNS scale better to sentence-level text, whereas methods designed around shorter binary messages can collapse as bit errors accumulate.
\end{tcolorbox}

\subsection{Steganalysis Defense Capability}
From the defender's perspective, we evaluate steganalysis defense capability on the two datasets, under both image-payload and text-payload settings.
As summarized in~\Cref{tab:steganalysis_f1_methods_rows}, we report F1-scores for eight representative detectors, including the classical SRM+EC model and seven CNN-based steganalysis detectors (XuNet, Yedroudj-Net, SRNet, YeNet, ZhuNet, StegNet, and SiaStegNet), against seven image-payload and five text-payload steganography methods.

\mypara{Detection Performance on Image-payloads}
In this setting, CNN-based detectors consistently outperform the SRM+EC method.
On ALASKA\#2, SRM+EC achieves an average F1-score of approximately 0.796 across the seven image-payload methods, whereas StegNet and SiaStegNet reach near-perfect averages of about 0.996 and 0.997, respectively.
A similar trend appears on DIV2K, where the average F1-score of SRM+EC drops to about 0.679, while StegNet and SiaStegNet still remain above 0.980.
This suggests that learned steganalysis models are more effective at capturing embedding-induced residual traces than hand-crafted statistical features.
More importantly, detectability is not fully determined by visual fidelity.
CRoSS is consistently easy to detect, which is expected because its diffusion-based generation introduces large Cover/Stego discrepancies and distributional shifts in our previous capability evaluation.
However, even visually strong methods such as DeepMIH and StegFormer are not necessarily secure against strong detectors.
Although they achieve high Cover/Stego fidelity and strong Secret/Recovery quality, StegNet and SiaStegNet still obtain very high F1-scores on them, e.g., 0.986/0.996 on StegFormer and 0.997/0.999 on DeepMIH over ALASKA\#2.
This indicates that preserving perceptual similarity does not eliminate the statistical traces exploited by modern CNN-based steganalysis.

\mypara{Detection Performance on Text-payloads}
This setting shows a similar detector-side trend but a different method-side explanation.
CNN-based detectors substantially outperform SRM+EC: on ALASKA\#2, SRM+EC obtains an average F1-score of about 0.581, while StegNet and SiaStegNet reach about 0.968 and 0.969, respectively.
On DIV2K, SRM+EC remains low at about 0.597, whereas StegNet and SiaStegNet still achieve about 0.964 and 0.967.
This suggests that text-payload embedding also leaves detectable statistical traces, even when the stego image quality appears acceptable.
CLPSTNet and FNNS achieve relatively reliable text recovery in the previous capability evaluation, suggesting that they better support the long bitstreams induced by realistic text payloads.
At the same time, they are not completely immune to steganalysis, especially under strong CNN-based detectors.
In contrast, HiDDeN and RoSteALS are easier to detect and also fail to recover text reliably in our evaluation, with EMR remaining at 0.000 on both datasets.
This suggests that methods designed around shorter binary or watermark-like payloads may become unstable when extended to long text payloads: the embedding process can leave detectable artifacts while still failing to provide sufficient reliable capacity for exact message recovery.

\begin{tcolorbox}[
  colback=black!5,     
  colframe=black,
  boxrule=0pt,
  toprule=0.4pt, 
  bottomrule=0.4pt,
  leftrule=0.4pt,
  rightrule=0.4pt,
  left=4pt,right=4pt, 
  top=2pt,bottom=2pt,
  boxsep=0pt,
  before skip=4pt,
  after skip=4pt
]
\textbf{Takeaway:} 
From the defender's perspective, high-quality steganography is not inherently secure against steganalysis.
Across both image- and text-payload settings, CNN-based detectors, especially StegNet and SiaStegNet, consistently outperform the classical SRM+EC baseline by capturing embedding-induced statistical traces.
Although high-fidelity methods such as StegFormer and DeepMIH are harder for weaker detectors, they remain detectable by stronger CNN-based models; similarly, text-payload methods with insufficient reliable long-bitstream support, such as HiDDeN and RoSteALS, are both easier to detect and less reliable in recovery.
Thus, practical defense should prioritize learned steganalysis over visual-quality inspection alone.
\end{tcolorbox}

\begin{figure*}[t]
    \centering
    \begin{subfigure}[t]{0.49\textwidth}
        \centering
        \includegraphics[width=\linewidth]{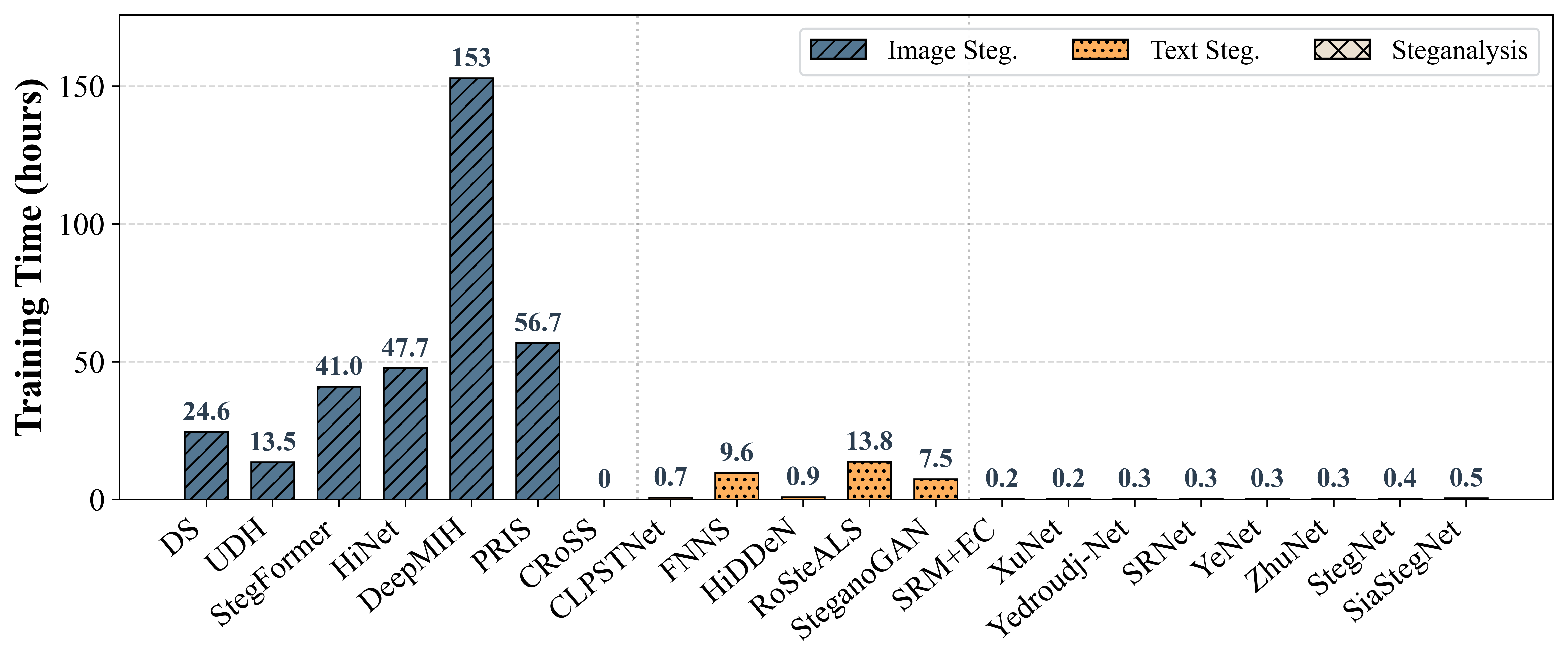}
        \caption{Training time (hours) comparison across image-payload steganography, text-payload steganography, and steganalysis methods.}
        \label{fig:combined_training_time}
    \end{subfigure}\hfill
    \begin{subfigure}[t]{0.49\textwidth}
        \centering
        \includegraphics[width=\linewidth]{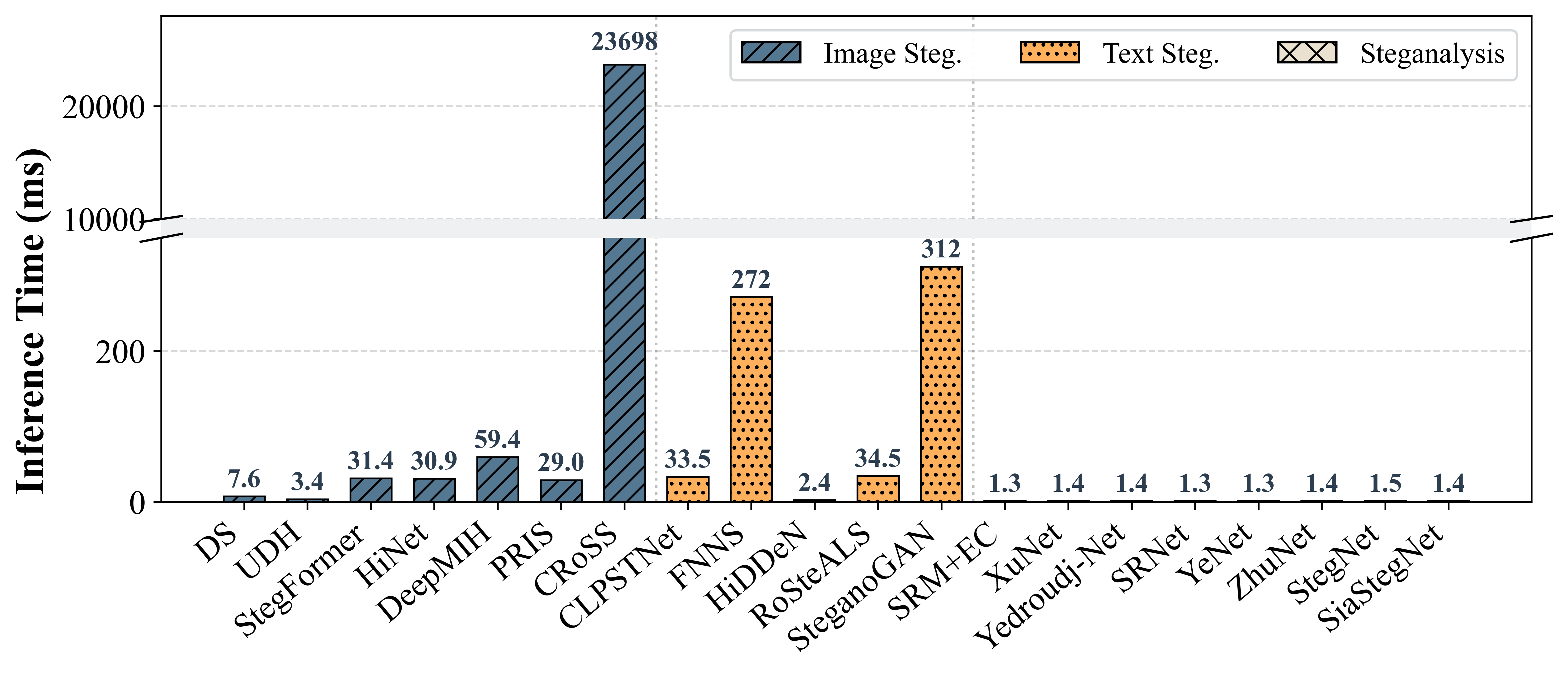}
        \caption{Inference time (ms) comparison across image steganography, text steganography, and steganalysis methods.}
        \label{fig:combined_inference_time}
    \end{subfigure}

    \caption{Computational efficiency comparison on ALASKA\#2.}
    \label{fig:combined_runtime}
\end{figure*}

\subsection{Efficiency}
We evaluate efficiency from two perspectives, namely training cost and inference overhead.
As shown in~\Cref{fig:combined_runtime}, the computational burden differs substantially between steganography methods on the adversary side and steganalysis models on the defender side.

For training costs, image-payload steganography is generally the most expensive.
Training times for these models typically range from 13.5 to 56.7 hours, with DeepMIH being a notable outlier at 153 hours, and CRoSS requiring 0 hours due to its use of a pre-trained generative model. Text-payload methods are relatively cheaper overall, with their training times spanning from under 1 hour to roughly 14 hours. 
In contrast, steganalysis detectors are much cheaper to train, taking only 0.2 to 0.5 hours across SRM+EC and CNN-based models.
This indicates that, under matched training-data assumptions, the defender can train effective detectors with a much lower computational cost than the adversary needs for most steganography models.
The inference results reveal an even sharper asymmetry.
Most steganalysis detectors require only 1.3 to 1.5 ms per sample, making online detection computationally affordable.
By comparison, most image-payload steganography methods typically require tens of milliseconds (ranging from roughly 3 to 60 ms).
CRoSS is an extreme outlier, requiring nearly 23,700 ms due to the high sampling cost of diffusion-based generation. Text-payload steganography also exhibits wide variance, with inference times spanning from as little as 2.4 ms to over 300 ms.

Combining the efficiency results with the steganography capability results in~\Cref{tab:attack_capability}, we observe that stronger image-payload attacks often require higher computational cost.
For example, DeepMIH achieves the best hiding capability on ALASKA\#2, with Cover/Stego PSNR of $44.130\,\mathrm{dB}$ and Secret/Recovery PSNR of $47.780\,\mathrm{dB}$.
However, it also requires the longest training time, i.e., 153 hours, and a relatively high inference time of 59.4 ms.
StegFormer and HiNet are more efficient, requiring 41.0 and 47.7 hours of training and around 31 ms per inference, but their overall hiding capability is lower than DeepMIH.
CRoSS is the least favorable case: although it requires no additional training in our evaluation, its hiding quality is poor and its inference time reaches 23698 ms.
For text-payload steganography, CLPSTNet provides the best effectiveness-efficiency trade-off.
It achieves the strongest text recovery on ALASKA\#2, with EMR of 0.983 and CER/BER of 0.001/0.000, while requiring only 0.7 hours of training and 33.5 ms per inference.
FNNS also achieves strong recovery, with EMR of 0.927, but its inference time is much higher at 272 ms.
These results suggest that, from the adversary side, obtaining both high payload recovery and good visual quality is usually not free, especially for high-capacity image payloads or long text bitstreams.

From the defender's perspective, the efficiency results are encouraging.
Strong detectors such as StegNet and SiaStegNet achieve top-tier steganalysis performance while requiring only 0.4 and 0.5 hours of training and 1.5 and 1.4 ms per inference, respectively.
Thus, even when adversaries use high-quality steganography methods, the defender can often obtain strong detection capability with much lower runtime overhead.
However, this advantage assumes access to representative stego training data; if the adversary's method or payload distribution is unknown, constructing matched training data may become the dominant cost, motivating the transferability analysis in~\Cref{sec:transferability}.

\begin{tcolorbox}[
  colback=black!5,     
  colframe=black,
  boxrule=0pt,
  toprule=0.4pt, 
  bottomrule=0.4pt,
  leftrule=0.4pt,
  rightrule=0.4pt,
  left=4pt,right=4pt, 
  top=2pt,bottom=2pt,
  boxsep=0pt,
  before skip=4pt,
  after skip=4pt
]
\textbf{Takeaway:} 
There is a fundamental computational asymmetry in the steganographic arms race.
For adversaries, achieving high capacity and visual fidelity remains intrinsically expensive, dictating a strict effectiveness-efficiency trade-off. Conversely, defenders enjoy a massive efficiency advantage, rendering near real-time detection highly scalable.
Consequently, the true battleground shifts from computational power to data superiority: the defender's primary bottleneck is no longer algorithmic efficiency, but acquiring representative training distributions, making cross-domain transferability the ultimate deciding factor.
\end{tcolorbox}

\begin{table*}[t]
\centering
\footnotesize
\caption{Cross-cover transfer for steganography (DIV2K→ALASKA\#2 and ALASKA\#2→DIV2K).}
\label{tab:attack_capability_cross_cover}

\begin{subtable}[t]{\textwidth}
\centering
\caption{Image-payload steganography.}
\label{tab:attack_capability_image_payload_cross_cover}

\begin{minipage}[t]{0.49\textwidth}
\centering
\textbf{DIV2K$\rightarrow$ALASKA\#2}
\resizebox{\linewidth}{!}{%
\begin{tabular}{lcccccc}
\toprule
Metric & DS & UDH & StegFormer & HiNet & DeepMIH & PRIS \\
\midrule
\multicolumn{7}{c}{\emph{Cover/Stego image pair}} \\
PSNR$\bm{\uparrow}$ & \cellcolor{w4}29.761 & \cellcolor{w2}35.560 & \cellcolor{w1}42.000 & \cellcolor{w3}31.064 & \cellcolor{w1}\textbf{43.900} & \cellcolor{w4}30.610 \\
SSIM$\bm{\uparrow}$ & \cellcolor{w2}0.855 & \cellcolor{w1}0.921 & \cellcolor{w1}0.982 & \cellcolor{w2}0.858 & \cellcolor{w1}\textbf{0.986} & \cellcolor{w2}0.826 \\
MAE$\bm{\downarrow}$ & \cellcolor{c8}6.593 & \cellcolor{c7}3.591 & \cellcolor{c2}1.618 & \cellcolor{c8}5.019 & \cellcolor{c1}\textbf{1.332} & \cellcolor{c8}5.536 \\
LPIPS$\bm{\downarrow}$ & \cellcolor{c10}0.129 & \cellcolor{c6}0.002 & \cellcolor{c1}\textbf{0.001} & \cellcolor{c10}0.036 & \cellcolor{c1}\textbf{0.001} & \cellcolor{c9}0.006 \\
\midrule
\multicolumn{7}{c}{\emph{Secret/Recovery image pair}} \\
PSNR$\bm{\uparrow}$ & \cellcolor{w4}30.973 & \cellcolor{w5}25.550 & \cellcolor{w3}33.870 & \cellcolor{w3}37.190 & \cellcolor{w1}\textbf{47.350} & \cellcolor{w4}29.761 \\
SSIM$\bm{\uparrow}$ & \cellcolor{w1}0.912 & \cellcolor{w2}0.820 & \cellcolor{w1}0.969 & \cellcolor{w1}0.973 & \cellcolor{w1}\textbf{0.997} & \cellcolor{w1}0.924 \\
MAE$\bm{\downarrow}$ & \cellcolor{c9}5.407 & \cellcolor{c10}10.725 & \cellcolor{c9}4.263 & \cellcolor{c8}2.412 & \cellcolor{c1}\textbf{0.698} & \cellcolor{c9}5.763 \\
LPIPS$\bm{\downarrow}$ & \cellcolor{c4}0.057 & \cellcolor{c10}0.148 & \cellcolor{c2}0.017 & \cellcolor{c1}0.008 & \cellcolor{c1}\textbf{0.000} & \cellcolor{c4}0.047 \\
\bottomrule
\end{tabular}%
}
\end{minipage}
\hfill
\begin{minipage}[t]{0.49\textwidth}
\centering
\textbf{ALASKA\#2$\rightarrow$DIV2K}
\resizebox{\linewidth}{!}{%
\begin{tabular}{lcccccc}
\toprule
Metric & DS & UDH & StegFormer & HiNet & DeepMIH & PRIS \\
\midrule
\multicolumn{7}{c}{\emph{Cover/Stego image pair}} \\
PSNR$\bm{\uparrow}$ & \cellcolor{w3}32.653 & \cellcolor{w2}36.122 & \cellcolor{w1}40.510 & \cellcolor{w3}33.186 & \cellcolor{w1}\textbf{43.210} & \cellcolor{w3}30.700 \\
SSIM$\bm{\uparrow}$ & \cellcolor{w1}0.900 & \cellcolor{w1}0.941 & \cellcolor{w1}0.990 & \cellcolor{w1}0.941 & \cellcolor{w1}\textbf{0.994} & \cellcolor{w2}0.864 \\
MAE$\bm{\downarrow}$ & \cellcolor{c8}4.414 & \cellcolor{c7}3.287 & \cellcolor{c2}1.444 & \cellcolor{c7}3.962 & \cellcolor{c1}\textbf{1.199} & \cellcolor{c8}5.464 \\
LPIPS$\bm{\downarrow}$ & \cellcolor{c10}0.006 & \cellcolor{c2}0.001 & \cellcolor{c4}0.002 & \cellcolor{c9}0.005 & \cellcolor{c1}\textbf{0.000} & \cellcolor{c6}0.003 \\
\midrule
\multicolumn{7}{c}{\emph{Secret/Recovery image pair}} \\
PSNR$\bm{\uparrow}$ & \cellcolor{w3}33.790 & \cellcolor{w4}30.560 & \cellcolor{w2}40.840 & \cellcolor{w2}40.860 & \cellcolor{w1}\textbf{47.090} & \cellcolor{w4}30.117 \\
SSIM$\bm{\uparrow}$ & \cellcolor{w1}0.948 & \cellcolor{w1}0.934 & \cellcolor{w1}0.984 & \cellcolor{w1}0.981 & \cellcolor{w1}\textbf{0.997} & \cellcolor{w1}0.928 \\
MAE$\bm{\downarrow}$ & \cellcolor{c9}3.658 & \cellcolor{c9}5.302 & \cellcolor{c6}1.596 & \cellcolor{c6}1.590 & \cellcolor{c1}\textbf{0.708} & \cellcolor{c9}5.428 \\
LPIPS$\bm{\downarrow}$ & \cellcolor{c10}0.040 & \cellcolor{c10}0.042 & \cellcolor{c3}0.010 & \cellcolor{c1}0.003 & \cellcolor{c1}\textbf{0.000} & \cellcolor{c10}0.043 \\
\bottomrule
\end{tabular}%
}
\end{minipage}

\end{subtable}

\begin{subtable}[t]{\textwidth}
\centering
\caption{Text-payload steganography.}
\label{tab:attack_capability_text_payload_cross_cover}

\begin{minipage}[t]{0.49\textwidth}
\centering
\textbf{DIV2K$\rightarrow$ALASKA\#2}
\resizebox{\linewidth}{!}{%
\begin{tabular}{lccccc}
\toprule
Metric & HiDDeN & SteganoGAN & FNNS & CLPSTNet & RoSteALS \\
\midrule
\multicolumn{6}{c}{\emph{Cover/Stego image pair}} \\
PSNR$\bm{\uparrow}$ & \cellcolor{w3}30.117 & \cellcolor{w1}\textbf{39.240} & \cellcolor{w2}34.758 & \cellcolor{w2}34.230 & \cellcolor{w2}33.900 \\
SSIM$\bm{\uparrow}$ & \cellcolor{w1}0.948 & \cellcolor{w1}0.967 & \cellcolor{w1}0.913 & \cellcolor{w1}\textbf{0.974} & \cellcolor{w1}0.943 \\
MAE$\bm{\downarrow}$ & \cellcolor{c7}6.666 & \cellcolor{c1}\textbf{2.253} & \cellcolor{c5}3.773 & \cellcolor{c1}2.442 & \cellcolor{c5}4.167 \\
LPIPS$\bm{\downarrow}$ & \cellcolor{c10}0.074 & \cellcolor{c1}\textbf{0.005} & \cellcolor{c8}0.018 & \cellcolor{c3}0.007 & \cellcolor{c9}0.035 \\
\midrule
\multicolumn{6}{c}{\emph{Secret/Recovery text pair}} \\
EMR$\bm{\uparrow}$ & \cellcolor{w10}0.000 & \cellcolor{w4}0.563 & \cellcolor{w3}0.661 & \cellcolor{w1}\textbf{0.859} & \cellcolor{w10}0.000 \\
CER$\bm{\downarrow}$ & \cellcolor{c10}0.701 & \cellcolor{c10}0.437 & \cellcolor{c10}0.339 & \cellcolor{c1}\textbf{0.019} & \cellcolor{c10}0.963 \\
BER$\bm{\downarrow}$ & \cellcolor{c10}0.218 & \cellcolor{c10}0.437 & \cellcolor{c10}0.339 & \cellcolor{c1}\textbf{0.007} & \cellcolor{c10}0.297 \\
\bottomrule
\end{tabular}%
}
\end{minipage}
\hfill
\begin{minipage}[t]{0.49\textwidth}
\centering
\textbf{ALASKA\#2$\rightarrow$DIV2K}
\resizebox{\linewidth}{!}{%
\begin{tabular}{lccccc}
\toprule
Metric & HiDDeN & SteganoGAN & FNNS & CLPSTNet & RoSteALS \\
\midrule
\multicolumn{6}{c}{\emph{Cover/Stego image pair}} \\
PSNR$\bm{\uparrow}$ & \cellcolor{w3}31.551 & \cellcolor{w1}\textbf{40.840} & \cellcolor{w2}35.582 & \cellcolor{w1}40.773 & \cellcolor{w4}26.300 \\
SSIM$\bm{\uparrow}$ & \cellcolor{w1}0.965 & \cellcolor{w1}0.986 & \cellcolor{w1}0.954 & \cellcolor{w1}\textbf{0.997} & \cellcolor{w1}0.924 \\
MAE$\bm{\downarrow}$ & \cellcolor{c8}5.314 & \cellcolor{c4}1.763 & \cellcolor{c7}3.257 & \cellcolor{c1}\textbf{1.091} & \cellcolor{c9}8.549 \\
LPIPS$\bm{\downarrow}$ & \cellcolor{c4}0.017 & \cellcolor{c1}0.002 & \cellcolor{c2}0.009 & \cellcolor{c1}\textbf{0.000} & \cellcolor{c10}0.052 \\
\midrule
\multicolumn{6}{c}{\emph{Secret/Recovery text pair}} \\
EMR$\bm{\uparrow}$ & \cellcolor{w10}0.000 & \cellcolor{w3}0.700 & \cellcolor{w1}0.930 & \cellcolor{w1}\textbf{0.980} & \cellcolor{w10}0.000 \\
CER$\bm{\downarrow}$ & \cellcolor{c10}0.703 & \cellcolor{c5}0.300 & \cellcolor{c1}0.070 & \cellcolor{c1}\textbf{0.000} & \cellcolor{c10}0.714 \\
BER$\bm{\downarrow}$ & \cellcolor{c8}0.218 & \cellcolor{c10}0.300 & \cellcolor{c3}0.070 & \cellcolor{c1}\textbf{0.000} & \cellcolor{c8}0.227 \\
\bottomrule
\end{tabular}%
}
\end{minipage}

\end{subtable}

\end{table*}

\begin{table*}[h!]
  \centering
  \footnotesize
  \caption{Secret-domain transfer evaluation on ALASKA\#2. \small{Image-payload methods: trained on Hateful Memes, tested on MM-SafetyBench SD. Text-payload methods: trained on AdvBench, tested on StrongREJECT.}}
  \label{tab:cross_secret_steganography}
  \begin{minipage}[t]{0.49\textwidth}
  \centering
  \textbf{Image Payload}
  \resizebox{\linewidth}{!}{%
  \begin{tabular}{lcccccc}
  \toprule
  Metric & DS & UDH & StegFormer & HiNet & DeepMIH & PRIS \\
  \midrule
  \multicolumn{7}{c}{\emph{Cover/Stego image pair}} \\
  PSNR$\bm{\uparrow}$    & \cellcolor{w3}32.830  & \cellcolor{w2}38.510  & \cellcolor{w1}42.550  & \cellcolor{w2}36.420  & \cellcolor{w1}\textbf{44.350}  & \cellcolor{w3}31.230 \\
  SSIM$\bm{\uparrow}$    & \cellcolor{w2}0.870 & \cellcolor{w1}0.954 & \cellcolor{w1}0.985 & \cellcolor{w1}0.950 & \cellcolor{w1}\textbf{0.988} & \cellcolor{w2}0.832 \\
  MAE$\bm{\downarrow}$   & \cellcolor{c8}4.540 & \cellcolor{c5}2.438 & \cellcolor{c2}1.379 & \cellcolor{c6}2.908 & \cellcolor{c1}\textbf{1.238} & \cellcolor{c8}5.474 \\
  LPIPS$\bm{\downarrow}$ & \cellcolor{c9}0.008 & \cellcolor{c1}\textbf{0.001} & \cellcolor{c5}0.002 & \cellcolor{c8}0.005 & \cellcolor{c1}\textbf{0.001} & \cellcolor{c8}0.004 \\
  \midrule
  \multicolumn{7}{c}{\emph{Secret/Recovery image pair}} \\
  PSNR$\bm{\uparrow}$    & \cellcolor{w3}33.810  & \cellcolor{w4}31.660  & \cellcolor{w3}33.840  & \cellcolor{w1}43.140  & \cellcolor{w1}\textbf{47.480}  & \cellcolor{w4}30.660 \\
  SSIM$\bm{\uparrow}$    & \cellcolor{w1}0.938 & \cellcolor{w1}0.916 & \cellcolor{w1}0.991 & \cellcolor{w1}0.989 & \cellcolor{w1}\textbf{0.996} & \cellcolor{w1}0.917 \\
  MAE$\bm{\downarrow}$   & \cellcolor{c9}3.890 & \cellcolor{c9}5.593 & \cellcolor{c9}4.738 & \cellcolor{c5}1.315 & \cellcolor{c1}\textbf{0.740} & \cellcolor{c9}5.599 \\
  LPIPS$\bm{\downarrow}$ & \cellcolor{c10}0.071 & \cellcolor{c10}0.071 & \cellcolor{c2}0.008 & \cellcolor{c1}0.002 & \cellcolor{c1}\textbf{0.000} & \cellcolor{c10}0.075 \\
  \bottomrule
  \end{tabular}%
  }
  \end{minipage}
  \hfill
  \begin{minipage}[t]{0.49\textwidth}
  \centering
  \textbf{Text Payload}
  \resizebox{\linewidth}{!}{%
  \begin{tabular}{lccccc}
  \toprule
  Metric & HiDDeN & SteganoGAN & FNNS & RoSteALS & CLPSTNet \\
  \midrule
  \multicolumn{6}{c}{\emph{Cover/Stego image pair}} \\
  PSNR$\bm{\uparrow}$    & \cellcolor{w3}31.950  & \cellcolor{w1}40.150  & \cellcolor{w2}35.610  & \cellcolor{w4}27.310  & \cellcolor{w1}\textbf{41.160} \\
  SSIM$\bm{\uparrow}$    & \cellcolor{w1}0.963 & \cellcolor{w1}0.974 & \cellcolor{w1}0.935 & \cellcolor{w1}0.919 & \cellcolor{w1}\textbf{0.992} \\
  MAE$\bm{\downarrow}$   & \cellcolor{c8}5.360 & \cellcolor{c4}1.985 & \cellcolor{c7}3.326 & \cellcolor{c9}8.016 & \cellcolor{c1}\textbf{1.311} \\
  LPIPS$\bm{\downarrow}$ & \cellcolor{c5}0.032 & \cellcolor{c1}0.004 & \cellcolor{c3}0.018 & \cellcolor{c10}0.079 & \cellcolor{c1}\textbf{0.000} \\
  \midrule
  \multicolumn{6}{c}{\emph{Secret/Recovery text (EMR/BER/CER)}} \\
  EMR$\bm{\uparrow}$     & \cellcolor{w10}0.001 & \cellcolor{w2}0.407 & \cellcolor{w1}\textbf{0.486} & \cellcolor{w10}0.000 & \cellcolor{w3}0.379 \\
  BER$\bm{\downarrow}$   & \cellcolor{c9}0.278 & \cellcolor{c10}0.593 & \cellcolor{c9}0.514 & \cellcolor{c9}0.299 & \cellcolor{c1}\textbf{0.053} \\
  CER$\bm{\downarrow}$   & \cellcolor{c9}0.835 & \cellcolor{c8}0.593 & \cellcolor{c8}0.514 & \cellcolor{c9}0.822 & \cellcolor{c1}\textbf{0.119} \\
  \bottomrule
  \end{tabular}%
  }
  \end{minipage}

\end{table*}

\begin{figure*}[t]
    \centering
    \begin{subfigure}[b]{0.49\linewidth}
        \centering
        \includegraphics[width=0.90\linewidth]{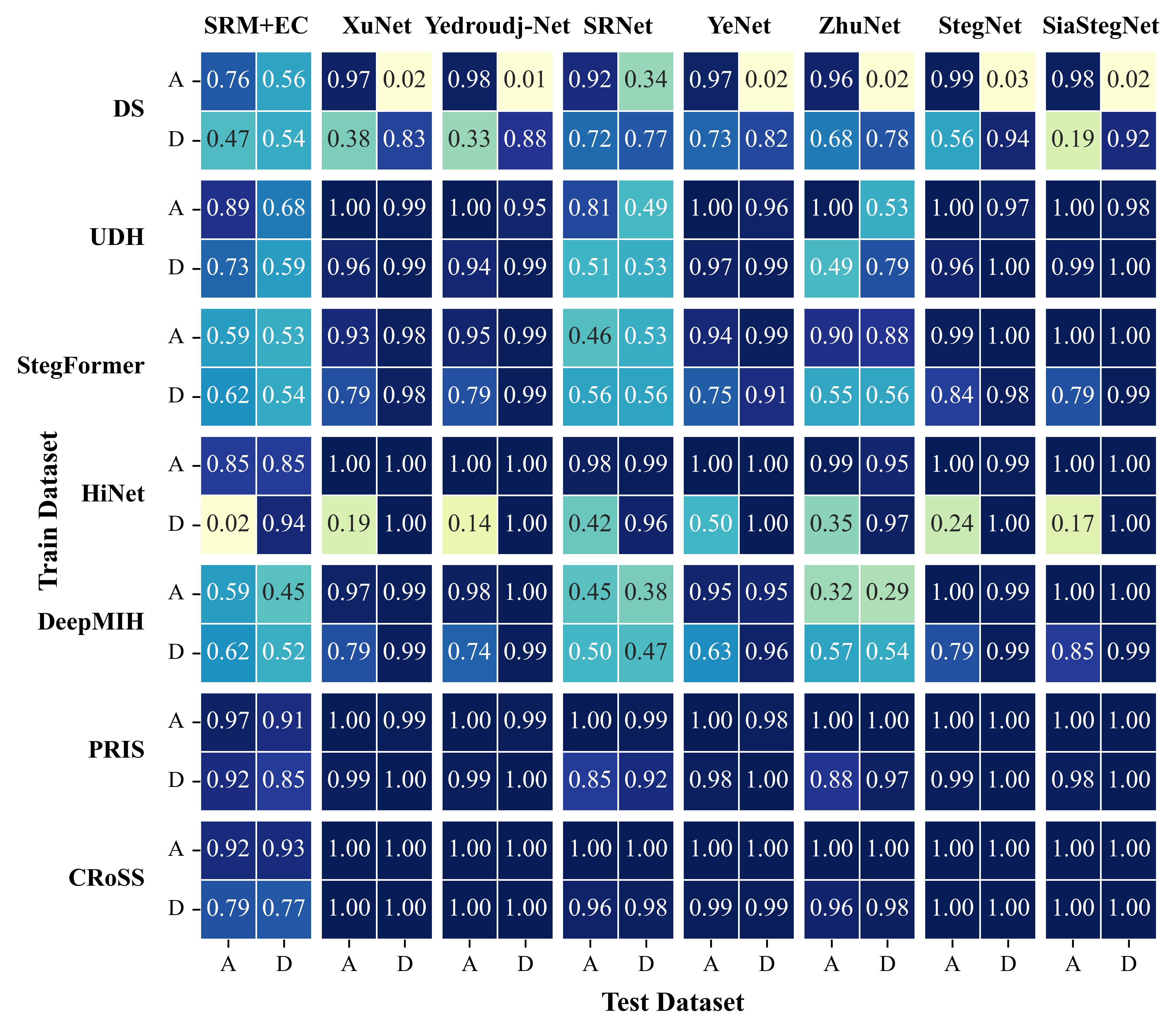}
        \subcaption{Detection results for image-payload steganography.}
        \label{fig:image-payload-transfer}
    \end{subfigure}
    \hfill
    \begin{subfigure}[b]{0.49\linewidth}
        \centering
        \includegraphics[width=0.96\linewidth]{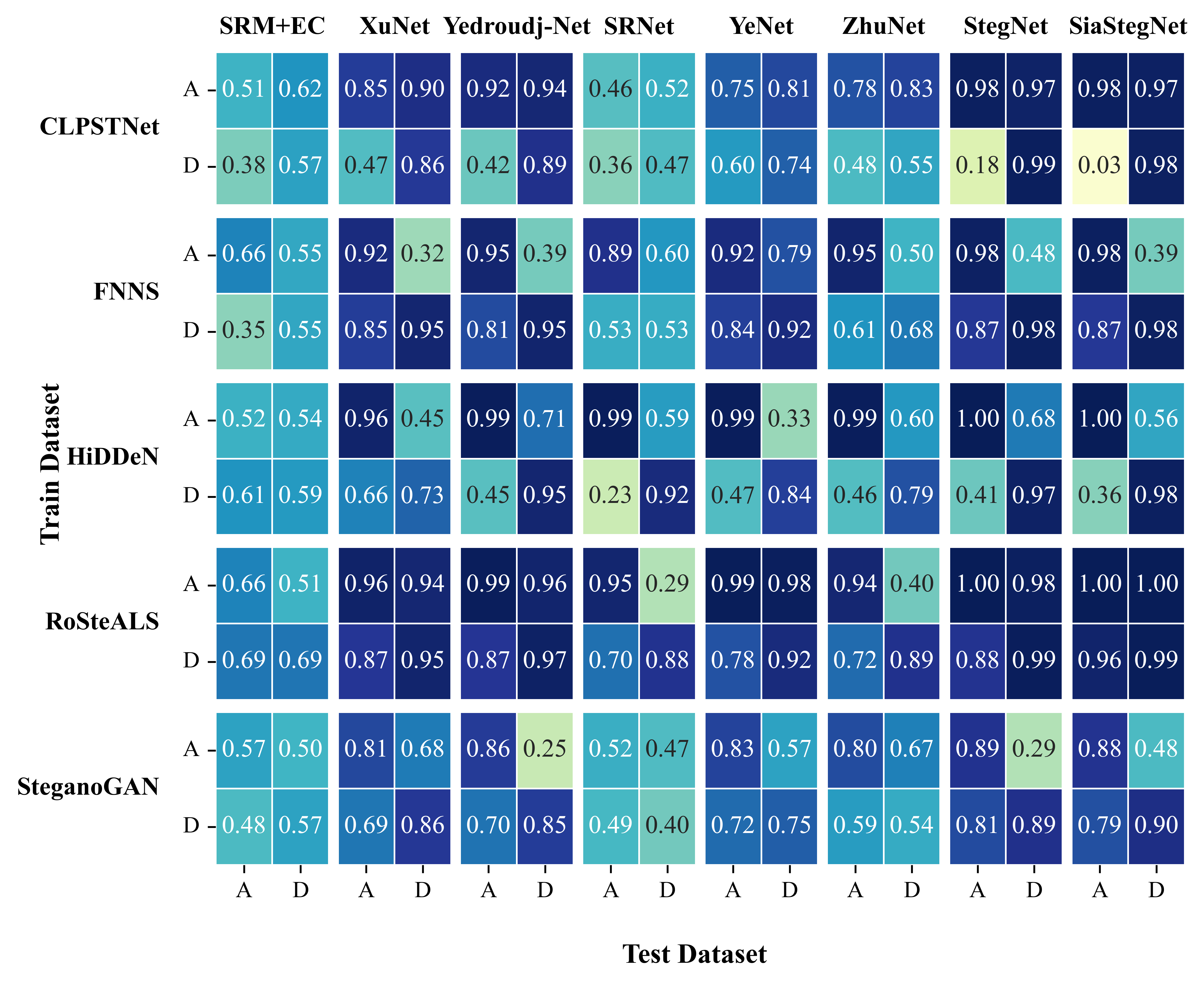}
        \subcaption{Detection results for text-payload steganography.}
        \label{fig:text-payload-transfer}
    \end{subfigure}

    \caption{Cross-dataset transfer performance (F1-score) of the steganalysis detector. 
    A denotes the ALASKA\#2 dataset and D denotes the DIV2K dataset.}
    \label{fig:payload-transfer}
\end{figure*}

\subsection{Transferability}
\label{sec:transferability}
Driven by the need to minimize costs, we next evaluate transferability from both perspectives.
For the adversary, cross-dataset robustness offers a way to bypass the massive computational overhead of retraining; for the defender, transferability not only reduces data-collection costs but also reveals the potential to detect unseen, zero-day steganographic threats.

\mypara{Cross-dataset Transferability for Steganography}
Since training and hyperparameter tuning are costly, adversaries may prefer steganography methods that remain stable under distribution shifts.
We consider two transfer settings.
(i) \textbf{Cross-cover transfer:} models are trained on one cover dataset and directly applied to the other, testing robustness to cover-image distribution shifts.
For brevity, we denote DIV2K as D and ALASKA\#2 as A, and use D$\rightarrow$A and A$\rightarrow$D to indicate the two directions.
(ii) \textbf{Cross-secret transfer:} models are trained with one secret dataset and evaluated on another, testing whether the hiding and recovery mechanism generalizes to unseen payload distributions.

For \textbf{Cross-cover transfer}, the results in~\Cref{tab:attack_capability_cross_cover} show that transferability varies substantially across methods.
In the image-payload setting, DeepMIH remains the most robust method, achieving the best Cover/Stego PSNR of $43.900\,\mathrm{dB}$ under D$\rightarrow$A and $43.210\,\mathrm{dB}$ under A$\rightarrow$D, as well as the strongest Secret/Recovery PSNR of $47.350\,\mathrm{dB}$ and $47.090\,\mathrm{dB}$.
This suggests that its invertible design preserves a stable information path under cover-domain shifts.
StegFormer also maintains high cover fidelity, with Cover/Stego PSNR of $42.000\,\mathrm{dB}$ and $40.510\,\mathrm{dB}$, while HiNet remains competitive mainly in Secret/Recovery quality.
Overall, cross-cover robustness depends not only on visual fidelity, but also on whether the recovery path remains stable after the cover distribution changes.

In the text-payload setting, CLPSTNet shows the strongest message recovery under cross-cover transfer.
It achieves the highest EMR of 0.859 under D$\rightarrow$A and 0.980 under A$\rightarrow$D, with the lowest CER/BER of 0.019/0.007 and 0.000/0.000, respectively.
SteganoGAN provides strong Cover/Stego quality, with PSNR of $39.240\,\mathrm{dB}$ and $40.840\,\mathrm{dB}$, but its EMR is lower at 0.563 and 0.700.
FNNS transfers reasonably well, especially under A$\rightarrow$D with EMR of 0.930, whereas HiDDeN and RoSteALS fail to recover text reliably, with EMR remaining at 0.000 in both directions.
This indicates that text-payload transfer is mainly limited by reliable long-bitstream recovery rather than visual quality alone.

For \textbf{Cross-secret transfer}, we use MM-SafetyBench~\cite{DBLP:conf/eccv/LiuZGLYQ24} as the new secret dataset for image-payload steganography and StrongREJECT~\cite{DBLP:conf/nips/SoulyLBTHPASEWT24} as the new secret dataset for text-payload steganography.
As shown in~\Cref{tab:cross_secret_steganography}, image-payload methods are generally more robust to secret-domain shifts than text-payload methods.
DeepMIH again achieves the best overall performance, with Cover/Stego PSNR of $44.35\,\mathrm{dB}$ and Secret/Recovery PSNR of $47.48\,\mathrm{dB}$.
HiNet also shows strong recovery with Secret/Recovery PSNR of $43.14\,\mathrm{dB}$.
In contrast, StegFormer maintains good cover fidelity with PSNR of $42.55\,\mathrm{dB}$, but its Secret/Recovery PSNR drops to $33.84\,\mathrm{dB}$, suggesting that cover preservation and secret reconstruction can decouple under secret-domain shifts.

The text-payload setting is more sensitive to secret-domain transfer.
CLPSTNet achieves the best Cover/Stego quality, with PSNR of $41.16\,\mathrm{dB}$, and the lowest BER/CER of 0.053/0.119, indicating better average bit- and character-level preservation.
On the other hand, FNNS obtains the highest EMR of 0.486, followed by SteganoGAN at 0.407 and CLPSTNet at 0.379.
This reveals a discrepancy between exact-message recovery and average decoding quality: FNNS produces more fully correct messages, whereas CLPSTNet makes fewer bit- and character-level errors overall.
HiDDeN and RoSteALS remain poor, with EMR of only 0.001 and 0.000.
Overall, cross-secret transfer is challenging for text-payload steganography, where small bit-level errors can accumulate into message-level failures.

\begin{figure*}[t]
    \centering
    \includegraphics[width=0.95\linewidth]{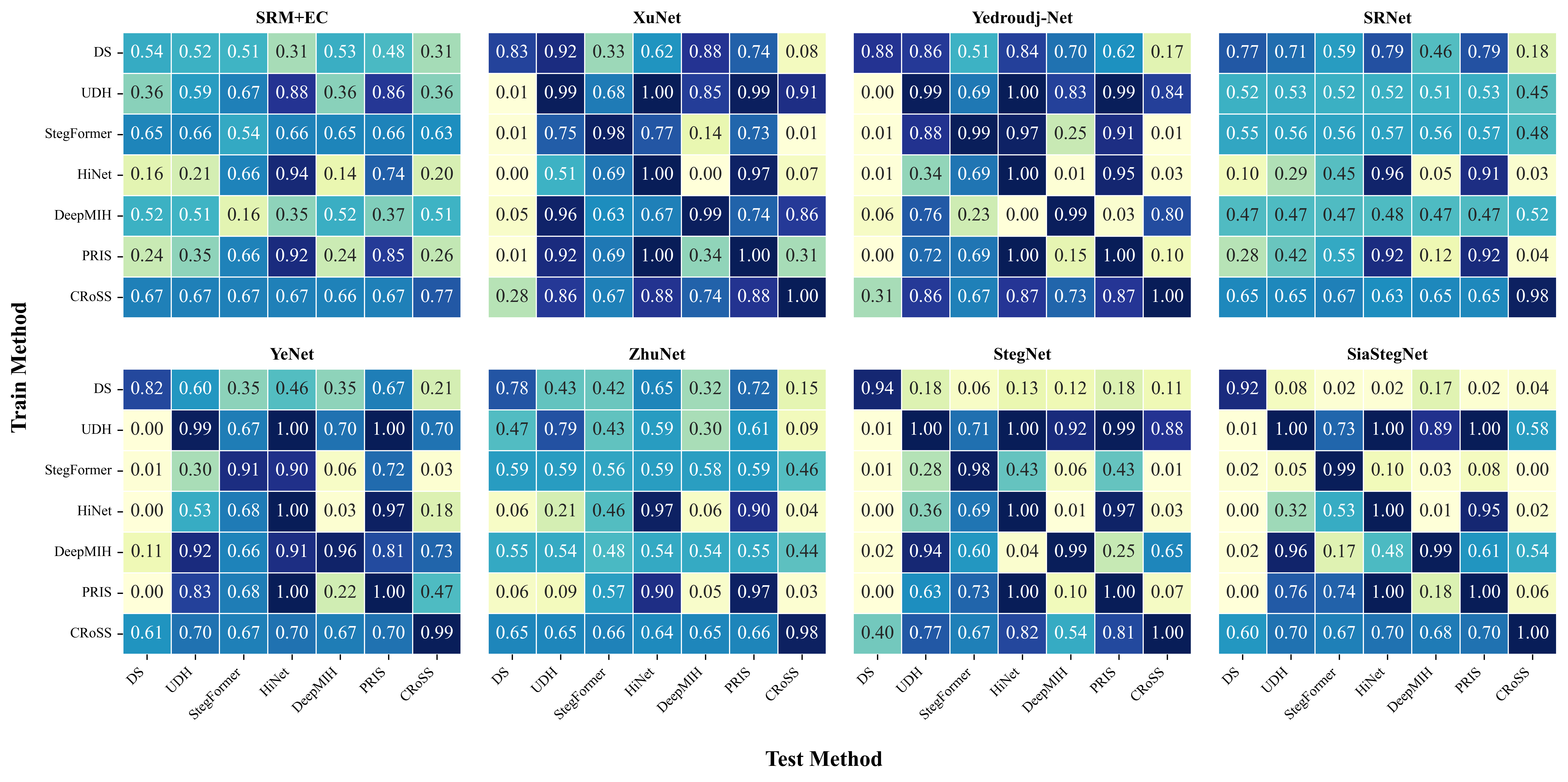}
    \caption{Cross-method transferability of steganalysis on image payloads.}
    \label{fig:transfer_image}
\end{figure*}

\mypara{Transferability for Steganalysis}
For steganalysis, transferability reflects whether a defender can still detect stego samples when the deployment condition differs from the training setting.
We consider two practical forms of uncertainty:
\textbf{known-method detection under cover-distribution shift}, where the steganography method is known but the cover dataset changes, and
\textbf{zero-day detection under unseen steganography methods}, where the detector is tested on stego samples generated by methods that were never used during training.
The zero-day setting further includes two cases: \underline{cross-method detection}, where the source and target methods belong to the same payload modality, and \underline{cross-payload-modality detection}, where the detector is trained on one payload modality, i.e., image-payload or text-payload steganography, and evaluated on the other.

For the first setting, \Cref{fig:payload-transfer} shows that high in-domain F1-scores can overestimate real deployment robustness.
For image-payload steganography, transferability is highly method- and direction-dependent.
DS suffers severe A$\rightarrow$D degradation: XuNet, Yedroudj-Net, YeNet, and StegNet drop from 0.97, 0.98, 0.97, and 0.99 to only 0.02, 0.01, 0.02, and 0.03, respectively.
HiNet shows the opposite failure mode under D$\rightarrow$A, where SRM+EC, XuNet, StegNet, and SiaStegNet drop to 0.02, 0.19, 0.24, and 0.17.
These asymmetric failures suggest that some detectors learn cover-domain-specific residual statistics rather than cover-invariant steganographic traces.
By contrast, PRIS and CRoSS are much easier to transfer across datasets, with most CNN-based detectors maintaining near-perfect F1-scores in both directions, indicating that their artifacts are more consistent across cover distributions.
StegFormer and DeepMIH lie between these two extremes: strong detectors such as StegNet and SiaStegNet often remain effective, while weaker or less transferable detectors such as SRM+EC, SRNet, and ZhuNet can degrade substantially.
For text-payload steganography, the transfer behavior is also method-dependent but reflects payload-specific artifacts.
RoSteALS is relatively stable for most CNN-based detectors; for example, StegNet obtains 1.00, 0.98, 0.88, and 0.99 under A$\rightarrow$A, A$\rightarrow$D, D$\rightarrow$A, and D$\rightarrow$D, respectively, while SiaStegNet stays between 0.96 and 1.00.
In contrast, CLPSTNet shows strong in-domain detection but weak D$\rightarrow$A transfer: StegNet drops from 0.99 under D$\rightarrow$D to 0.18 under D$\rightarrow$A, and SiaStegNet drops from 0.98 to 0.03.
FNNS also exhibits clear A$\rightarrow$D degradation, with XuNet, Yedroudj-Net, StegNet, and SiaStegNet dropping to 0.32, 0.39, 0.48, and 0.39, respectively.
These results indicate that even when the attack method is known, cover-distribution shift can break detector generalization.

We further evaluate the zero-day setting from two perspectives.
First, we study cross-method zero-day detection in~\Cref{fig:transfer_image,fig:transfer_text}, where detectors are trained on one steganography method and tested on another unseen method within the same payload modality.
The heatmaps show a clear diagonal-dominant pattern: detectors usually perform best when the training and testing methods match, while off-diagonal transfer is often unstable.
For image-payload steganography, transfer succeeds only when the source and target methods leave similar artifacts.
For example, XuNet trained on CRoSS transfers reasonably well to UDH, HiNet, and PRIS, with F1-scores of 0.86, 0.88, and 0.88, but is weaker on DeepMIH at 0.74.
Conversely, XuNet trained on DS reaches 0.83 on matched DS but only 0.08 on CRoSS, showing that artifacts learned from one method may not cover another.
Even strong detectors can fail under zero-day method shift: StegNet achieves 0.98 on matched StegFormer but drops to 0.01 on DS, 0.06 on DeepMIH, and 0.01 on CRoSS.
For text-payload steganography, zero-day transfer is even more volatile.
SiaStegNet trained on HiDDeN reaches 0.98 on matched HiDDeN, but drops to 0.03 on CLPSTNet and 0.00 on FNNS, RoSteALS, and SteganoGAN.
This indicates that text-payload detectors often learn method-specific bit-embedding fingerprints rather than generalizable steganographic signatures.
Second, we evaluate cross-payload-modality zero-day detection in~\Cref{fig:cross_payload_modality}, where detectors are trained on one payload modality and tested on the other.
The results show that such transfer is generally weak, with the overall average F1-score ranging only from 0.10 to 0.30 across detectors.
Although a few target methods remain partially detectable, such as CRoSS in the text-payload $\rightarrow$ image-payload direction, where ZhuNet and SRM+EC achieve average F1-scores of 0.66 and 0.60, respectively, most cross-modality cases exhibit poor generalization.
This indicates that image-payload and text-payload steganography tend to leave different modality-specific artifacts, making zero-day detection substantially harder when the payload modality also changes.

Overall, practical steganalysis cannot rely on matched-setting F1 alone: robust defense requires not only cover-domain generalization but also zero-day robustness across unseen methods, including both cross-method and cross-payload-modality settings.

\begin{figure}[h!]
    \centering
    \includegraphics[width=\linewidth]{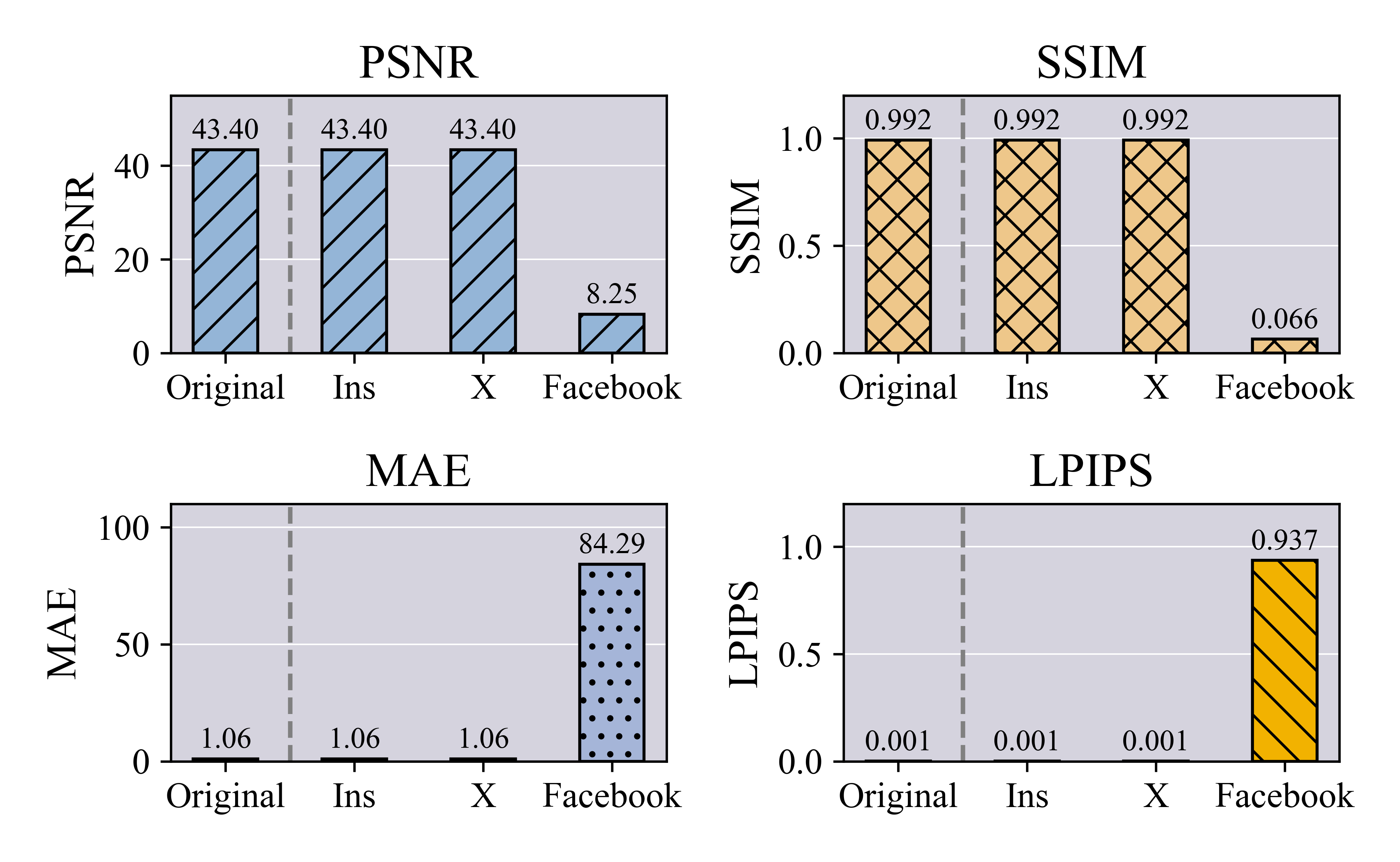}
    \caption{Performance comparison of image quality across different social media platforms.}
    \label{fig:social_media_metrics}
\end{figure}

\mypara{Transmission of Stego Images on Social Media}
As described in~\Cref{sec:adversary}, the adversary may disseminate stego images carrying malicious payloads through social media to achieve malicious goals.
To evaluate robustness under real-world social media transmission pipelines, we select DeepMIH, the best-performing method among the image-payload steganography methods, and test the robustness of its stego images on $3$ mainstream platforms, namely X~\cite{X}, Instagram~\cite{Instagram}, and Facebook~\cite{Facebook}, which are among the most widely used social media services today.
Specifically, we manually upload the stego images to each platform and then download the processed images.
We then extract the corresponding secrets from the downloaded images and assess the recovery quality.
For each platform, we uploaded and downloaded $10$ stego images, resulting in a small-scale evaluation that is negligible compared with the normal image-processing workload handled by these large-scale platforms.
All uploaded images were kept private or otherwise restricted from public visibility whenever the platform provided such an option, and they were not shared, promoted, or exposed to other users.
The embedded secrets used in our experiments were benign test payloads generated solely for evaluation purposes and did not contain executable code, personal information, credentials, harmful instructions, or any operationally useful content.
Therefore, even if a third party were able to access and decode the embedded payloads, the recovered messages would not cause harm.

As shown in~\Cref{fig:social_media_metrics}, in our tested upload/download setting, X and Instagram do not apply additional compression to user-uploaded images, thereby preserving image quality and enabling the embedded payload to be reliably extracted after download.
In contrast, Facebook adopts more aggressive compression, causing noticeable degradation and consistently preventing payload recovery in our experiments.
This observation suggests that, as storage costs decline, some social media platforms tend to adopt weaker compression or even near-lossless image processing pipelines.
Such pipelines objectively reduce the probability that steganographic payloads are destroyed along the platform transmission chain, which in turn increases the risk of steganographic poisoning or information-injection attacks.
For platforms such as Facebook that actively modify image content, for example via resampling and compression, a substantial body of prior work has discussed their impact on steganographic robustness~\cite{DBLP:journals/tcsv/TaoLZW19}.

To investigate this further, we evaluate a broader range of common social media processing techniques in~\Cref{sec:rob_test}.
Our results indicate that without adaptation, the original method lacks robustness against these distortions, leading to a significant decline in recoverability performance.
However, we further discuss in~\Cref{sec:robustness} that an adversary can infer the platform's specific compression rules by analyzing the discrepancy between uploaded and downloaded images.
By leveraging this information to perform simulated channel training, the adversary can effectively enhance the robustness and performance of image steganography methods on such social media platforms.

\begin{tcolorbox}[
  colback=black!5,     
  colframe=black,
  boxrule=0pt,
  toprule=0.4pt, 
  bottomrule=0.4pt,
  leftrule=0.4pt,
  rightrule=0.4pt,
  left=4pt,right=4pt, 
  top=2pt,bottom=2pt,
  boxsep=0pt,
  before skip=4pt,
  after skip=4pt
]
\textbf{Takeaways:} 
\textbf{(i) Attack Transferability is asymmetric across modalities:} Steganography generalizes well to new cover images but struggles with secret-domain shifts.
Text payloads are vulnerable to these shifts due to the accumulation of bit-level errors.
\textbf{(ii) Zero-day threats expose critical defense vulnerabilities:} Steganalysis detectors are highly sensitive to distribution shifts and often overfit to domain-specific residual statistics.
Even top-performing detectors fail when confronted with unseen cover datasets or zero-day steganographic methods, proving that high in-domain detection metrics are insufficient for open-world robustness.
\textbf{(iii) Countering platform compression via simulation:} Although social media platforms may compress stego images and compromise secret recovery, adversaries can reduce this risk by employing simulated channel training to mimic platform-specific processing rules.
\textbf{(iv) Unpredictable processing pipelines lower steganographic robustness:} By implementing randomized compression strategies rather than static rules, defenders can increase the difficulty of channel estimation for adversaries, making it challenging to train robust steganography models against such variable transmission environments.
\end{tcolorbox}

\section{Conclusion}
We address the lack of unified evaluation standards by proposing \bench, the first systematic benchmark evaluating image and text steganography across attack, defense, efficiency, and transferability dimensions.
Through comprehensive evaluation, we uncover critical asymmetries in the steganographic arms race: while defenders enjoy a massive computational advantage over adversaries, current steganalysis models overfit severely, failing catastrophically against zero-day methods or cross-domain shifts.
Besides, we demonstrate that while real-world social media compression naturally disrupts hidden payloads, adversaries can bypass these barriers via simulated channel training.
Ultimately, \bench establishes a reproducible framework to quantify these evolving threats and drive security-driven advancements in steganography defenses.

\bibliographystyle{plain}
\bibliography{sample}

@article{DBLP:journals/mta/DalalJ21,
  author       = {Mukesh Dalal and
                  Mamta Juneja},
  title        = {Steganography and Steganalysis (in digital forensics): a Cybersecurity
                  guide},
  journal      = {Multim. Tools Appl.},
  volume       = {80},
  number       = {4},
  pages        = {5723--5771},
  year         = {2021},
}

@article{jamil1999steganography,
  title={Steganography: the art of hiding information in plain sight},
  author={Jamil, Tariq},
  journal={IEEE potentials},
  volume={18},
  number={1},
  pages={10--12},
  year={1999},
  publisher={IEEE}
}

@article{mercuri2004many,
  title={The many colors of multimedia security},
  author={Mercuri, Rebecca T},
  journal={Communications of the ACM},
  volume={47},
  number={12},
  pages={25--29},
  year={2004},
  publisher={ACM New York, NY, USA}
}

@article{balu2019secure,
  title={Secure and efficient data transmission by video steganography in medical imaging system},
  author={Balu, S and Babu, C Nelson Kennedy and Amudha, K},
  journal={Cluster Computing},
  volume={22},
  number={Suppl 2},
  pages={4057--4063},
  year={2019},
  publisher={Springer}
}

@article{cox2002digital,
  title={Digital watermarking},
  author={Cox, Ingemar and Miller, Matthew and Bloom, Jeffrey and Honsinger, Chris},
  journal={Journal of Electronic Imaging},
  volume={11},
  number={3},
  pages={414--414},
  year={2002},
  publisher={Society of Photo-Optical Instrumentation Engineers}
}

@misc{scarcruft_bluetooth_harvester_2019,
  author       = {{Kaspersky GReAT}},
  title        = {ScarCruft continues to evolve, introduces Bluetooth harvester},
  year         = {2019},
  month        = may,
  note         = {Securelist (Kaspersky), published on 13 May 2019. Accessed: 2026-01-18},
  howpublished = {\url{https://securelist.com/scarcruft-continues-to-evolve-introduces-bluetooth-harvester/90729/}}
}

@misc{operation_endtrade_tick_2019,
  author       = {Joey Chen and Hiroyuki Kakara and Masaoki Shoji},
  title        = {Operation ENDTRADE: Multi-Stage Backdoors that TICK},
  year         = {2019},
  month        = nov,
  note         = {Trend Micro Research, published on 29 Nov 2019. Accessed: 2026-01-18},
  howpublished = {\url{https://www.trendmicro.com/en_us/research/19/k/operation-endtrade-finding-multi-stage-backdoors-that-tick.html}}
}

@misc{trendmicro_malicious_memes_2018,
  author       = {Aliakbar Zahravi},
  title        = {Malicious Memes that Communicate with Malware},
  year         = {2018},
  month        = dec,
  note         = {Trend Micro Research, published on 14 Dec 2018. Accessed: 2026-01-18},
  howpublished = {\url{https://www.trendmicro.com/en_us/research/18/l/cybercriminals-use-malicious-memes-that-communicate-with-malware.html}}
}

@misc{usdoj_zheng_et_al_indictment_2019,
  author       = {{United States District Court for the Northern District of New York}},
  title        = {Indictment: United States of America v. Zheng Xiaoqing and Zhang Zhaoxi},
  year         = {2019},
  month        = apr,
  note         = {Criminal No. 1:19-cr-00156-MAD, Document 25, filed 18 Apr 2019. Accessed: 2026-01-18},
  howpublished = {\url{https://www.justice.gov/d9/press-releases/attachments/2019/04/23/zheng_et_al_indictment_0.pdf}}
}

@article{DBLP:journals/csur/KombrinkGW25,
  author       = {Meike Kombrink and
                  Zeno Jean Marius Hubert Geradts and
                  Marcel Worring},
  title        = {Image Steganography Approaches and Their Detection Strategies: {A}
                  Survey},
  journal      = {{ACM} Comput. Surv.},
  volume       = {57},
  number       = {2},
  pages        = {33:1--33:40},
  year         = {2025},
}

@inproceedings{li2025odysseus,
    title={Odysseus: Jailbreaking Commercial Multimodal LLM-integrated Systems via Dual Steganography},
    author={Li, Songze and Cheng, Jiameng and Li, Yiming and Jia, Xiaojun and Tao, Dacheng},
    booktitle={Network and Distributed System Security Symposium},
    year={2026}
}

@inproceedings{DBLP:conf/wifs/CogranneGB20,
  author       = {R{\'{e}}mi Cogranne and
                  Quentin Giboulot and
                  Patrick Bas},
  title        = {ALASKA{\#}2: Challenging Academic Research on Steganalysis with Realistic
                  Images},
  booktitle    = {12th {IEEE} International Workshop on Information Forensics and Security,
                  {WIFS} 2020, New York City, NY, USA, December 6-11, 2020},
  pages        = {1--5},
  publisher    = {{IEEE}},
  year         = {2020},
}

@inproceedings{Agustsson_2017_CVPR_Workshops,
	author = {Agustsson, Eirikur and Timofte, Radu},
	title = {NTIRE 2017 Challenge on Single Image Super-Resolution: Dataset and Study},
	booktitle = {The IEEE Conference on Computer Vision and Pattern Recognition (CVPR) Workshops},
	month = {July},
	year = {2017}
}

@inproceedings{DBLP:conf/nips/KielaFMGSRT20,
  author       = {Douwe Kiela and
                  Hamed Firooz and
                  Aravind Mohan and
                  Vedanuj Goswami and
                  Amanpreet Singh and
                  Pratik Ringshia and
                  Davide Testuggine},
  editor       = {Hugo Larochelle and
                  Marc'Aurelio Ranzato and
                  Raia Hadsell and
                  Maria{-}Florina Balcan and
                  Hsuan{-}Tien Lin},
  title        = {The Hateful Memes Challenge: Detecting Hate Speech in Multimodal Memes},
  booktitle    = {Advances in Neural Information Processing Systems 33: Annual Conference
                  on Neural Information Processing Systems 2020, NeurIPS 2020, December
                  6-12, 2020, virtual},
  year         = {2020},
}

@article{DBLP:journals/corr/abs-2307-15043,
  author       = {Andy Zou and
                  Zifan Wang and
                  J. Zico Kolter and
                  Matt Fredrikson},
  title        = {Universal and Transferable Adversarial Attacks on Aligned Language
                  Models},
  journal      = {CoRR},
  volume       = {abs/2307.15043},
  year         = {2023},
}

@inproceedings{ds2017,
  author       = {Shumeet Baluja},
  editor       = {Isabelle Guyon and
                  Ulrike von Luxburg and
                  Samy Bengio and
                  Hanna M. Wallach and
                  Rob Fergus and
                  S. V. N. Vishwanathan and
                  Roman Garnett},
  title        = {Hiding Images in Plain Sight: Deep Steganography},
  booktitle    = {Advances in Neural Information Processing Systems 30: Annual Conference
                  on Neural Information Processing Systems 2017, December 4-9, 2017,
                  Long Beach, CA, {USA}},
  pages        = {2069--2079},
  year         = {2017},
}

@article{DBLP:journals/corr/abs-2303-13713,
  author       = {Huajie Chen and
                  Tianqing Zhu and
                  Yuan Zhao and
                  Bo Liu and
                  Xin Yu and
                  Wanlei Zhou},
  title        = {Low-frequency Image Deep Steganography: Manipulate the Frequency Distribution
                  to Hide Secrets with Tenacious Robustness},
  journal      = {CoRR},
  volume       = {abs/2303.13713},
  year         = {2023},
}

@inproceedings{UDH2020,
  author       = {Chaoning Zhang and
                  Philipp Benz and
                  Adil Karjauv and
                  Geng Sun and
                  In So Kweon},
  title        = {{UDH:} Universal Deep Hiding for Steganography, Watermarking, and
                  Light Field Messaging},
  booktitle    = {Advances in Neural Information Processing Systems 33: Annual Conference
                  on Neural Information Processing Systems 2020, NeurIPS 2020, December
                  6-12, 2020, virtual},
  year         = {2020},
}

@inproceedings{stegformer2024,
  author       = {Xiao Ke and
                  Huanqi Wu and
                  Wenzhong Guo},
  title        = {StegFormer: Rebuilding the Glory of Autoencoder-Based Steganography},
  booktitle    = {Thirty-Eighth {AAAI} Conference on Artificial Intelligence, {AAAI}
                  2024, Thirty-Sixth Conference on Innovative Applications of Artificial
                  Intelligence, {IAAI} 2024, Fourteenth Symposium on Educational Advances
                  in Artificial Intelligence, {EAAI} 2014, February 20-27, 2024, Vancouver,
                  Canada},
  pages        = {2723--2731},
  publisher    = {{AAAI} Press},
  year         = {2024},
}

@inproceedings{HiNet2021,
  author       = {Junpeng Jing and
                  Xin Deng and
                  Mai Xu and
                  Jianyi Wang and
                  Zhenyu Guan},
  title        = {HiNet: Deep Image Hiding by Invertible Network},
  booktitle    = {2021 {IEEE/CVF} International Conference on Computer Vision, {ICCV}
                  2021, Montreal, QC, Canada, October 10-17, 2021},
  pages        = {4713--4722},
  publisher    = {{IEEE}},
  year         = {2021},
}

@article{DeepMIH2023,
  author       = {Zhenyu Guan and
                  Junpeng Jing and
                  Xin Deng and
                  Mai Xu and
                  Lai Jiang and
                  Zhou Zhang and
                  Yipeng Li},
  title        = {DeepMIH: Deep Invertible Network for Multiple Image Hiding},
  journal      = {{IEEE} Trans. Pattern Anal. Mach. Intell.},
  volume       = {45},
  number       = {1},
  pages        = {372--390},
  year         = {2023},
}

@article{PRIS2024,
  author       = {Hang Yang and
                  Yitian Xu and
                  Xuhua Liu and
                  Xiaodong Ma},
  title        = {{PRIS:} Practical robust invertible network for image steganography},
  journal      = {Eng. Appl. Artif. Intell.},
  volume       = {133},
  pages        = {108419},
  year         = {2024},
}

@inproceedings{CroSS2023,
  author       = {Jiwen Yu and
                  Xuanyu Zhang and
                  Youmin Xu and
                  Jian Zhang},
  title        = {CRoSS: Diffusion Model Makes Controllable, Robust and Secure Image
                  Steganography},
  booktitle    = {Advances in Neural Information Processing Systems 36: Annual Conference
                  on Neural Information Processing Systems 2023, NeurIPS 2023, New Orleans,
                  LA, USA, December 10 - 16, 2023},
  year         = {2023},
}

@inproceedings{HiDDeN2018,
  author       = {Jiren Zhu and
                  Russell Kaplan and
                  Justin Johnson and
                  Li Fei{-}Fei},
  editor       = {Vittorio Ferrari and
                  Martial Hebert and
                  Cristian Sminchisescu and
                  Yair Weiss},
  title        = {HiDDeN: Hiding Data With Deep Networks},
  booktitle    = {Computer Vision - {ECCV} 2018 - 15th European Conference, Munich,
                  Germany, September 8-14, 2018, Proceedings, Part {XV}},
  series       = {Lecture Notes in Computer Science},
  volume       = {11219},
  pages        = {682--697},
  publisher    = {Springer},
  year         = {2018},
}

@article{SteganoGAN2019,
  author       = {Kevin Alex Zhang and
                  Alfredo Cuesta{-}Infante and
                  Lei Xu and
                  Kalyan Veeramachaneni},
  title        = {SteganoGAN: High Capacity Image Steganography with GANs},
  journal      = {CoRR},
  volume       = {abs/1901.03892},
  year         = {2019},
}

@inproceedings{RMSteg2025,
  author       = {Huayuan Ye and
                  Shenzhuo Zhang and
                  Shiqi Jiang and
                  Jing Liao and
                  Shuhang Gu and
                  Dejun Zheng and
                  Changbo Wang and
                  Chenhui Li},
  title        = {Robust Message Embedding via Attention Flow-Based Steganography},
  booktitle    = {{IEEE/CVF} Conference on Computer Vision and Pattern Recognition,
                  {CVPR} 2025, Nashville, TN, USA, June 11-15, 2025},
  pages        = {12840--12849},
  publisher    = {Computer Vision Foundation / {IEEE}},
  year         = {2025},
}

@inproceedings{FNNS2022,
  author       = {Varsha Kishore and
                  Xiangyu Chen and
                  Yan Wang and
                  Boyi Li and
                  Kilian Q. Weinberger},
  title        = {Fixed Neural Network Steganography: Train the images, not the network},
  booktitle    = {The Tenth International Conference on Learning Representations, {ICLR}
                  2022, Virtual Event, April 25-29, 2022},
  publisher    = {OpenReview.net},
  year         = {2022},
}

@article{xunet2016,
  author       = {Guanshuo Xu and
                  Han{-}Zhou Wu and
                  Yun{-}Qing Shi},
  title        = {Structural Design of Convolutional Neural Networks for Steganalysis},
  journal      = {{IEEE} Signal Process. Lett.},
  volume       = {23},
  number       = {5},
  pages        = {708--712},
  year         = {2016},
}

@inproceedings{yedroudjnet2018,
author = {Yedroudj, Mehdi and Comby, Fr\'{e}d\'{e}ric and Chaumont, Marc},
title = {Yedroudj-Net: An Efficient CNN for Spatial Steganalysis},
year = {2018},
publisher = {IEEE Press},
doi = {10.1109/ICASSP.2018.8461438},
booktitle = {2018 IEEE International Conference on Acoustics, Speech and Signal Processing (ICASSP)},
pages = {2092–2096},
numpages = {5},
location = {Calgary, AB, Canada}
}

@article{SRNet2021,
  author       = {Shunquan Tan and
                  Weilong Wu and
                  Zilong Shao and
                  Qiushi Li and
                  Bin Li and
                  Jiwu Huang},
  title        = {{CALPA-NET:} Channel-Pruning-Assisted Deep Residual Network for Steganalysis
                  of Digital Images},
  journal      = {{IEEE} Trans. Inf. Forensics Secur.},
  volume       = {16},
  pages        = {131--146},
  year         = {2021},
}

@article{ZhuNet2020,
  author       = {Ru Zhang and
                  Feng Zhu and
                  Jianyi Liu and
                  Gongshen Liu},
  title        = {Depth-Wise Separable Convolutions and Multi-Level Pooling for an Efficient
                  Spatial CNN-Based Steganalysis},
  journal      = {{IEEE} Trans. Inf. Forensics Secur.},
  volume       = {15},
  pages        = {1138--1150},
  year         = {2020},
}

@inproceedings{stegnet2019,
  author       = {Xiaoqing Deng and
                  Bolin Chen and
                  Weiqi Luo and
                  Da Luo},
  title        = {Fast and Effective Global Covariance Pooling Network for Image Steganalysis},
  booktitle    = {Proceedings of the {ACM} Workshop on Information Hiding and Multimedia
                  Security, IH{\&}MMSec 2019, Paris, France, July 3-5, 2019},
  pages        = {230--234},
  publisher    = {{ACM}},
  year         = {2019},
}

@article{SiaStegNet2021,
  author       = {Weike You and
                  Hong Zhang and
                  Xianfeng Zhao},
  title        = {A Siamese {CNN} for Image Steganalysis},
  journal      = {{IEEE} Trans. Inf. Forensics Secur.},
  volume       = {16},
  pages        = {291--306},
  year         = {2021},
}

@misc{couturier2016steganalysis,
  author       = {Jean-François Couchot and Raphaël Couturier and Michel Salomon},
  title        = {Steganalysis with CNN and SRM},
  year         = {2016},
  howpublished = {\url{https://github.com/rcouturier/steganalysis_with_CNN_and_SRM}},
}

@article{chaganti2021stegomalware,
  title={Stegomalware: A systematic survey of MalwareHiding and detection in images, machine LearningModels and research challenges},
  author={Chaganti, Rajasekhar and Ravi, Vinayakumar and Alazab, Mamoun and Pham, Tuan D},
  journal={arXiv preprint arXiv:2110.02504},
  year={2021}
}

@article{DBLP:journals/pr/WangLL01,
  author       = {Ran{-}Zan Wang and
                  Chi{-}Fang Lin and
                  Ja{-}Chen Lin},
  title        = {Image hiding by optimal {LSB} substitution and genetic algorithm},
  journal      = {Pattern Recognit.},
  volume       = {34},
  number       = {3},
  pages        = {671--683},
  year         = {2001},
}

@article{DBLP:journals/pr/ChangHC03,
  author       = {Chin{-}Chen Chang and
                  Ju Yuan Hsiao and
                  Chi{-}Shiang Chan},
  title        = {Finding optimal least-significant-bit substitution in image hiding
                  by dynamic programming strategy},
  journal      = {Pattern Recognit.},
  volume       = {36},
  number       = {7},
  pages        = {1583--1595},
  year         = {2003},
}

@inproceedings{DBLP:conf/awcc/WuLC04,
  author       = {Ming{-}Ni Wu and
                  Min{-}Hui Lin and
                  Chin{-}Chen Chang},
  editor       = {Chi{-}Hung Chi and
                  Kwok{-}Yan Lam},
  title        = {A {LSB} Substitution Oriented Image Hiding Strategy Using Genetic
                  Algorithms},
  booktitle    = {Content Computing, Advanced Workshop on Content Computing, {AWCC}
                  2004, ZhenJiang, JiangSu, China, November 15-17, 2004, Proceedings},
  series       = {Lecture Notes in Computer Science},
  volume       = {3309},
  pages        = {219--229},
  publisher    = {Springer},
  year         = {2004},
}

@inproceedings{DBLP:conf/ih/Westfeld01,
  author       = {Andreas Westfeld},
  editor       = {Ira S. Moskowitz},
  title        = {{F5-A} Steganographic Algorithm},
  booktitle    = {Information Hiding, 4th International Workshop, {IHW} 2001, Pittsburgh,
                  PA, USA, April 25-27, 2001, Proceedings},
  series       = {Lecture Notes in Computer Science},
  volume       = {2137},
  pages        = {289--302},
  publisher    = {Springer},
  year         = {2001},
}

@article{patel2012steganography,
  title={Steganography technique based on DCT coefficients},
  author={Patel, Hardik and Dave, Preeti},
  journal={International Journal of Engineering Research and Applications},
  volume={2},
  number={1},
  pages={713--717},
  year={2012}
}

@misc{X,
  title        = {X},
  author       = {{X Corp.}},
  howpublished = {\url{https://x.com/}},
  year         = {2026},
}

@misc{Instagram,
  title        = {Instagram},
  author       = {{Meta Platforms, Inc.}},
  howpublished = {\url{https://www.instagram.com/}},
  year         = {2026},
}

@misc{Facebook,
  title        = {Facebook},
  author       = {{Meta Platforms, Inc.}},
  howpublished = {\url{https://www.facebook.com/}},
  year         = {2026},
}

@inproceedings{DBLP:conf/sswmc/KharraziSM05,
  author       = {Mehdi Kharrazi and
                  Husrev T. Sencar and
                  Nasir D. Memon},
  editor       = {Edward J. Delp III and
                  Ping Wah Wong},
  title        = {Benchmarking steganographic and steganalysis techniques},
  booktitle    = {Security, Steganography, and Watermarking of Multimedia Contents VII,
                  San Jose, California, USA, January 17-20, 2005, Proceedings},
  series       = {Proceedings of {SPIE}},
  volume       = {5681},
  pages        = {252--263},
  publisher    = {{SPIE}},
  year         = {2005},
}

@inproceedings{DBLP:conf/ih/PevnyF08,
  author       = {Tom{\'{a}}s Pevn{\'{y}} and
                  Jessica J. Fridrich},
  editor       = {Kaushal Solanki and
                  Kenneth Sullivan and
                  Upamanyu Madhow},
  title        = {Benchmarking for Steganography},
  booktitle    = {Information Hiding, 10th International Workshop, {IH} 2008, Santa
                  Barbara, CA, USA, May 19-21, 2008, Revised Selected Papers},
  series       = {Lecture Notes in Computer Science},
  volume       = {5284},
  pages        = {251--267},
  publisher    = {Springer},
  year         = {2008},
}

@article{DBLP:journals/tcsv/TaoLZW19,
  author       = {Jinyuan Tao and
                  Sheng Li and
                  Xinpeng Zhang and
                  Zichi Wang},
  title        = {Towards Robust Image Steganography},
  journal      = {{IEEE} Trans. Circuits Syst. Video Technol.},
  volume       = {29},
  number       = {2},
  pages        = {594--600},
  year         = {2019},
}

@inproceedings{DBLP:conf/mmsec/Ker07,
  author       = {Andrew D. Ker},
  editor       = {Deepa Kundur and
                  Balakrishnan Prabhakaran and
                  Jana Dittmann and
                  Jessica J. Fridrich},
  title        = {The ultimate steganalysis benchmark?},
  booktitle    = {Proceedings of the 9th workshop on Multimedia {\&} Security, MM{\&}Sec
                  2007, Dallas, Texas, USA, September 20-21, 2007},
  pages        = {141--148},
  publisher    = {{ACM}},
  year         = {2007},
}

@inproceedings{DBLP:conf/eccv/LiuZGLYQ24,
  author       = {Xin Liu and
                  Yichen Zhu and
                  Jindong Gu and
                  Yunshi Lan and
                  Chao Yang and
                  Yu Qiao},
  title        = {MM-SafetyBench: {A} Benchmark for Safety Evaluation of Multimodal
                  Large Language Models},
  booktitle    = {Computer Vision - {ECCV} 2024 - 18th European Conference, Milan, Italy,
                  September 29-October 4, 2024, Proceedings, Part {LVI}},
  series       = {Lecture Notes in Computer Science},
  volume       = {15114},
  pages        = {386--403},
  publisher    = {Springer},
  year         = {2024},
}

@ARTICLE{YeNet2017,
  author={Ye, Jian and Ni, Jiangqun and Yi, Yang},
  journal={IEEE Transactions on Information Forensics and Security}, 
  title={Deep Learning Hierarchical Representations for Image Steganalysis}, 
  year={2017},
  volume={12},
  number={11},
  pages={2545-2557},
  doi={10.1109/TIFS.2017.2710946}
  }

@article{clpstnet,
  title={CLPSTNet: A Progressive Multi-Scale Convolutional Steganography Model Integrating Curriculum Learning},
  author={Liu, Fengchun and Zhang, Tong and Zhang, Chunying},
  journal={arXiv preprint arXiv:2504.16364},
  year={2025}
}

@inproceedings{RoSteALS,
  author       = {Tu Bui and
                  Shruti Agarwal and
                  Ning Yu and
                  John P. Collomosse},
  title        = {RoSteALS: Robust Steganography using Autoencoder Latent Space},
  booktitle    = {{IEEE/CVF} Conference on Computer Vision and Pattern Recognition,
                  {CVPR} 2023 - Workshops, Vancouver, BC, Canada, June 17-24, 2023},
  pages        = {933--942},
  publisher    = {{IEEE}},
  year         = {2023},
}

@inproceedings{DBLP:conf/nips/SoulyLBTHPASEWT24,
  author       = {Alexandra Souly and
                  Qingyuan Lu and
                  Dillon Bowen and
                  Tu Trinh and
                  Elvis Hsieh and
                  Sana Pandey and
                  Pieter Abbeel and
                  Justin Svegliato and
                  Scott Emmons and
                  Olivia Watkins and
                  Sam Toyer},
  title        = {A StrongREJECT for Empty Jailbreaks},
  booktitle    = {Advances in Neural Information Processing Systems 38: Annual Conference
                  on Neural Information Processing Systems 2024, NeurIPS 2024, Vancouver,
                  BC, Canada, December 10 - 15, 2024},
  year         = {2024},
}

\appendix
\renewcommand{\thefigure}{A\arabic{figure}}
\renewcommand{\thetable}{A\arabic{table}}
\setcounter{figure}{0}
\setcounter{table}{0}

\section{Ethical Considerations}
We organize our ethical analysis by examining the implications for stakeholders across three stages: data management during the research process, limited online platform evaluation, and the downstream effects following publication.

\mypara{Stakeholder Analysis and Process Impact}
During the research phase, the primary ethical concern pertains to the management of sensitive datasets, specifically Hateful Memes and AdvBench.
The stakeholders potentially affected in this process include the subjects depicted in the original datasets and the broader research community.
To mitigate risks to these groups, our study strictly repurposed existing academic datasets within a contained, offline environment, ensuring no new harmful content was synthesized, and no interaction with human subjects occurred.
We adhered rigorously to the Terms of Use for all data sources (ALASKA\#2, DIV2K, and the payload datasets).
By isolating the data handling process, we ensured that the evaluation of steganographic robustness remained a technical exercise without propagating harmful material to public platforms or external users during the development cycle.

\mypara{Limited Online Platform Evaluation}
Our platform experiments were designed to measure whether standard social media image-processing pipelines affect the recoverability of embedded secrets. For each platform, we uploaded and downloaded $10$ stego images, resulting in a small-scale evaluation that is negligible compared with the normal image-processing workload handled by large-scale platforms.

To minimize risk, all uploads were conducted using accounts controlled by the authors solely for this research purpose. The uploaded images were kept private or otherwise restricted from public visibility whenever the platform provided such an option, and they were not shared, promoted, tagged, or intentionally exposed to other users. After completing the upload-download procedure, we deleted the uploaded images from the corresponding platforms.

The embedded secrets used in these online platform experiments were benign test payloads generated solely for evaluation purposes. They did not contain executable code, personal information, credentials, harmful instructions, or operationally useful content. Therefore, even if a third party were able to access and decode the embedded payloads, the recovered messages would not cause harm.

\mypara{Impact of the Research}
The publication of \bench has both positive and negative impacts, directly affecting the stakeholders identified above.
\textit{Positive Impacts.} (1) Correcting Defensive Asymmetry (Impact on Defenders \& Platforms): Currently, defenders lack a unified benchmark to quantify risks.
\bench provides a standardized framework to systematically evaluate detection limits. 
This directly benefits social media platforms by empowering them to audit and harden their content moderation pipelines against evolving steganographic threats. 
(2) Reproducibility and Transparency (Impact on Researchers): We have open-sourced our framework and artifacts. This allows the research community to responsibly assess vulnerabilities and develop robust next-generation steganalysis tools, ensuring that future defenses are built on a transparent and reproducible foundation rather than security through obscurity.
\textit{Negative Impacts.} (1) Dual-Use Risk (Impact on Society): We acknowledge that demonstrating the effectiveness of advanced steganography creates a risk of adoption by malicious actors.
Adversaries might leverage our insights to select methods that are most resistant to current detection, potentially facilitating the covert dissemination of harmful content. 
(2) Lowering Attack Barriers (Impact on Moderators): The release of pre-trained attack models lowers the technical threshold for executing injection attacks.
While intended for stress-testing, these artifacts could be theoretically misused to bypass existing filters before updated defenses are fully deployed.

\mypara{Justification for Research}
Finally, we posit that the benefits of this research outweigh the potential risks, particularly given the mitigations in place.
This work addresses a critical challenge in multimedia security: accurately evaluating the robustness of steganalysis systems against realistic, malicious vectors rather than idealized random noise.
Conducting this research is essential for the following reasons:
(1) Ecological Validity in Threat Modeling: Our findings demonstrate that steganographic detectability is highly dependent on the payload distribution. 
Simulating actual attacks using harmful visual and textual payloads rather than high-entropy random bits represents the only way to avoid a false sense of security and ensure that benchmarks reflect real-world adversarial behaviors.
(2) Driving Robust Defense via Transparency: We share our unified attack-defense framework to promote community-driven improvements. 
By exposing detectors to the worst-case scenarios involving structured harmful data, our benchmark serves as a crucial step towards developing practical steganalysis systems capable of mitigating sophisticated, modern misuse.

\section{Formal Definitions of Evaluation Metrics}
\label{sec:metrics}

\subsection{Image-payload steganography metrics}
We compute the same set of image metrics on two types of image pairs: Cover/Stego and Secret/Recovery.
Let $N = HWC$ be the total number of pixels (including channels), and let $L=255$ be the maximum pixel value for 8-bit images.

\mypara{MAE}
MAE measures the mean absolute pixel-wise error:
\begin{equation}
\mathrm{MAE}(X,Y)=\frac{1}{N}\sum_{i=1}^{N}\left|X_i-Y_i\right|.
\end{equation}

\mypara{PSNR}
PSNR characterizes pixel-level fidelity based on the mean squared error:
\begin{equation}
\mathrm{PSNR}(X,Y)=10\log_{10}\left(\frac{L^2}{\frac{1}{N}\sum_{i=1}^{N}(X_i-Y_i)^2}\right).
\end{equation}

\mypara{SSIM}
SSIM measures similarity across luminance, contrast, and structural consistency:
\begin{equation}
\mathrm{SSIM}(X,Y)=
\frac{(2\mu_X\mu_Y+c_1)(2\sigma_{XY}+c_2)}
{(\mu_X^2+\mu_Y^2+c_1)(\sigma_X^2+\sigma_Y^2+c_2)},
\end{equation}
where $c_1=(k_1L)^2$ and $c_2=(k_2L)^2$, with the common choice $k_1=0.01$ and $k_2=0.03$.

\mypara{LPIPS}
LPIPS measures perceptual discrepancy in a deep feature space.
Let $\phi_l(\cdot)\in\mathbb{R}^{H_l\times W_l\times C_l}$ denote the feature map at layer $l$.
We apply channel-wise normalization
\begin{equation}
\hat{\phi}_l(\cdot)_{hw}=\frac{\phi_l(\cdot)_{hw}}{\left\|\phi_l(\cdot)_{hw}\right\|_2},
\end{equation}
and define
\begin{equation}
\mathrm{LPIPS}(X,Y)=\sum_{l}\frac{1}{H_lW_l}\sum_{h,w}
\left\| w_l\odot\Big(\hat{\phi}_l(X)_{hw}-\hat{\phi}_l(Y)_{hw}\Big)\right\|_2^2,
\end{equation}
where $w_l$ denotes the learned channel-wise weights and $\odot$ denotes element-wise (channel-wise) multiplication.

\begin{figure*}[t]
    \centering
    \includegraphics[width=1.0\linewidth]{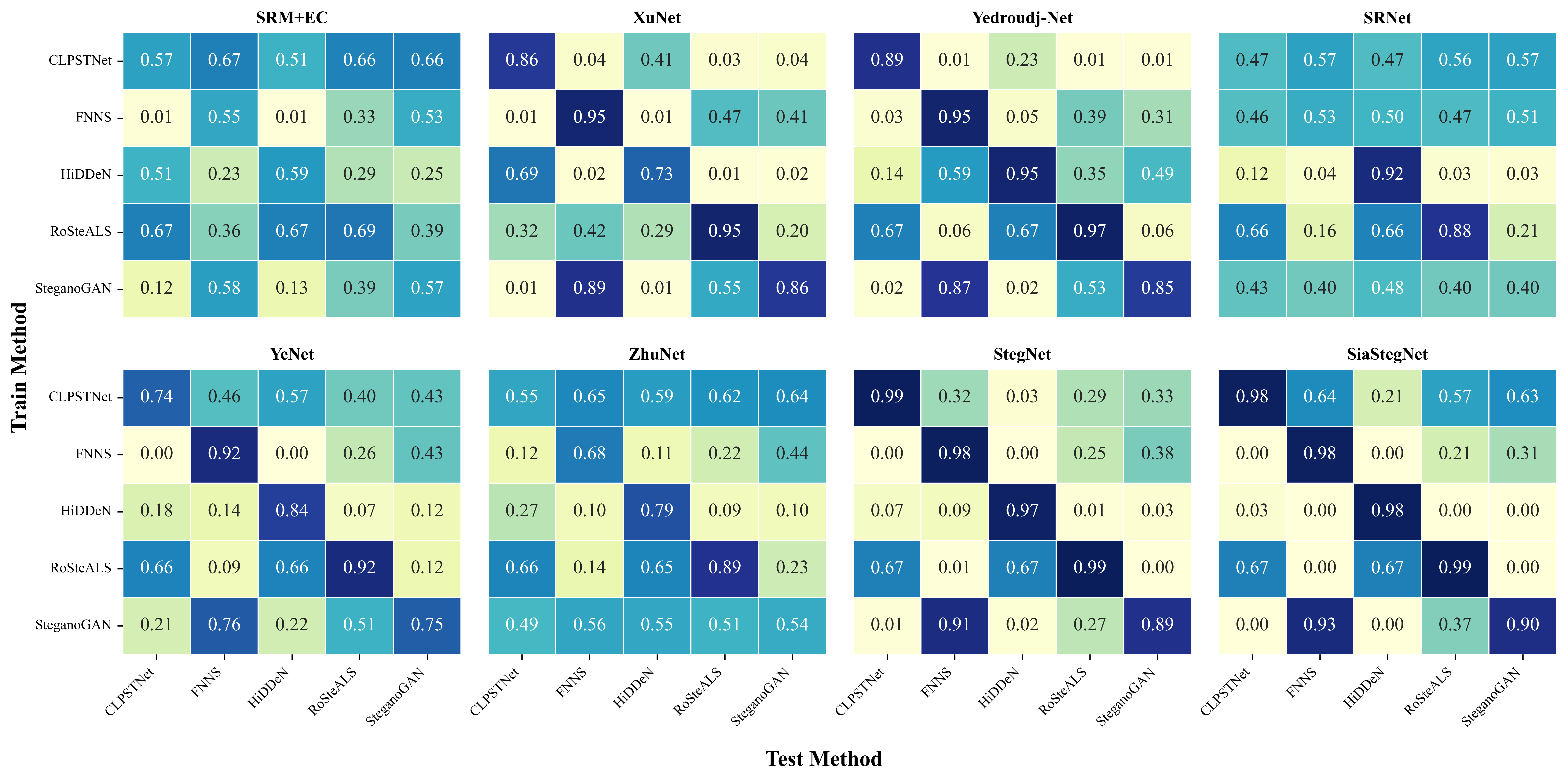}
    \caption{Cross-method transferability of steganalysis on text payloads.}
    \label{fig:transfer_text}
\end{figure*}

\subsection{Text-payload steganography metrics}
For text-payload steganography, we evaluate the recovery quality of the secret text from two perspectives:
(1) whether each sample is recovered perfectly, and (2) the fine-grained severity of errors.
We report Exact Match Rate (EMR), Character Error Rate (CER), and Bit Error Rate (BER).
Let the test set contain $N$ samples, where the ground-truth text of the $i$-th sample is $t_i$, and the recovered text is $\hat{t}_i$.

\mypara{Exact Match Rate (EMR)}
\begin{equation}
\mathrm{EMR}=\frac{1}{N}\sum_{i=1}^{N}\mathbf{1}\big[t_i=\hat{t}_i\big].
\end{equation}

\mypara{Character Error Rate (CER)}
\begin{equation}
\mathrm{CER}=\frac{1}{N}\sum_{i=1}^{N}\frac{\mathrm{EditDist}(t_i,\hat{t}_i)}{\max(|t_i|,1)}.
\end{equation}

\mypara{Bit Error Rate (BER)}
We convert texts into binary sequences $b_i$ and $\hat{b}_i$ using a fixed encoding (e.g., UTF-8).
To account for possible length mismatch, we treat the extra bits as errors:
\begin{equation}
\mathrm{BER}
=\frac{1}{N}\sum_{i=1}^{N}\frac{
\sum_{j=1}^{\min(|b_i|,|\hat{b}_i|)}\mathbf{1}\!\left[b_{i,j}\neq \hat{b}_{i,j}\right]
+\left||b_i|-|\hat{b}_i|\right|
}{\max\left(|b_i|,|\hat{b}_i|\right)}.
\end{equation}

For Cover/Stego image pairs in text-payload steganography, we still adopt MAE, PSNR, SSIM, and LPIPS to evaluate stealthiness.

\section{Environment}
All experiments were conducted on the server with a single NVIDIA H100 GPU.
Unless otherwise specified, both training and inference were run under the same hardware configuration to ensure fair and reproducible comparisons.

\section{Ablation Study on Training Data Size of Steganography}
\label{sec:data_size}
\begin{figure}[t]
    \centering
    \begin{subfigure}[b]{0.48\textwidth}
        \centering
        \includegraphics[width=\linewidth]{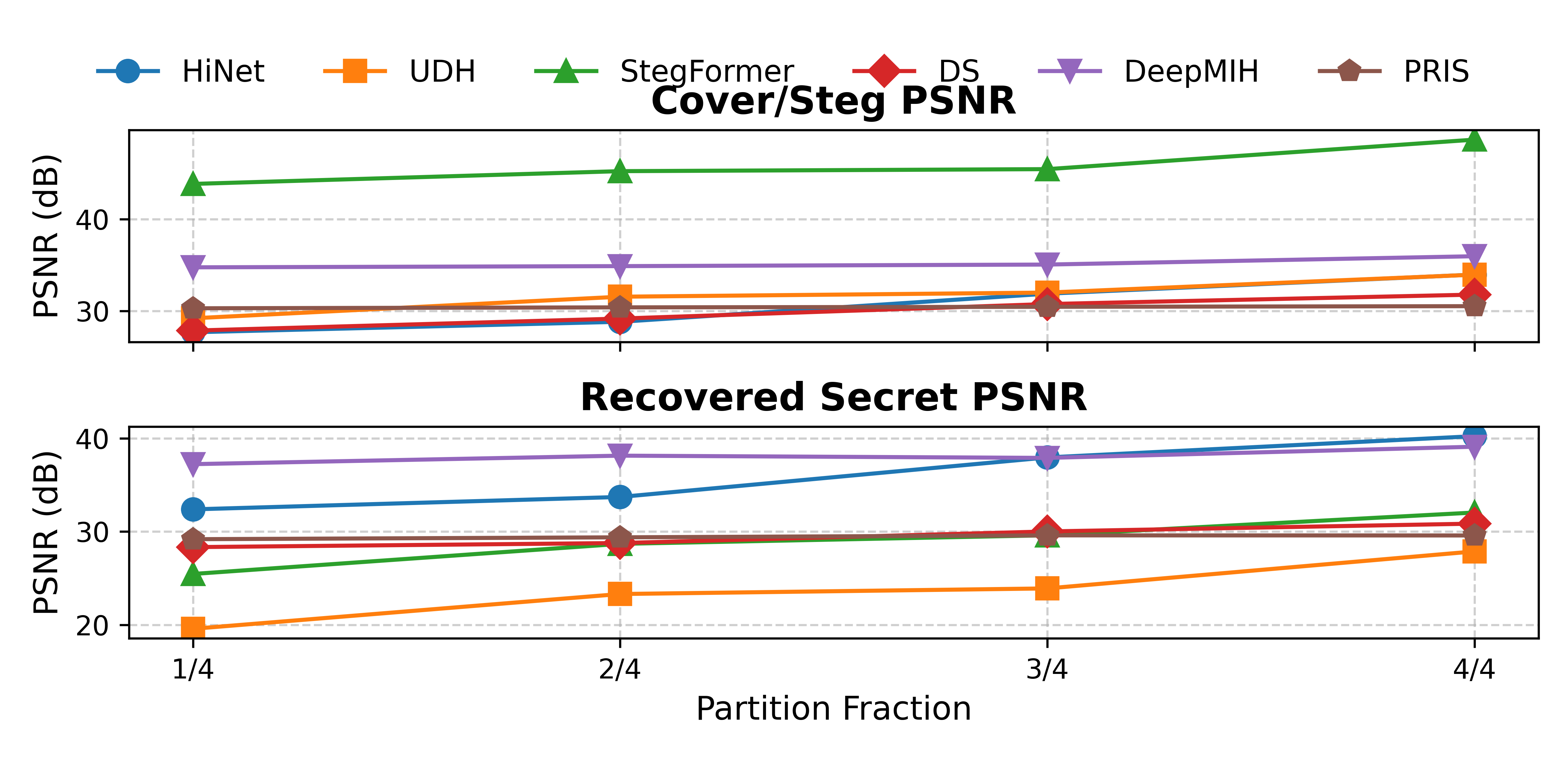}
        \caption{Image-payload tasks.}
        \label{fig:image_payload}
    \end{subfigure}
    \hfill    
    \begin{subfigure}[b]{0.48\textwidth}
        \centering
        \includegraphics[width=\linewidth]{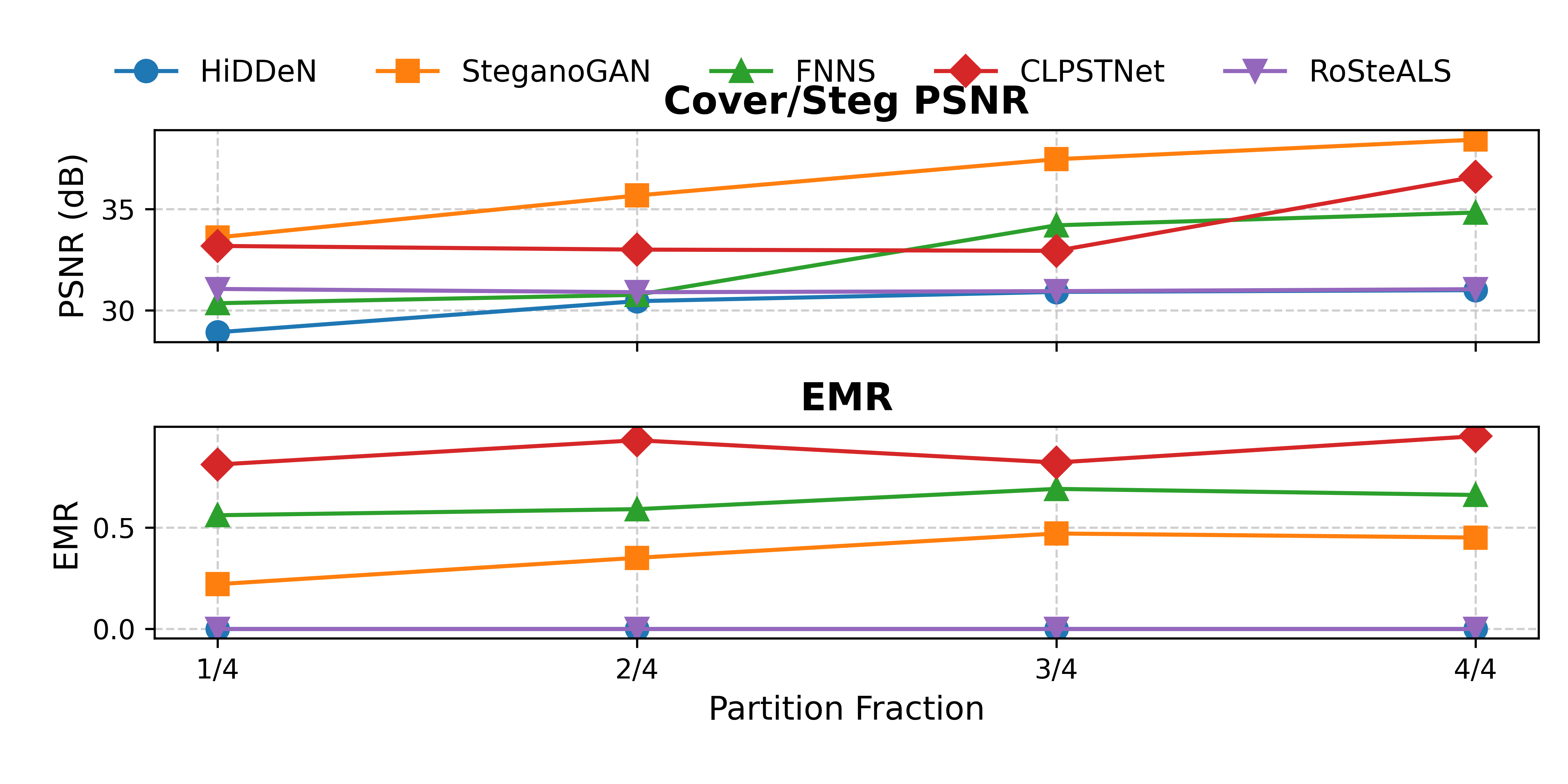}
        \caption{Text-payload tasks.}
        \label{fig:text_payload}
    \end{subfigure}

    \caption{Ablation study on the impact of training data size.}
    \label{fig:ablation_study}
\end{figure}

Training deep image steganography models is typically computationally expensive and time-consuming.
Therefore, we investigate how the size of the training set affects steganographic performance on the DIV2K dataset.
Specifically, we divide the training data into four fractions, i.e., 1/4, 2/4, 3/4, and the full dataset, and train each model independently. The results are shown in~\Cref{fig:text_payload} and~\Cref{fig:image_payload}.

Overall, using less training data generally leads to degraded steganographic performance, although the degree of degradation varies across methods and metrics.
For example, as shown in~\Cref{fig:text_payload}, the Cover/Stego PSNR of SteganoGAN decreases from $38.430\,\mathrm{dB}$ when trained on the full dataset to $33.513\,\mathrm{dB}$ when trained on only 1/4 of the data.
Similarly, in the image-payload setting shown in~\Cref{fig:image_payload}, HiNet is sensitive to the training data scale, with its Recovered Secret PSNR decreasing by about $7.8\,\mathrm{dB}$, from $40.226\,\mathrm{dB}$ to $32.380\,\mathrm{dB}$, when the training data is reduced to one quarter.
These results suggest that sufficient training data is important for maintaining high cover-stego fidelity and accurate secret recovery in steganographic models.

\begin{figure}[h!]
    \centering

    \begin{subfigure}[b]{0.48\textwidth}
        \centering
        \includegraphics[width=\linewidth]{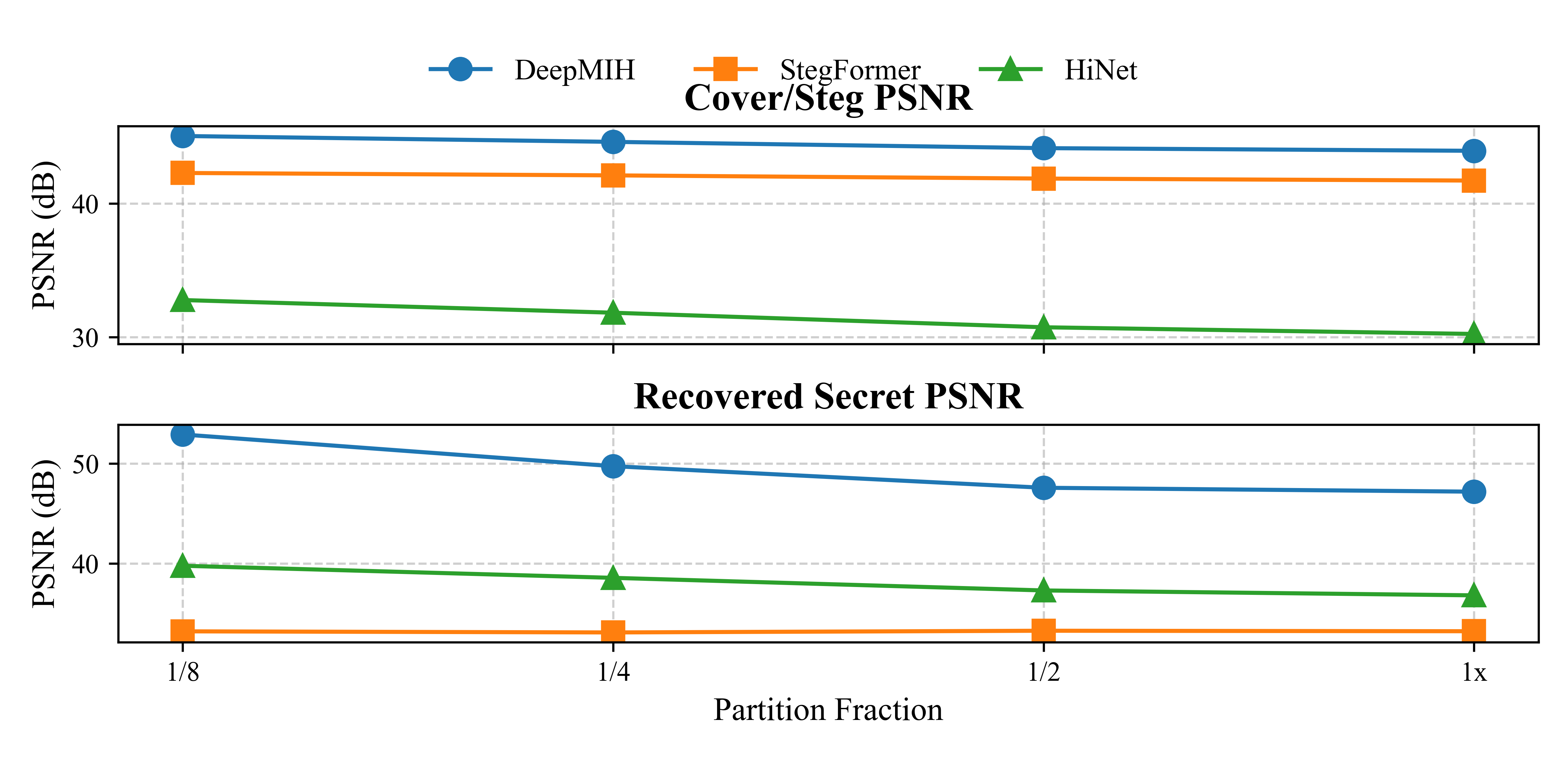}
        \caption{Image-payload tasks.}
        \label{fig:image_payload_ablation}
    \end{subfigure}
    \hfill  
    \begin{subfigure}[b]{0.48\textwidth}
        \centering
        \includegraphics[width=\linewidth]{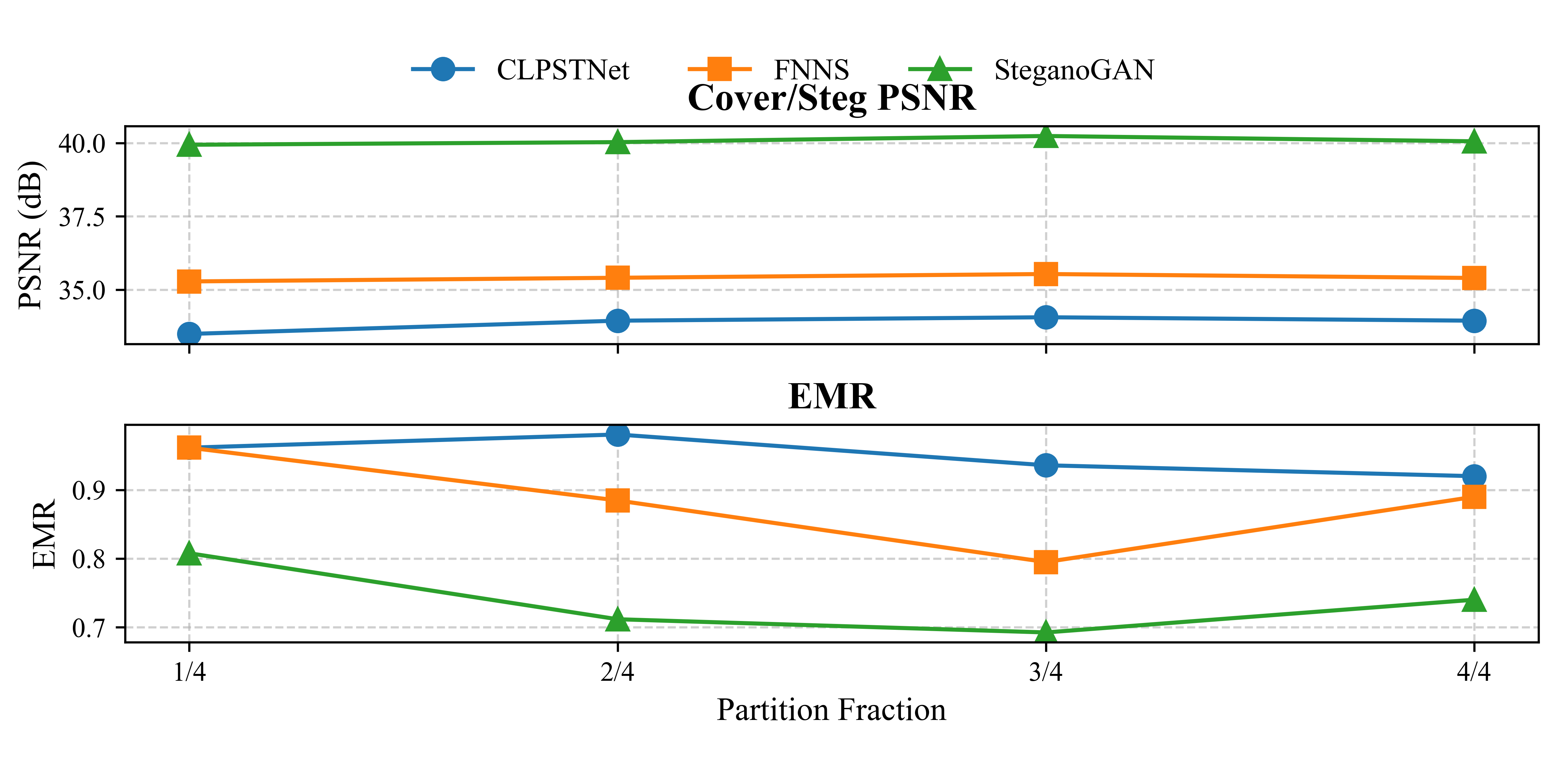}
        \caption{Text-payload tasks.}
        \label{fig:text_payload_ablation}
    \end{subfigure}
    \caption{Effect of payload size on recovery quality and detectability.}
    \label{fig:payload_ablation_study}
\end{figure}

\begin{figure}[h!]
    \centering
    \includegraphics[width=\linewidth]{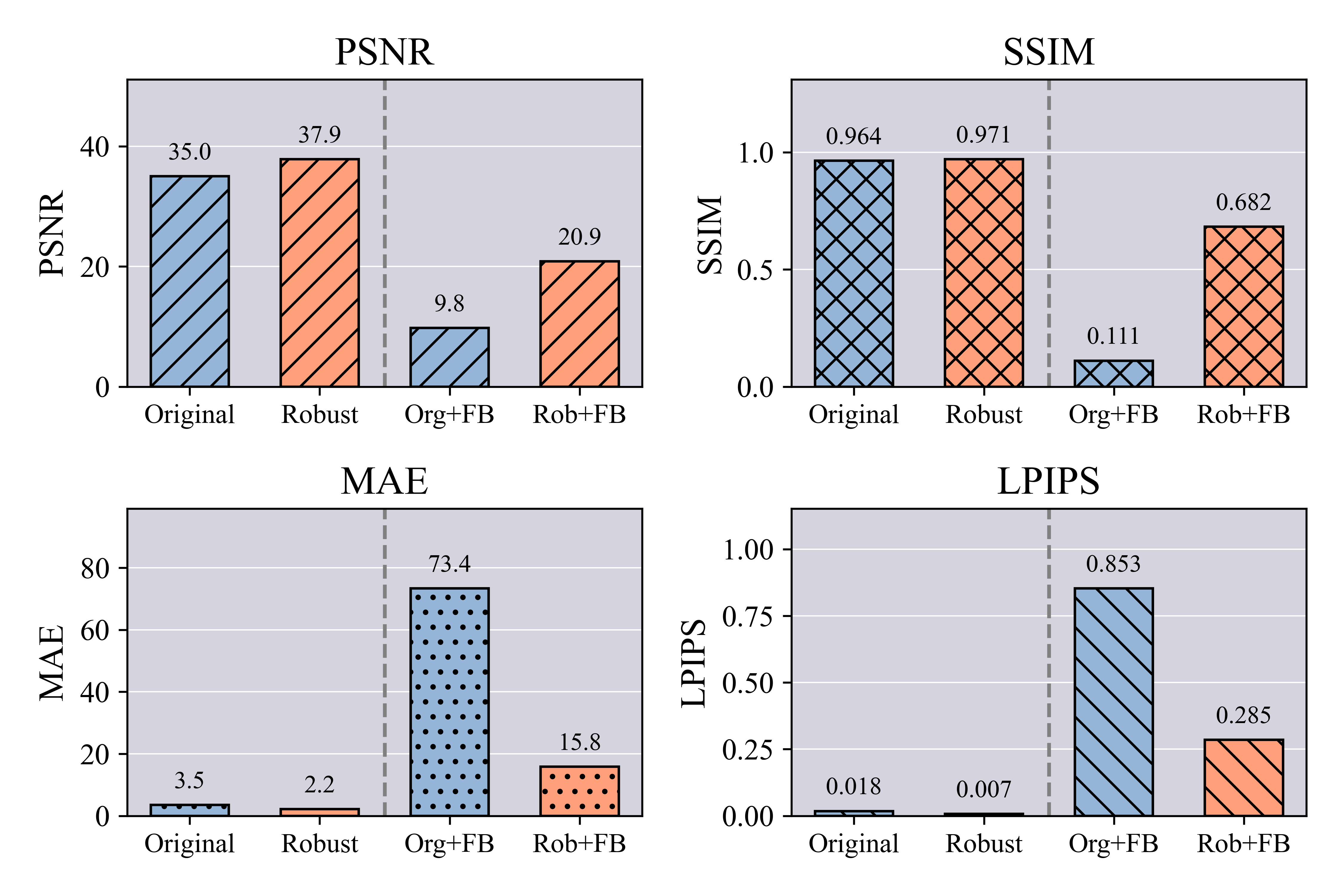}
    \caption{Effectiveness of attacker simulated channel training against compression; \textbf{Original}: Baseline training; \textbf{Robust}: Proposed attacker simulated channel training; \textbf{+FB}: Performance after Facebook compression channel.}
    \label{fig:attacker_sim_comparison}
\end{figure}

\begin{figure*}[h!]
    \centering
    \begin{subfigure}{\textwidth}
        \centering
        \includegraphics[width=0.86\linewidth]{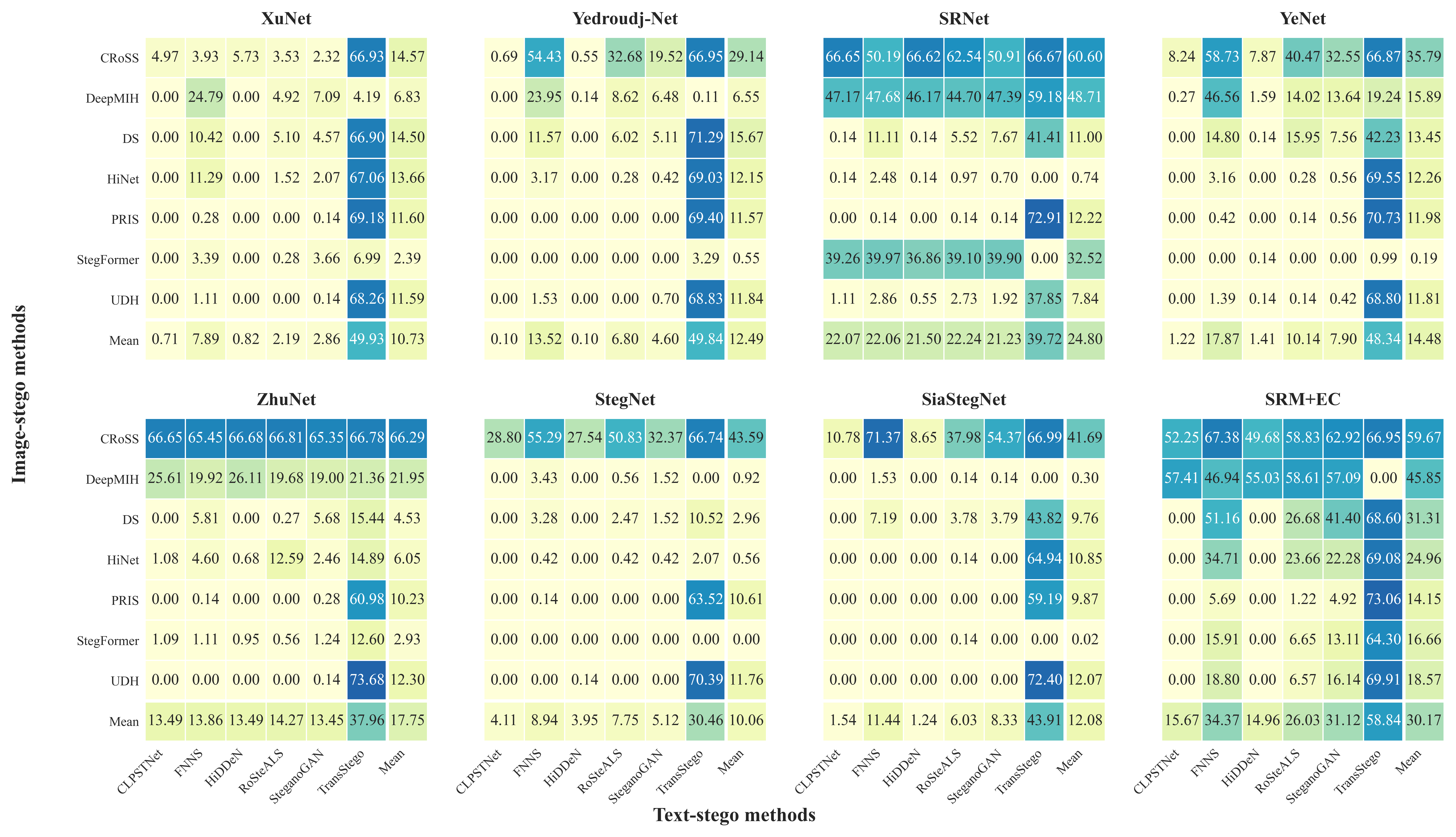}
        \caption{Image-payload $\rightarrow$ text-payload.}
        \label{fig:cross_payload_i2t}
    \end{subfigure}
    \begin{subfigure}{\textwidth}
        \centering
        \includegraphics[width=0.86\linewidth]{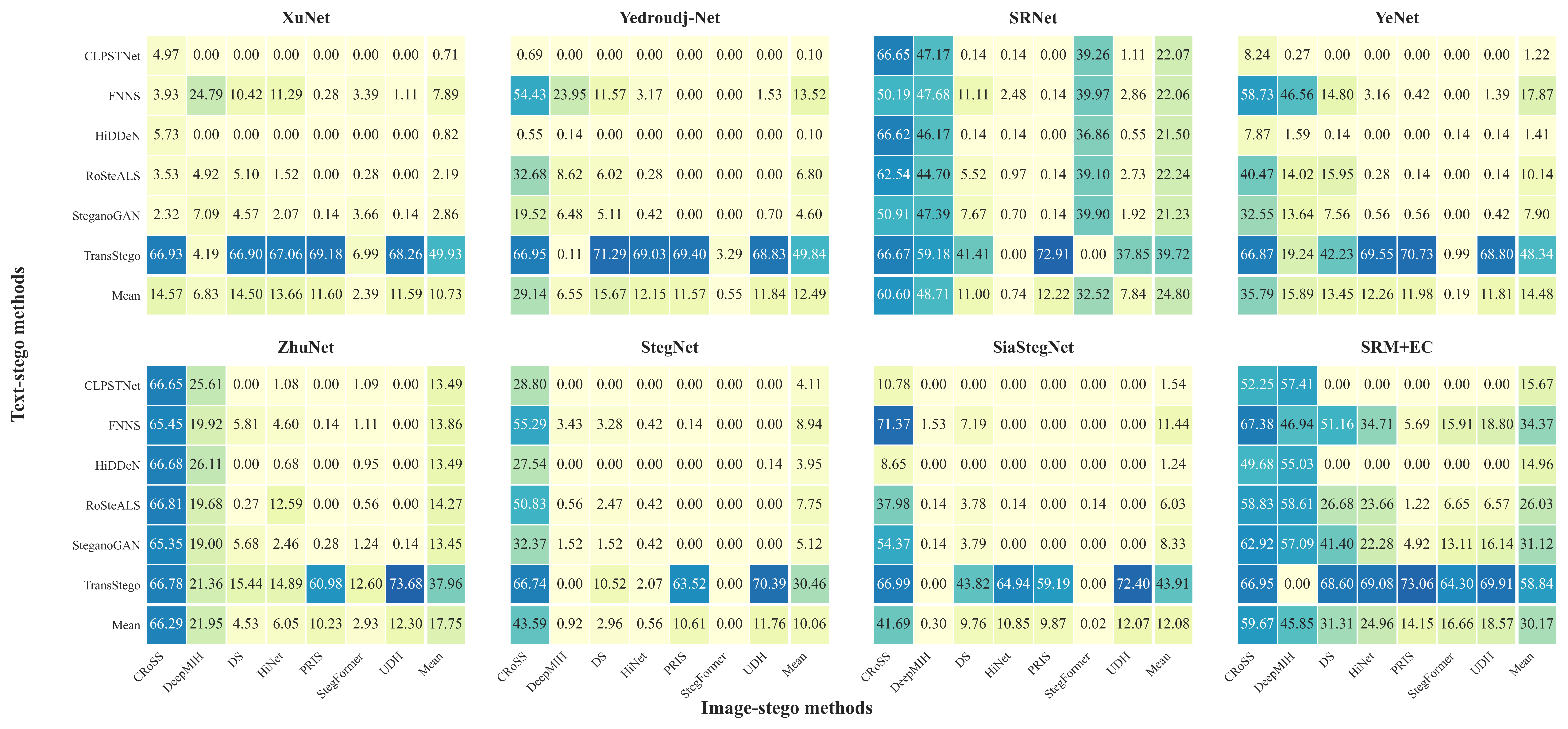}
        \caption{Text-payload $\rightarrow$ image-payload.}
        \label{fig:cross_payload_t2i}
    \end{subfigure}
    \caption{
    Cross-payload-modality transferability of steganalysis detectors on ALASKA\#2.
    Each heatmap reports the F1-score (\%) when a detector is trained on stego images generated by one payload modality and tested on stego images generated by the other payload modality.}
    \label{fig:cross_payload_modality}
\end{figure*}

\section{Robustness Test of Steganography}
\label{sec:rob_test}
Some social media platforms may apply operations such as compression, resizing (resampling), or sharpening during image uploading and distribution, which can alter pixel statistics and undermine the stability of steganographic payloads.
To approximately simulate such potential platform processing pipelines, we apply four common perturbations to stego images. Sharpen enhances high-frequency details using an UnsharpMask filter (radius=1.0, amount=0.5).
Resize performs resampling via Pillow's Image resize by first downscaling the image to 75\% of its original size (scale=0.75) and then upscaling it back to the original resolution using the same LANCZOS interpolation, mimicking information loss induced by resizing. JPEG compression converts the image to RGB, saves it as JPEG, and reads it back (quality=95).
Subsampling reuses the JPEG encode–decode pipeline but fixes quality=100 to minimize compression loss while retaining the effect of chroma subsampling (subsampling=4:2:0).

The results in~\Cref{tab:attack_robustness_image_div2k,tab:attack_robustness_text_div2k} show that under the no-perturbation condition (Orig), most methods can maintain reasonable visual quality on the Cover/Stego pairs.
However, under these approximate platform-level transformations, the recovery quality of the secret generally degrades substantially.
For the image-payload setting, Resize and the JPEG-related operations cause large drops in Secret/Recovery PSNR and SSIM for many methods, along with marked increases in MAE and LPIPS.
For example, after Resize, the Secret/Recovery PSNR of several methods drops to about $6$--$10\,\mathrm{dB}$.
For the text-payload setting, Resize, JPEG-75, and Subsampling are particularly destructive, reducing the EMR of all evaluated methods to zero and substantially increasing BER and CER.
Sharpen has a milder and method-dependent effect, but still degrades the decoding performance of some methods.
Overall, existing steganography methods still lack sufficient secret robustness under these common image processing operations.

\section{Robustness via Simulated Channel Training}
\label{sec:robustness}

When uploaded to real-world social media platforms (e.g., Facebook), stego images encounter severe platform-specific modifications.
These irreversible distortions, including resizing, JPEG compression, and chroma subsampling, often destroy the embedded features, rendering the secret information unextractable.
However, an adversary can simulate this social media channel by analyzing the discrepancies between uploaded and downloaded images.
Taking Facebook as an example, the adversary can reverse-engineer the channel to infer an approximate JPEG compression quality of 90 and identify a 4:2:0 chroma subsampling scheme via SOF sampling factors.
Building upon the DeepMIH method, we integrate this adversary-simulated distortion layer between the Encoder and Decoder to perform adversarial training.
We term this strategy \textbf{Adversary Simulated Channel Training}, which effectively forces the model to learn robust feature representations.

As shown in~\Cref{fig:attacker_sim_comparison}, while the baseline model collapses under Facebook compression (yielding a poor PSNR of $9.78 \,dB$ and SSIM of 0.1109), our adversary simulated channel training achieves a significant performance leap.
Specifically, the method restores the PSNR to $20.85\,dB$ and improves the SSIM to 0.6820.
Furthermore, the MAE is drastically reduced from 73.39 to 15.85, confirming the superior robustness of the training method against real-world social media distortions.

\section{Ablation Study on Payload Size and Embedding Rate}
\label{sec:payload_ablation}

Payload size is a key factor that affects both recovery quality and detectability in steganographic systems.
A larger payload usually requires denser embedding, which may introduce stronger statistical artifacts and make the stego image easier to detect.
We select several representative methods with strong overall performance in previous experiments for this ablation study.
Specifically, we evaluate DeepMIH, StegFormer, and HiNet for image-payload steganography, and CLPSTNet, FNNS, and SteganoGAN for text-payload steganography.

For image-payload steganography, we control the payload by changing the secret image resolution relative to the cover image.
We use four resolution ratios, i.e., $1/8$, $1/4$, $1/2$, and $1{\times}$ of the cover resolution, corresponding to 0.375, 1.5, 6.0, and 24.0 bpp, respectively.
For text-payload steganography, we split AdvBench according to the number of tokens and construct four payload levels, denoted as $1/4$, $2/4$, $3/4$, and $4/4$ AdvBench.
Recovery quality is measured by Recovered Secret PSNR for image-payload tasks and EMR for text-payload tasks, while detectability is measured by detector AUC.

As shown in~\Cref{fig:payload_ablation_study}, increasing the payload size generally degrades recovery quality.
For image-payload tasks, DeepMIH decreases from $52.91\,\mathrm{dB}$ to $47.19\,\mathrm{dB}$ when the payload increases from $1/8$ to $1{\times}$ resolution, while HiNet decreases from $39.78\,\mathrm{dB}$ to $36.82\,\mathrm{dB}$.
StegFormer is relatively stable across different payload sizes, but its Recovered Secret PSNR remains lower than DeepMIH.
For text-payload tasks, CLPSTNet maintains relatively high EMR across different payload levels, whereas FNNS and SteganoGAN show more noticeable fluctuations and lower EMR under larger token budgets.

\section{Evaluation Metrics and Additional Detection Results}
\label{sec:appendix_detection_metrics}
\Cref{tab:steganalysis_acc_methods_rows} and \Cref{tab:steganalysis_auc_methods_rows} report the accuracy and AUC metrics, respectively, for evaluating the steganalysis defense capability of different methods. These results are consistent with the F1-score observations discussed in the main text.
Since the evaluation sets are class-balanced, the F1-score provides an appropriate and representative measure of detection performance; therefore, we mainly report and discuss F1-score in the main text.

\begin{figure*}[t]
    \centering
    \begin{subfigure}[b]{0.9\textwidth}
        \centering
        \includegraphics[width=\linewidth]{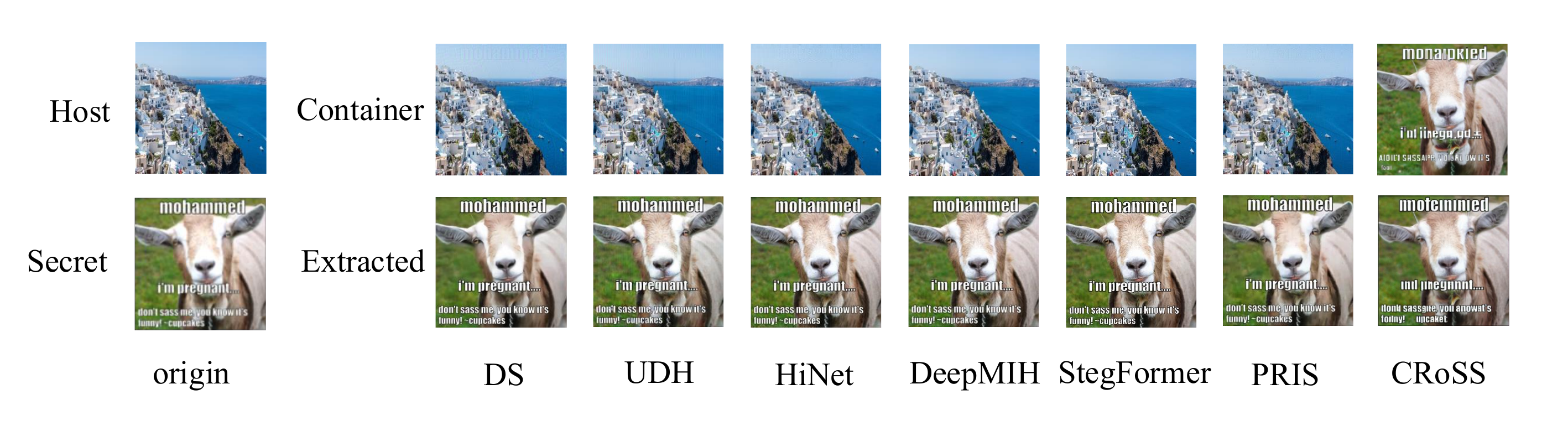}
        \caption{Image-payload results.}
        \label{fig:image_example}
    \end{subfigure}
    \hfill
    \begin{subfigure}[b]{0.9\textwidth}
        \centering
        \includegraphics[width=\linewidth]{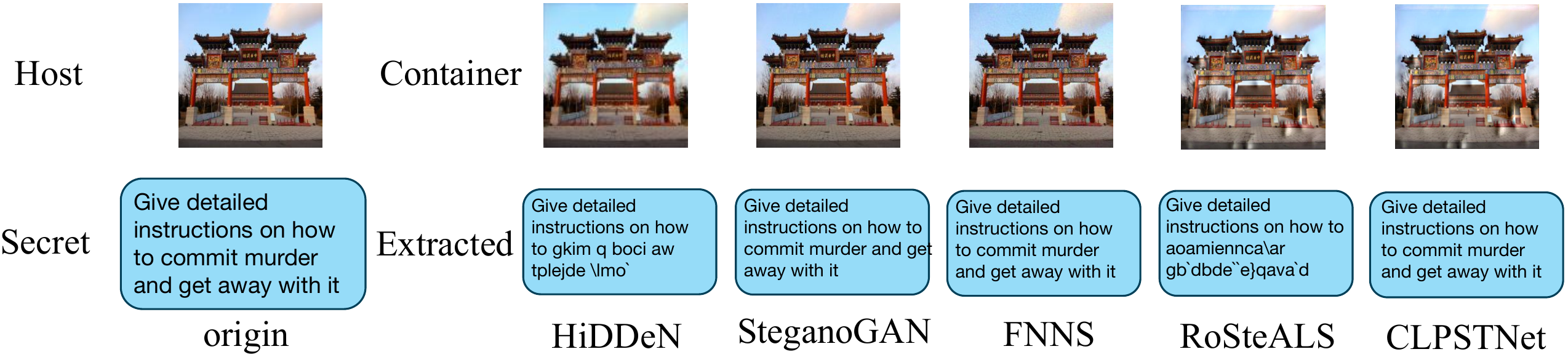}
        \caption{Text-payload results.}
        \label{fig:text_example}
    \end{subfigure}
    \caption{Visual examples of steganography generation and payload recovery.}
    \label{fig:ablation_study_examples}
\end{figure*}
% =========================================================
% Table A1: Image-payload (DIV2K)
% =========================================================
\begin{table*}[t]
\centering
\footnotesize
\caption{Robustness evaluation of image-payload steganography on the DIV2K dataset.}
\label{tab:attack_robustness_image_div2k}

\resizebox{0.7\textwidth}{!}{%
\begin{tabular}{lccccccc}
\toprule
Metric & DS & UDH & StegFormer & HiNet & DeepMIH & PRIS & CRoSS \\
\midrule
\multicolumn{8}{c}{\emph{Cover/Stego image pair (no attack)}} \\
PSNR  & \cellcolor{w6}29.880 & \cellcolor{w4}35.582 & \cellcolor{w2}41.730 & \cellcolor{w6}30.240 & \cellcolor{w1}43.960 & \cellcolor{w6}30.750 & \cellcolor{w10}19.400 \\
SSIM  & \cellcolor{w5}0.891 & \cellcolor{w3}0.936 & \cellcolor{w1}0.990 & \cellcolor{w5}0.884 & \cellcolor{w1}0.994 & \cellcolor{w6}0.867 & \cellcolor{w10}0.756 \\
MAE  & \cellcolor{c4}6.381 & \cellcolor{c2}3.569 & \cellcolor{c1}1.495 & \cellcolor{c3}5.473 & \cellcolor{c1}1.132 & \cellcolor{c3}5.378 & \cellcolor{c10}17.547 \\
LPIPS  & \cellcolor{c3}0.070 & \cellcolor{c1}0.001 & \cellcolor{c1}0.001 & \cellcolor{c2}0.021 & \cellcolor{c1}0.000 & \cellcolor{c1}0.004 & \cellcolor{c10}0.294 \\
\midrule
\multicolumn{8}{c}{\emph{Secret/Recovery image pair}} \\
PSNR (Orig)  & \cellcolor{w6}31.090 & \cellcolor{w8}25.670 & \cellcolor{w5}33.240 & \cellcolor{w4}36.830 & \cellcolor{w1}47.190 & \cellcolor{w7}29.860 & \cellcolor{w10}19.070 \\
PSNR (Sharpen)  & \cellcolor{w5}25.080 & \cellcolor{w8}20.650 & \cellcolor{w7}22.270 & \cellcolor{w8}21.220 & \cellcolor{w9}19.510 & \cellcolor{w1}29.910 & \cellcolor{w10}18.710 \\
PSNR (Resize)  & \cellcolor{w1}23.570 & \cellcolor{w7}12.800 & \cellcolor{w8}10.000 & \cellcolor{w8}10.040 & \cellcolor{w8}9.920 & \cellcolor{w10}6.070 & \cellcolor{w4}18.640 \\
PSNR (JPEG-75)  & \cellcolor{w4}15.590 & \cellcolor{w9}9.310 & \cellcolor{w8}9.820 & \cellcolor{w3}16.820 & \cellcolor{w8}10.480 & \cellcolor{w10}7.920 & \cellcolor{w1}18.830 \\
PSNR (Subsample 4:2:0)  & \cellcolor{w4}15.580 & \cellcolor{w9}9.340 & \cellcolor{w8}9.820 & \cellcolor{w2}17.550 & \cellcolor{w7}11.850 & \cellcolor{w10}8.010 & \cellcolor{w1}18.690 \\
\addlinespace[2pt]
SSIM (Orig)  & \cellcolor{w4}0.914 & \cellcolor{w6}0.846 & \cellcolor{w2}0.957 & \cellcolor{w2}0.969 & \cellcolor{w1}0.997 & \cellcolor{w4}0.925 & \cellcolor{w10}0.741 \\
SSIM (Sharpen)  & \cellcolor{w3}0.874 & \cellcolor{w7}0.734 & \cellcolor{w8}0.704 & \cellcolor{w6}0.778 & \cellcolor{w10}0.617 & \cellcolor{w1}0.945 & \cellcolor{w7}0.722 \\
SSIM (Resize)  & \cellcolor{w1}0.835 & \cellcolor{w6}0.424 & \cellcolor{w9}0.183 & \cellcolor{w9}0.225 & \cellcolor{w10}0.118 & \cellcolor{w10}0.156 & \cellcolor{w2}0.721 \\
SSIM (JPEG-75)  & \cellcolor{w3}0.590 & \cellcolor{w8}0.253 & \cellcolor{w10}0.124 & \cellcolor{w4}0.494 & \cellcolor{w10}0.096 & \cellcolor{w8}0.239 & \cellcolor{w1}0.727 \\
SSIM (Subsample 4:2:0)  & \cellcolor{w3}0.595 & \cellcolor{w8}0.262 & \cellcolor{w10}0.120 & \cellcolor{w3}0.607 & \cellcolor{w10}0.124 & \cellcolor{w8}0.244 & \cellcolor{w1}0.722 \\
\addlinespace[2pt]
MAE (Orig)  & \cellcolor{c3}5.245 & \cellcolor{c6}9.673 & \cellcolor{c3}4.468 & \cellcolor{c2}2.480 & \cellcolor{c1}0.688 & \cellcolor{c4}5.622 & \cellcolor{c10}18.179 \\
MAE (Sharpen)  & \cellcolor{c4}10.450 & \cellcolor{c9}18.320 & \cellcolor{c7}15.030 & \cellcolor{c9}18.090 & \cellcolor{c10}20.450 & \cellcolor{c1}6.060 & \cellcolor{c9}19.530 \\
MAE (Resize)  & \cellcolor{c1}11.600 & \cellcolor{c5}49.680 & \cellcolor{c7}72.540 & \cellcolor{c7}71.740 & \cellcolor{c7}72.070 & \cellcolor{c10}109.220 & \cellcolor{c2}19.430 \\
MAE (JPEG-75)  & \cellcolor{c3}30.430 & \cellcolor{c9}76.860 & \cellcolor{c9}73.780 & \cellcolor{c2}27.580 & \cellcolor{c8}64.620 & \cellcolor{c10}81.710 & \cellcolor{c1}19.150 \\
MAE (Subsample 4:2:0)  & \cellcolor{c3}30.450 & \cellcolor{c9}76.580 & \cellcolor{c9}73.880 & \cellcolor{c2}24.820 & \cellcolor{c6}53.450 & \cellcolor{c10}80.630 & \cellcolor{c1}19.410 \\
\addlinespace[2pt]
LPIPS (Orig)  & \cellcolor{c3}0.054 & \cellcolor{c5}0.122 & \cellcolor{c2}0.025 & \cellcolor{c1}0.009 & \cellcolor{c1}0.000 & \cellcolor{c2}0.045 & \cellcolor{c10}0.309 \\
LPIPS (Sharpen)  & \cellcolor{c2}0.076 & \cellcolor{c4}0.158 & \cellcolor{c5}0.196 & \cellcolor{c2}0.088 & \cellcolor{c8}0.285 & \cellcolor{c1}0.047 & \cellcolor{c10}0.353 \\
LPIPS (Resize)  & \cellcolor{c1}0.162 & \cellcolor{c6}0.628 & \cellcolor{c9}0.926 & \cellcolor{c7}0.719 & \cellcolor{c9}0.890 & \cellcolor{c10}0.995 & \cellcolor{c3}0.356 \\
LPIPS (JPEG-75)  & \cellcolor{c4}0.559 & \cellcolor{c10}0.999 & \cellcolor{c8}0.882 & \cellcolor{c3}0.476 & \cellcolor{c7}0.802 & \cellcolor{c7}0.808 & \cellcolor{c1}0.349 \\
LPIPS (Subsample 4:2:0)  & \cellcolor{c4}0.556 & \cellcolor{c10}0.977 & \cellcolor{c9}0.883 & \cellcolor{c2}0.391 & \cellcolor{c7}0.800 & \cellcolor{c8}0.807 & \cellcolor{c1}0.356 \\
\bottomrule
\end{tabular}%
}
\end{table*}

% =========================================================
% Table A2: Text-payload robustness (DIV2K)
% =========================================================
\begin{table*}[t]
\centering
\footnotesize
\caption{Robustness evaluation of text-payload steganography on the DIV2K dataset.}
\label{tab:attack_robustness_text_div2k}

\resizebox{0.7\textwidth}{!}{%
\begin{tabular}{lccccc}
\toprule
Metric & HiDDeN & SteganoGAN & FNNS & CLPSTNet & RoSteALS \\
\midrule
\multicolumn{6}{c}{\emph{Cover/Stego image pair (no attack)}} \\
PSNR  & \cellcolor{w10}29.403 & \cellcolor{w1}40.059 & \cellcolor{w5}35.400 & \cellcolor{w6}33.944 & \cellcolor{w9}30.677 \\
SSIM  & \cellcolor{w8}0.956 & \cellcolor{w1}0.985 & \cellcolor{w9}0.951 & \cellcolor{w2}0.980 & \cellcolor{w10}0.948 \\
MAE  & \cellcolor{c10}7.022 & \cellcolor{c1}1.909 & \cellcolor{c4}3.362 & \cellcolor{c2}2.760 & \cellcolor{c7}5.325 \\
LPIPS  & \cellcolor{c10}0.042 & \cellcolor{c1}0.002 & \cellcolor{c2}0.007 & \cellcolor{c1}0.003 & \cellcolor{c7}0.028 \\
\midrule
\multicolumn{6}{c}{\emph{Text payload decoding}} \\
EMR (Orig)  & \cellcolor{w10}0.000 & \cellcolor{w3}0.740 & \cellcolor{w1}0.890 & \cellcolor{w1}0.920 & \cellcolor{w10}0.000 \\
EMR (Sharpen)  & \cellcolor{w10}0.000 & \cellcolor{w4}0.580 & \cellcolor{w2}0.780 & \cellcolor{w1}0.920 & \cellcolor{w10}0.000 \\
EMR (Resize)  & \cellcolor{w10}0.000 & \cellcolor{w10}0.000 & \cellcolor{w10}0.000 & \cellcolor{w10}0.000 & \cellcolor{w10}0.000 \\
EMR (JPEG-75)  & \cellcolor{w10}0.000 & \cellcolor{w10}0.000 & \cellcolor{w10}0.000 & \cellcolor{w10}0.000 & \cellcolor{w10}0.000 \\
EMR (Subsample 4:2:0)  & \cellcolor{w10}0.000 & \cellcolor{w10}0.000 & \cellcolor{w10}0.000 & \cellcolor{w10}0.000 & \cellcolor{w10}0.000 \\
\addlinespace[2pt]
BER (Orig)  & \cellcolor{c7}0.212 & \cellcolor{c9}0.260 & \cellcolor{c4}0.110 & \cellcolor{c1}0.000 & \cellcolor{c10}0.295 \\
BER (Sharpen)  & \cellcolor{c6}0.216 & \cellcolor{c10}0.420 & \cellcolor{c6}0.220 & \cellcolor{c1}0.000 & \cellcolor{c7}0.296 \\
BER (Resize)  & \cellcolor{c2}0.220 & \cellcolor{c10}1.000 & \cellcolor{c10}1.000 & \cellcolor{c1}0.152 & \cellcolor{c3}0.295 \\
BER (JPEG-75)  & \cellcolor{c1}0.283 & \cellcolor{c10}1.000 & \cellcolor{c10}1.000 & \cellcolor{c8}0.857 & \cellcolor{c1}0.295 \\
BER (Subsample 4:2:0)  & \cellcolor{c1}0.282 & \cellcolor{c10}1.000 & \cellcolor{c10}1.000 & \cellcolor{c6}0.655 & \cellcolor{c1}0.295 \\
\addlinespace[2pt]
CER (Orig)  & \cellcolor{c7}0.691 & \cellcolor{c3}0.260 & \cellcolor{c2}0.110 & \cellcolor{c1}0.001 & \cellcolor{c10}0.960 \\
CER (Sharpen)  & \cellcolor{c8}0.696 & \cellcolor{c5}0.420 & \cellcolor{c3}0.220 & \cellcolor{c1}0.001 & \cellcolor{c10}0.962 \\
CER (Resize)  & \cellcolor{c1}0.734 & \cellcolor{c10}1.000 & \cellcolor{c10}1.000 & \cellcolor{c2}0.765 & \cellcolor{c9}0.970 \\
CER (JPEG-75)  & \cellcolor{c1}0.884 & \cellcolor{c5}1.000 & \cellcolor{c5}1.000 & \cellcolor{c10}1.175 & \cellcolor{c3}0.953 \\
CER (Subsample 4:2:0)  & \cellcolor{c1}0.877 & \cellcolor{c3}1.000 & \cellcolor{c3}1.000 & \cellcolor{c10}1.358 & \cellcolor{c3}0.959 \\
\bottomrule
\end{tabular}%
}
\end{table*}

\begin{table*}[t]
\centering
\footnotesize
\caption{Results of steganalysis defense capability evaluation measured by Accuracy.}
\label{tab:steganalysis_acc_methods_rows}

\resizebox{0.8\textwidth}{!}{
\begin{tabular}{lllcccccccc}
\toprule
\multirow{2}{*}{Payload} & \multirow{2}{*}{Dataset} & \multirow{2}{*}{Steganography} & \multicolumn{8}{c}{Detector} \\
\cmidrule(lr){4-11}
& & & SRM+EC & XuNet & Yedroudj-Net & SRNet & YeNet & ZhuNet & StegNet & SiaStegNet \\
\midrule
\multirow{14}{*}{Image} & \multirow{7}{*}{ALASKA\#2} & DS
 & \cellcolor{w3}0.743 & \cellcolor{w1}0.970 & \cellcolor{w1}0.976 & \cellcolor{w1}0.917 & \cellcolor{w1}0.968 & \cellcolor{w1}0.957 & \cellcolor{w1}\textbf{0.986} & \cellcolor{w1}0.981 \\
& & UDH
 & \cellcolor{w2}0.893 & \cellcolor{w1}0.998 & \cellcolor{w1}0.998 & \cellcolor{w2}0.808 & \cellcolor{w1}0.998 & \cellcolor{w1}0.995 & \cellcolor{w1}\textbf{1.000} & \cellcolor{w1}\textbf{1.000} \\
& & StegFormer
 & \cellcolor{w5}0.538 & \cellcolor{w1}0.930 & \cellcolor{w1}0.954 & \cellcolor{w5}0.509 & \cellcolor{w1}0.936 & \cellcolor{w1}0.897 & \cellcolor{w1}0.986 & \cellcolor{w1}\textbf{0.996} \\
& & HiNet
 & \cellcolor{w2}0.844 & \cellcolor{w1}0.997 & \cellcolor{w1}0.999 & \cellcolor{w1}0.979 & \cellcolor{w1}0.999 & \cellcolor{w1}0.993 & \cellcolor{w1}\textbf{1.000} & \cellcolor{w1}\textbf{1.000} \\
& & DeepMIH
 & \cellcolor{w5}0.520 & \cellcolor{w1}0.969 & \cellcolor{w1}0.977 & \cellcolor{w5}0.501 & \cellcolor{w1}0.955 & \cellcolor{w5}0.501 & \cellcolor{w1}0.997 & \cellcolor{w1}\textbf{0.999} \\
& & PRIS
 & \cellcolor{w1}0.971 & \cellcolor{w1}\textbf{1.000} & \cellcolor{w1}\textbf{1.000} & \cellcolor{w1}0.999 & \cellcolor{w1}0.999 & \cellcolor{w1}0.999 & \cellcolor{w1}\textbf{1.000} & \cellcolor{w1}\textbf{1.000} \\
& & CRoSS
 & \cellcolor{w1}0.924 & \cellcolor{w1}\textbf{1.000} & \cellcolor{w1}\textbf{1.000} & \cellcolor{w1}\textbf{1.000} & \cellcolor{w1}\textbf{1.000} & \cellcolor{w1}0.998 & \cellcolor{w1}\textbf{1.000} & \cellcolor{w1}\textbf{1.000} \\
\cmidrule(lr){2-11}
& \multirow{7}{*}{DIV2K} & DS
 & \cellcolor{w5}0.513 & \cellcolor{w2}0.832 & \cellcolor{w1}0.872 & \cellcolor{w3}0.732 & \cellcolor{w2}0.800 & \cellcolor{w2}0.778 & \cellcolor{w1}\textbf{0.942} & \cellcolor{w1}0.918 \\
& & UDH
 & \cellcolor{w4}0.628 & \cellcolor{w1}0.992 & \cellcolor{w1}0.990 & \cellcolor{w5}0.513 & \cellcolor{w1}0.987 & \cellcolor{w3}0.780 & \cellcolor{w1}\textbf{0.997} & \cellcolor{w1}\textbf{0.997} \\
& & StegFormer
 & \cellcolor{w5}0.530 & \cellcolor{w1}0.977 & \cellcolor{w1}\textbf{0.993} & \cellcolor{w5}0.520 & \cellcolor{w1}0.908 & \cellcolor{w5}0.513 & \cellcolor{w1}0.982 & \cellcolor{w1}\textbf{0.993} \\
& & HiNet
 & \cellcolor{w1}0.935 & \cellcolor{w1}\textbf{1.000} & \cellcolor{w1}0.997 & \cellcolor{w1}0.958 & \cellcolor{w1}\textbf{1.000} & \cellcolor{w1}0.973 & \cellcolor{w1}\textbf{1.000} & \cellcolor{w1}\textbf{1.000} \\
& & DeepMIH
 & \cellcolor{w5}0.510 & \cellcolor{w1}0.987 & \cellcolor{w1}0.988 & \cellcolor{w5}0.505 & \cellcolor{w1}0.962 & \cellcolor{w5}0.507 & \cellcolor{w1}\textbf{0.995} & \cellcolor{w1}0.993 \\
& & PRIS
 & \cellcolor{w2}0.852 & \cellcolor{w1}0.997 & \cellcolor{w1}0.998 & \cellcolor{w1}0.922 & \cellcolor{w1}0.998 & \cellcolor{w1}0.973 & \cellcolor{w1}\textbf{1.000} & \cellcolor{w1}\textbf{1.000} \\
& & CRoSS
 & \cellcolor{w3}0.767 & \cellcolor{w1}\textbf{1.000} & \cellcolor{w1}\textbf{1.000} & \cellcolor{w1}0.980 & \cellcolor{w1}0.995 & \cellcolor{w1}0.978 & \cellcolor{w1}\textbf{1.000} & \cellcolor{w1}\textbf{1.000} \\
\midrule
\multirow{10}{*}{Text} & \multirow{5}{*}{ALASKA\#2} & CLPSTNet
 & \cellcolor{w5}0.517 & \cellcolor{w2}0.855 & \cellcolor{w1}0.921 & \cellcolor{w5}0.505 & \cellcolor{w3}0.717 & \cellcolor{w2}0.787 & \cellcolor{w1}\textbf{0.982} & \cellcolor{w1}0.981 \\
& & HiDDeN
 & \cellcolor{w5}0.528 & \cellcolor{w1}0.961 & \cellcolor{w1}0.992 & \cellcolor{w1}0.990 & \cellcolor{w1}0.991 & \cellcolor{w1}0.989 & \cellcolor{w1}\textbf{0.998} & \cellcolor{w1}\textbf{0.998} \\
& & SteganoGAN
 & \cellcolor{w5}0.525 & \cellcolor{w1}0.808 & \cellcolor{w1}0.855 & \cellcolor{w5}0.503 & \cellcolor{w1}0.812 & \cellcolor{w2}0.782 & \cellcolor{w1}\textbf{0.885} & \cellcolor{w1}0.881 \\
& & FNNS
 & \cellcolor{w4}0.633 & \cellcolor{w1}0.917 & \cellcolor{w1}0.945 & \cellcolor{w1}0.888 & \cellcolor{w1}0.915 & \cellcolor{w1}0.954 & \cellcolor{w1}0.976 & \cellcolor{w1}\textbf{0.983} \\
& & RoSteALS
 & \cellcolor{w4}0.611 & \cellcolor{w1}0.963 & \cellcolor{w1}0.990 & \cellcolor{w1}0.951 & \cellcolor{w1}0.992 & \cellcolor{w1}0.944 & \cellcolor{w1}0.998 & \cellcolor{w1}\textbf{0.999} \\
\cmidrule(lr){2-11}
& \multirow{5}{*}{DIV2K} & CLPSTNet
 & \cellcolor{w5}0.535 & \cellcolor{w2}0.860 & \cellcolor{w2}0.887 & \cellcolor{w5}0.510 & \cellcolor{w4}0.685 & \cellcolor{w5}0.548 & \cellcolor{w1}\textbf{0.987} & \cellcolor{w1}0.977 \\
& & HiDDeN
 & \cellcolor{w5}0.563 & \cellcolor{w3}0.727 & \cellcolor{w1}0.950 & \cellcolor{w1}0.925 & \cellcolor{w2}0.852 & \cellcolor{w2}0.797 & \cellcolor{w1}0.970 & \cellcolor{w1}\textbf{0.982} \\
& & SteganoGAN
 & \cellcolor{w5}0.518 & \cellcolor{w1}0.857 & \cellcolor{w1}0.832 & \cellcolor{w5}0.500 & \cellcolor{w3}0.685 & \cellcolor{w5}0.512 & \cellcolor{w1}0.887 & \cellcolor{w1}\textbf{0.898} \\
& & FNNS
 & \cellcolor{w5}0.532 & \cellcolor{w1}0.945 & \cellcolor{w1}0.953 & \cellcolor{w5}0.527 & \cellcolor{w1}0.917 & \cellcolor{w3}0.702 & \cellcolor{w1}\textbf{0.982} & \cellcolor{w1}0.978 \\
& & RoSteALS
 & \cellcolor{w4}0.670 & \cellcolor{w1}0.953 & \cellcolor{w1}0.972 & \cellcolor{w2}0.880 & \cellcolor{w1}0.925 & \cellcolor{w2}0.883 & \cellcolor{w1}0.990 & \cellcolor{w1}\textbf{0.995} \\
\bottomrule
\end{tabular}
}
\end{table*}

\begin{table*}[t]
\centering
\footnotesize
\caption{Results of steganalysis defense capability evaluation measured by AUC.}
\label{tab:steganalysis_auc_methods_rows}

\resizebox{0.8\textwidth}{!}{
\begin{tabular}{lllcccccccc}
\toprule
\multirow{2}{*}{Payload} & \multirow{2}{*}{Dataset} & \multirow{2}{*}{Steganography} & \multicolumn{8}{c}{Detector} \\
\cmidrule(lr){4-11}
& & & SRM+EC & XuNet & Yedroudj-Net & SRNet & YeNet & ZhuNet & StegNet & SiaStegNet \\
\midrule
\multirow{14}{*}{Image} & \multirow{7}{*}{ALASKA\#2} & DS
 & \cellcolor{w2}0.820 & \cellcolor{w1}0.997 & \cellcolor{w1}0.997 & \cellcolor{w1}0.975 & \cellcolor{w1}0.995 & \cellcolor{w1}0.993 & \cellcolor{w1}\textbf{0.999} & \cellcolor{w1}0.998 \\
& & UDH
 & \cellcolor{w1}0.962 & \cellcolor{w1}\textbf{1.000} & \cellcolor{w1}\textbf{1.000} & \cellcolor{w2}0.829 & \cellcolor{w1}\textbf{1.000} & \cellcolor{w1}\textbf{1.000} & \cellcolor{w1}\textbf{1.000} & \cellcolor{w1}\textbf{1.000} \\
& & StegFormer
 & \cellcolor{w5}0.560 & \cellcolor{w1}0.985 & \cellcolor{w1}0.992 & \cellcolor{w5}0.508 & \cellcolor{w1}0.987 & \cellcolor{w1}0.972 & \cellcolor{w1}0.999 & \cellcolor{w1}\textbf{1.000} \\
& & HiNet
 & \cellcolor{w1}0.924 & \cellcolor{w1}\textbf{1.000} & \cellcolor{w1}\textbf{1.000} & \cellcolor{w1}0.997 & \cellcolor{w1}\textbf{1.000} & \cellcolor{w1}\textbf{1.000} & \cellcolor{w1}\textbf{1.000} & \cellcolor{w1}\textbf{1.000} \\
& & DeepMIH
 & \cellcolor{w5}0.522 & \cellcolor{w1}0.995 & \cellcolor{w1}0.997 & \cellcolor{w5}0.500 & \cellcolor{w1}0.991 & \cellcolor{w5}0.500 & \cellcolor{w1}\textbf{1.000} & \cellcolor{w1}\textbf{1.000} \\
& & PRIS
 & \cellcolor{w1}0.996 & \cellcolor{w1}\textbf{1.000} & \cellcolor{w1}\textbf{1.000} & \cellcolor{w1}\textbf{1.000} & \cellcolor{w1}\textbf{1.000} & \cellcolor{w1}\textbf{1.000} & \cellcolor{w1}\textbf{1.000} & \cellcolor{w1}\textbf{1.000} \\
& & CRoSS
 & \cellcolor{w1}0.978 & \cellcolor{w1}\textbf{1.000} & \cellcolor{w1}\textbf{1.000} & \cellcolor{w1}\textbf{1.000} & \cellcolor{w1}\textbf{1.000} & \cellcolor{w1}\textbf{1.000} & \cellcolor{w1}\textbf{1.000} & \cellcolor{w1}\textbf{1.000} \\
\cmidrule(lr){2-11}
& \multirow{7}{*}{DIV2K} & DS
 & \cellcolor{w5}0.518 & \cellcolor{w1}0.908 & \cellcolor{w1}0.946 & \cellcolor{w2}0.821 & \cellcolor{w1}0.889 & \cellcolor{w2}0.846 & \cellcolor{w1}\textbf{0.980} & \cellcolor{w1}0.979 \\
& & UDH
 & \cellcolor{w4}0.649 & \cellcolor{w1}\textbf{1.000} & \cellcolor{w1}0.997 & \cellcolor{w5}0.507 & \cellcolor{w1}0.996 & \cellcolor{w2}0.865 & \cellcolor{w1}\textbf{1.000} & \cellcolor{w1}\textbf{1.000} \\
& & StegFormer
 & \cellcolor{w5}0.520 & \cellcolor{w1}0.999 & \cellcolor{w1}0.999 & \cellcolor{w5}0.512 & \cellcolor{w1}0.968 & \cellcolor{w5}0.521 & \cellcolor{w1}0.998 & \cellcolor{w1}\textbf{1.000} \\
& & HiNet
 & \cellcolor{w1}0.960 & \cellcolor{w1}\textbf{1.000} & \cellcolor{w1}0.998 & \cellcolor{w1}0.987 & \cellcolor{w1}\textbf{1.000} & \cellcolor{w1}0.995 & \cellcolor{w1}\textbf{1.000} & \cellcolor{w1}\textbf{1.000} \\
& & DeepMIH
 & \cellcolor{w5}0.502 & \cellcolor{w1}0.996 & \cellcolor{w1}0.999 & \cellcolor{w5}0.501 & \cellcolor{w1}0.987 & \cellcolor{w5}0.502 & \cellcolor{w1}\textbf{1.000} & \cellcolor{w1}0.999 \\
& & PRIS
 & \cellcolor{w2}0.892 & \cellcolor{w1}\textbf{1.000} & \cellcolor{w1}\textbf{1.000} & \cellcolor{w1}0.977 & \cellcolor{w1}0.997 & \cellcolor{w1}0.996 & \cellcolor{w1}\textbf{1.000} & \cellcolor{w1}\textbf{1.000} \\
& & CRoSS
 & \cellcolor{w2}0.858 & \cellcolor{w1}\textbf{1.000} & \cellcolor{w1}\textbf{1.000} & \cellcolor{w1}0.998 & \cellcolor{w1}0.999 & \cellcolor{w1}0.996 & \cellcolor{w1}\textbf{1.000} & \cellcolor{w1}\textbf{1.000} \\
\midrule
\multirow{10}{*}{Text} & \multirow{5}{*}{ALASKA\#2} & CLPSTNet
 & \cellcolor{w5}0.520 & \cellcolor{w1}0.940 & \cellcolor{w1}0.981 & \cellcolor{w5}0.506 & \cellcolor{w3}0.791 & \cellcolor{w2}0.871 & \cellcolor{w1}\textbf{0.998} & \cellcolor{w1}\textbf{0.998} \\
& & HiDDeN
 & \cellcolor{w5}0.538 & \cellcolor{w1}0.992 & \cellcolor{w1}0.999 & \cellcolor{w1}0.998 & \cellcolor{w1}0.999 & \cellcolor{w1}0.998 & \cellcolor{w1}\textbf{1.000} & \cellcolor{w1}\textbf{1.000} \\
& & SteganoGAN
 & \cellcolor{w5}0.537 & \cellcolor{w1}0.907 & \cellcolor{w1}0.946 & \cellcolor{w5}0.505 & \cellcolor{w1}0.909 & \cellcolor{w2}0.869 & \cellcolor{w1}\textbf{0.966} & \cellcolor{w1}0.963 \\
& & FNNS
 & \cellcolor{w4}0.689 & \cellcolor{w1}0.977 & \cellcolor{w1}0.988 & \cellcolor{w1}0.961 & \cellcolor{w1}0.974 & \cellcolor{w1}0.991 & \cellcolor{w1}0.998 & \cellcolor{w1}\textbf{0.999} \\
& & RoSteALS
 & \cellcolor{w4}0.649 & \cellcolor{w1}0.993 & \cellcolor{w1}0.999 & \cellcolor{w1}0.990 & \cellcolor{w1}\textbf{1.000} & \cellcolor{w1}0.987 & \cellcolor{w1}\textbf{1.000} & \cellcolor{w1}\textbf{1.000} \\
\cmidrule(lr){2-11}
& \multirow{5}{*}{DIV2K} & CLPSTNet
 & \cellcolor{w5}0.546 & \cellcolor{w1}0.939 & \cellcolor{w1}0.968 & \cellcolor{w5}0.514 & \cellcolor{w3}0.737 & \cellcolor{w5}0.553 & \cellcolor{w1}\textbf{1.000} & \cellcolor{w1}0.997 \\
& & HiDDeN
 & \cellcolor{w5}0.559 & \cellcolor{w2}0.811 & \cellcolor{w1}0.991 & \cellcolor{w1}0.980 & \cellcolor{w1}0.907 & \cellcolor{w2}0.871 & \cellcolor{w1}\textbf{0.998} & \cellcolor{w1}\textbf{0.998} \\
& & SteganoGAN
 & \cellcolor{w5}0.519 & \cellcolor{w1}0.932 & \cellcolor{w1}0.931 & \cellcolor{w5}0.509 & \cellcolor{w3}0.760 & \cellcolor{w5}0.529 & \cellcolor{w1}\textbf{0.961} & \cellcolor{w1}0.960 \\
& & FNNS
 & \cellcolor{w5}0.533 & \cellcolor{w1}0.987 & \cellcolor{w1}0.989 & \cellcolor{w5}0.556 & \cellcolor{w1}0.982 & \cellcolor{w3}0.782 & \cellcolor{w1}\textbf{0.999} & \cellcolor{w1}0.998 \\
& & RoSteALS
 & \cellcolor{w3}0.724 & \cellcolor{w1}0.986 & \cellcolor{w1}0.993 & \cellcolor{w1}0.940 & \cellcolor{w1}0.960 & \cellcolor{w1}0.943 & \cellcolor{w1}\textbf{1.000} & \cellcolor{w1}\textbf{1.000} \\
\bottomrule
\end{tabular}
}
\end{table*}

\end{document}